\documentclass{cccg24}

\usepackage{graphicx,amssymb,amsmath}
\usepackage[utf8]{inputenc}
\usepackage[T1]{fontenc}
\usepackage{url}
\usepackage{hyperref}

\usepackage{xargs}

\usepackage[pdftex,dvipsnames]{xcolor} 
\usepackage[colorinlistoftodos,prependcaption,textsize=tiny]{todonotes}

\usepackage{pgfplots,tikz,tikz-cd}
\pgfplotsset{compat=1.15}
\usepackage{mathrsfs}
\usepackage{soul}
\usepackage[normalem]{ulem}
\usetikzlibrary{arrows}

\usepackage{nicefrac}
\usepackage{float}
\usepackage{paralist}
\usepackage[all]{xy}
\usepackage{comment}
\usepackage{xspace}
\usepackage{ulem}

\newcommand{\Real}{\mathbb{R}}

\newcommand{\fakef}{\varrho}
\newcommand{\fakeg}{\chi}

\newcommand{\complexsize}{N}

\newcommand{\radmeasure}{\ell_2\text{--radius}}

\usepackage{cite}
\usepackage{csquotes}

\usepackage{sectsty}
\sectionfont{\fontsize{12}{15}\selectfont}

\newcommand\de{\mathrel{\bullet\mkern-2.5mu{\rightarrow}}}

\usepackage{amsthm,mathtools,extpfeil}

\newcommand{\pself}{\phi}
\newcommand{\qself}{\psi}

\newcommand{\pmodule}{\mathbb{P}}

\newcommand{\qmodule}{\mathbb{Q}}

\newcommand{\rhomb}{\mathrm{Rho}}

\newcommand{\interval}{\mathbb{I}}

\newcommand{\hyperarrange}{\mathrm{Arr}}

\newcommand{\fpbarcode}{B_p(\mathcal{F})}

\newcommand{\boldM}{\mathbf{M}}

\newcommand{\boldH}{\mathbf{H}}

\newcommand{\boldW}{\mathbf{W}}

\newcommand{\surjects}{\twoheadrightarrow}

\newcommand{\maxsmplx}{\kappa}

\usepackage{framed}

\newcommand{\filtration}{\mathcal{F}}
\newcommand{\filtrationv}{\mathcal{D}}
\newcommand{\minhombasis}{\mathscr{M}}
\newcommand{\matching}{\mathcal{M}}

\newcommand{\injects}{\xhookrightarrow{}}

\newcommand{\pointsetone}{P}
\newcommand{\pointsettwo}{Q}
\DeclareMathOperator{\Cech}{\check{\mathsf{C}}}
\DeclareMathOperator{\cechfull}{\check{C}{ech}}
\DeclareMathOperator{\id}{id}

\DeclareMathOperator{\Rips}{V}
\DeclareMathOperator{\ripsfull}{Rips}

\DeclareMathOperator{\diam}{diam}

\DeclareMathOperator{\level}{level}

\DeclareMathOperator{\vertexset}{Vert}

\DeclareMathOperator{\low}{low}

\DeclareMathOperator{\high}{high}

\DeclareMathOperator{\middling}{mid}

\newcommand{\PCC}{\mathcal{P}}

\newcommand{\found}{\textnormal{Found}}

\newcommand{\property}{\Pi}

\newcommand{\RBB}{\mathbb{R}}
\newcommand{\NBB}{\mathbb{N}}
\newcommand{\ZBB}{\mathbb{Z}}

\newcommand{\eps}{\varepsilon}
\newcommand{\dd}{\mathrm d}

\newcommand{\hombasis}{\mathcal{H}}

\newcommand{\represents}{\mathcal{R}}

\newcommand{\ibd}{{\interval}^{[b,d)}}

\newcommand{\complex}{\mathsf{K}}

\newcommand{\subcomplex}{\mathsf{L}}

\newcommand{\barcode}{\mathcal{B}}

\newcommand{\toomuch}{\text{end}}

\newcommand{\pointset}{P}

\newcommand{\startshere}{\text{start}}

\newcommand{\sphere}[2]{S_{{#1},{#2}}}

\newcommand{\ssphere}{S}

\newcommand{\iinterval}{\mathbb{I}^{\mathbf{i}}}

\newcommand{\jinterval}{\mathbb{I}^{\mathbf{j}}}

\newcommand{\intervalone}{\mathbb{I}^{[b_1,d_1)}}

\newcommand{\Intervalone}{\mathbb{I}^{\mathbf{i}_1}}

\newcommand{\filfam}{\mathfrak{F}}

\newcommand{\basisset}{\Omega}

\newcommand{\oneinterval}{{[b_1,d_1)}}

\newcommand{\intervaltwo}{\mathbb{I}^{[b_2,d_2)}}

\newcommand{\Intervaltwo}{\mathbb{I}^{\mathbf{i}_2}}

\newcommand{\Intervalthree}{\mathbb{I}^{\mathbf{i}_3}}

\newcommand{\twointerval}{{[b_2,d_2)}}

\newcommand{\threeinterval}{{[b_3,d_3)}}

\newcommand{\MBC}{\textsc{Minimum Bounding Chain}\xspace}

\newcommand{\pMHB}{\textsc{Minimum $p$-Homology Basis}\xspace}

\newcommand{\persMHB}{\textsc{Minimum persistent homology Basis}\xspace}

\newcommand{\boldi}{\mathbf{i}}

\newcommand{\boldj}{\mathbf{j}}

\newcommand{\rhombvertices}[1]{{\mathrm{VR}}_{#1}(P)}

\newcommand{\mycomment}[1]{}
\newcommand{\cancel}[1]

\newcommandx{\unsure}[2][1=]{\todo[linecolor=red,backgroundcolor=red!25,bordercolor=red,#1]{#2}}
\newcommandx{\change}[2][1=]{\todo[linecolor=blue,backgroundcolor=blue!25,bordercolor=blue,#1]{#2}}
\newcommandx{\info}[2][1=]{\todo[linecolor=OliveGreen,backgroundcolor=OliveGreen!25,bordercolor=OliveGreen,#1]{#2}}
\newcommandx{\improvement}[2][1=]{\todo[linecolor=Plum,backgroundcolor=Plum!25,bordercolor=Plum,#1]{#2}}

\newcommand{\abhishek}[1] {{\sf\textcolor{red}{{#1}}}}
\newcommand{\amritendu}[1] {{\sf\textcolor{blue}{{#1}}}}

\usepackage[vlined, ruled,linesnumbered]{algorithm2e}
\usepackage{algpseudocode}

\usepackage{cleveref}

\theoremstyle{plain}
\newtheorem{thm}{Theorem}
\newtheorem{proposition}[theorem]{Proposition}
\newtheorem{corollary}[theorem]{Corollary}
\newtheorem{fact}[theorem]{Fact}

\theoremstyle{definition}
\newtheorem{definition}[theorem]{Definition}

\newtheorem*{proof*}{Proof}

\theoremstyle{remark}
\newtheorem{remark}{Remark}[section]
\newtheorem*{notation*}{Notation}

\usepackage{float}

\usepackage{babel}

\definecolor{dtsfsf}{rgb}{0.8274509803921568,0.1843137254901961,0.1843137254901961}
\definecolor{wrwrwr}{rgb}{0.3803921568627451,0.3803921568627451,0.3803921568627451}
\definecolor{rvwvcq}{rgb}{0.08235294117647059,0.396078431372549,0.7529411764705882}

\title{Geometric Localization of Homology Cycles}

\author{Amritendu Dhar\thanks{Department of Computer Science and Automation, Indian Institute of Science, Bangalore, {\tt amritendud@iisc.ac.in}}
\and 
Vijay Natarajan\thanks{Department of Computer Science and Automation, Indian Institute of Science, Bangalore and Zuse Institute, Berlin, {\tt vijayn@iisc.ac.in}}
\and 
Abhishek Rathod\thanks{Department of Computer Science, Ben Gurion University of the Negev, {\tt arathod@post.bgu.ac.il}}}

\begin{document}
\thispagestyle{empty}

\maketitle

\begin{abstract}
   Computing an optimal cycle in a given homology class, also referred to as the homology localization problem, is known to be an NP-hard problem in general. Furthermore, there is currently no known optimality criterion that localizes
    classes geometrically and admits a stability property under the setting of persistent homology. We present a geometric optimization of the cycles that is computable in polynomial time and is stable in an approximate sense.  Tailoring our search criterion to different settings, we obtain various optimization problems like optimal homologous cycle, minimum homology basis, and minimum persistent homology basis. 
    In practice, the (trivial) exact algorithm is computationally expensive despite having a worst case polynomial runtime. Therefore, we design approximation algorithms for the above problems and study their performance experimentally. These algorithms have reasonable runtimes for moderate sized datasets and the cycles computed by these algorithms are consistently of high quality as demonstrated via experiments on multiple datasets.
\end{abstract}

\section{Introduction}
Homology groups and their persistent version called \emph{persistent homology} play a central role in topological data analysis (TDA), a thriving research field of equal interest to computer scientists, mathematicians and data scientists~\cite{EH10,DW22}. The ranks for homology groups and the barcodes for persistent homology groups have been extensively studied both from algorithmic and mathematical perspectives. With the growth of TDA in applications, there is an increasing need for computing homology cycles that localize given homology classes or constitute a basis for the homology group. Often applications require these cycles to be tightest possible or geometry-aware in some sense rather than being completely oblivious of the embedding space. This demand has led to studying homologous or basis cycles under various optimization criteria.
A number of optimization results in this direction have now appeared in the literature both in persistent and non-persistent settings~\cite{borradaile2017minimum,BMN20,chambers2009minimum,cohen2022lexicographic,chen2011hardness,CPS22,CLV20,dey2011optimal,dey2010approximating,Rathod20}. 

The quality of the optimal cycles depends on the choice of a weight function. For instance, one may choose a weight for each $p$-cycle $\zeta$ to be the sum
of non-negative weights assigned to each $p$-simplex in $\zeta$. 
Optimizing this measure over a class of a given cycle $\zeta$ localizes the class $[\zeta]$ in the sense that it selects a cycle in the class with the least weight. Unfortunately, this problem is known to be NP-hard in general~\cite{chambers2009minimum,chen2011hardness} except
for some special cases~\cite{dey2011optimal,deyperstwo}.
Polynomial time algorithms are known for certain optimization criteria~\cite{chen2010measuring,deyperstwo} or in lower dimensions~\cite{borradaile2017minimum,chambers2009minimum,dey2010approximating,erickson2005greedy}. 

\paragraph*{Outline and Contributions.}
Precisely, we achieve the following. Given a simplicial complex $\complex$ with the vertices in a point set $P\subset \mathbb{R}^d$ and linearly embedded simplices, we define the weight of a  cycle $\zeta$ as the radius of the smallest $(d-1)$-sphere that encloses $\zeta$. This measure, in some sense, captures the locality of $\zeta$ with respect to its geometry.
In \Cref{sec:homloc}, we study how homology localization serves as an archetype application.
Then, we solve other versions of the optimal cycle problem including \emph{minimum homology basis} in \Cref{sec:opt_homologybasis} and \emph{minimum persistent homology basis} in \Cref{sec:phbasis}. For the persistent version, in \Cref{sec:repmatch}, we show optimal persistent homology bases are stable in an approximate sense.
For previous results on optimal persistent cycles~\cite{CLV20,deyperstwo} such stability is not known.
The approximation algorithms described in this paper have been implemented. 
In \Cref{sec:results}, we report experimental results for the approximate algorithms. In our experiments, we found that even the approximate algorithms return cycles of consistently high quality confirming the value of $\radmeasure$ as an optimization criterion.
We further compare experimental results on persistent homology with that of PersLoop \cite{Persloop}, which is a state of the art software for computing optimal persistent 1-cycles.
We visually infer that our cycles are "tighter" than those of PersLoop on multiple datasets of practical importance.

\paragraph*{Related work.}
A criterion related to ours was considered by Chen and Freedman~\cite{chen2010measuring} who proposed to compute a minimum homology basis while optimizing the  shortest path radius of the geodesic balls containing the basis cycles. With an embedding in the Euclidean space, the $\radmeasure$ of the geometric balls capture locality more concisely than the shortest path radius. 
Yet another measures of optimality for cycles that is tractable, namely lexicographically optimality~\cite{CLV20},  suffers from the drawback that it requires a parameter: a total order on simplices.   In applications, it is sometimes desirable that the optimal cycles be stable with regard to the change in the input data~\cite{barbensi2022hypergraphs}. An optimization criterion that is geometry-aware, polynomial time computable, and results in some kind of stability is \emph{volume optimal cycles} by Obayashi~\cite{obayashione,obayashitwo}. However, unlike our measure, the approach described in~\cite{obayashione} works only for computing representatives of finite bars.
 In another related work, Li et al.~\cite{li2021} obtain minimal representatives using linear programming for a variety of optimization criteria with impressive runtimes. However, their software does not work for arbitrary filtrations yet~\cite{pc}.
In summary, our key contribution in this work is that we introduce a natural measure of optimality of cycles that has good theoretical properties and is well-behaved in practice. 


\cancel{
In the last two decades, computational topology has burgeoned from a niche area for specialists to a thriving research field of equal interest to computer scientists, mathematicians and data scientists~\cite{}. The rise of topological methods for data analysis has been concomitant with the concurrent emergence of machine learning and data analysis as one of the principal fields in computer science.
Largely driven by applications of topological methods, there  is an increased demand for good algorithms that translate to good software. For the set of problems that admit polynomial time solutions, there have been some noteworthy  success stories. These include the development of fast algorithms and efficient software for computing single and multiparameter persistent homology~\cite{}, vectorizations of persistent homology, Reeb graphs~\cite{}, mapper~\cite{}, and a wide range of tools for scalar and vector field data analysis~\cite{}.

On the other hand, as in all the other application domains, some of the most fundamental questions in computational topology have natural combinatorial formulations, and phrased this way, many of these questions are NP-hard, NP-hard to approximate within a constant factor, and intractable with respect to natural parameterizations. The most striking examples here are computation of:
\begin{inparaenum}
\item optimal (persistent) homology basis, and 
\item optimal homologous cycle.
\end{inparaenum}

In its usual formulation, the problem of computing optimal $p$-th homology basis can be described as follows: given a weighted simplicial complex, compute 
}

\cancel{Perhaps here we can talk about  optimization problems and how geometry makes problems tractable. This is discussed at length, for instance, in the article by Agarwal et al.~\cite{Agarwalgeom}.}

\cancel{I suggest we name our radius measure as "$\ell_2$-radius". This is to avoid confusion with the radius measure considered by Chen and Freedman, which is the radius defined by the shortest path metric on the $1$-skeleton of $\complex$.}

\section{Background and preliminaries}
\cancel{
\amritendu{Tentative contents of sections, \\ Background : PErsistent Homology,\\
l2 Metric: Definition,\\
Computing optimal homologous cycle problem defn and algorithm, \\ 
Computing minimum homology basis: problem, defn and algorithm \\ 
Computing optimal persistent cycle reprs.: Definition and algorithm \\
Stability  \\
\\ Discussion about the title of the paper \\
Experiments \\
optional. Alternate geodesic ball metric
}
}

In this section, we recall some preliminaries on persistent homology.
For the rest of this section, we work only with simplexwise filtrations: That is, we have a filtration $\mathcal{F}$ on $\mathbb{R}$ where the complexes change
only at finite set of values $a_1<a_2<\ldots<a_n$ and every change involves addition of 
a unique simplex $\sigma_{a_i}$ for $i \in [n]$. 
\[
{\mathcal F}: \emptyset=\complex_{a_0} \stackrel{\sigma_{a_0}}{\hookrightarrow}\complex_{a_1} \stackrel{\sigma_{a_1}}{\hookrightarrow } \complex_{a_2} \stackrel{\sigma_{a_2}}{\hookrightarrow} \dots \stackrel{\sigma_{a_{n-1}}}\hookrightarrow \complex_n=\complex \]

Using $p$-th homology groups of the complexes over the field $\ZBB_2$, we get a sequence 
of vector spaces connected by inclusion-induced linear maps:
\[
{H_p\mathcal F}:   H_p(\complex_{a_0}) \rightarrow H_p(\complex_{a_1}) \rightarrow  H_p(\complex_{a_2}) \rightarrow \dots \]

The sequence $H_p\mathcal F$ with the linear maps is called a \emph{persistence
module}. There is a special persistence module
called
the \emph{interval module} ${\interval}^{[b,d)}$ associated to the interval $[b,d)$. Denoting the vector space indexed at $a\in \Real $ as ${\interval}_a$, this interval module 
is given by 

\begin{equation*} 
{\interval}_a^{[b,d)} = 
\begin{cases} \mathbb{Z}_2 &\text{if $a\in [b,d)$} \\ 
0 &\text{otherwise} 
\end{cases} 
\end{equation*} 

together with identity maps $\id_{a,a'}: {\interval}_a^{[b,d)} \to {\interval}_{a'}^{[b,d)}$
for all $a,a' \in [b,d)$ with $a\leq a'$.

It is known due to a result of Gabriel~\cite{Gabriel72} that
a persistence module defined with finite complexes admits a decomposition
\begin{equation*} \label{eq:decomp}
   H_p{\mathcal F}\cong \bigoplus_\alpha {\interval}^{[b_\alpha,d_\alpha)} 
\end{equation*}
 which is unique up to isomorphism and permutation of the intervals. The intervals $[b_\alpha,d_\alpha)$ are called the
\emph{bars}. The multiset of bars forms the
\emph{barcode} of the persistence module
$H_p\mathcal F$, denoted by $\barcode_p(\mathcal F)$.
The following two definitions are taken from~\cite{deypersone}.
\cancel{
Of course, one can extend such a filtration to entire $\mathbb{Z}$
by assuming all complexes indexed to the left of $0$ to be empty and all complexes to the
right of $n$ to be $\complex$ with no additions at all. Furthermore, one can then extend
the index set from $\mathbb{Z}$ to $\mathbb{R}$ by assuming that $K_{a'}=K_a$ for every $a\in\mathbb{Z}$ and for each
$a'$ in the half-open interval
$[a,b)$ where $(a,b)\in \mathbb{R}\setminus\mathbb{Z}$. We will assume such extension
in next section where we define interleaving between two persistence modules.}

\begin{definition}
For an interval $[b,d)$, we say that  $\zeta$ is a \emph{representative cycle} for $[b,d)$, or  simply  $\zeta$ \emph{represents} $[b,d)$, if  one of the following holds:
\begin{itemize}
\item $d\neq +\infty$, $\zeta$ is a cycle in $\complex_{b}$ containing $\sigma_{b}$, and $\zeta$ is not a boundary in $\complex_{d-1}$ but becomes one in $\complex_{d}$.
\item $d= +\infty$, and $\zeta$ is a cycle in $\complex_{b}$ containing $\sigma_{b}$.
\end{itemize}
\end{definition}

\begin{definition}[Persistent cycles]
\label{defn:pers_cyc_strong}
A $p$-cycle  $\zeta$  that represents an interval $[b,d)\in \barcode_p(\mathcal{F})$ is called a  \emph{persistent $p$-cycle} for $[b,d)$.
\end{definition}

For a bar $[b,d)$, $\sigma_b$ is said to be a \emph{creator simplex} and $\sigma_d$ is called a \emph{destroyer simplex}. 

It is easy to check that if $\zeta$ is a representative cycle for  $[b_i,d_i)$ and $\xi$ is a representative cycle for $ [b_j,d_j)$, where $b_j < b_i$ and $d_j < d_i$, then $\zeta + \xi $ is also a representative cycle for  $[b_i,d_i)$.
The set of representative cycles for interval $[b_i,d_i)$ is denoted by $\represents([b_i,d_i))$.
Representatives of bars of the form $[b,\infty)$ are called \textit{essential cycles}.

\begin{definition}[Persistent basis] \label{defn:pers_basis}
Let $J$ be the indexing set for the intervals in the barcode  $\barcode_p(\mathcal{F})$ of filtration $\mathcal{F}$. 
That is, for every $j\in J$, $[b_j,d_j)$ is an interval in $\barcode_p(\mathcal{F})$.
Then a set of $p$-cycles $\{\zeta_j \mid j\in J\}$ is called a \emph{persistent $p$-basis} for $\mathcal{F}$ if 
\[H_p{\mathcal F} \,\, = \,\, \bigoplus_{j\in J} {\interval}^{\zeta_j} 
\text{ where ${\interval}^{\zeta_j}$ is defined by } \]
\[{\interval}_a^{\zeta_j} = 
\begin{cases} [\zeta_j] &\text{if $a\in [b_j,d_j)$} \\ 
0 &\text{otherwise.} 
\end{cases} \]

Here, for every $j\in J$ and  every $a,a' \in [b_j,d_j)$ with $a\leq a'$ the maps ${\interval}_a^{\zeta_j} \to {\interval}_{a'}^{\zeta_j}$
 are the induced maps on homology restricted to $[\zeta_j]$, respectively.
\end{definition}

The following theorem by Dey et al.~\cite{deypersone} relates persistent cycles to persistent bases. 
\begin{thm}[\protect{\cite[Theorem 1]{deypersone}}]
\label{thm:deypersone}
Let $J$ be the indexing set for the intervals in the barcode  $\barcode_p(\mathcal{F})$ of filtration $\mathcal{F}$.
Then, an indexed set of $p$-cycles $\{ \zeta_j \mid j \in J\}$ is a persistent $p$-basis for a filtration $\mathcal{F}$ if and only if $\zeta_j \in \represents([b_j,d_j))$ for every $j\in J$.
\end{thm}

\label{sec:back}

\cancel{
\section{Generic binary search on the Rhomboid tiling} \label{sec:binsearch}
In this section, we give a generic algorithm that probes the Rhomboid tiling
with a binary search to find a cycle 
contained in a smallest possible
sphere while satisfying a property. 
\cancel{
Suppose that we are given a point set $\pointset$ of cardinality $n$ embedded in a Euclidean space of a fixed dimension $d$. The hyperplane arrangement $\hyperarrange(P)$ associated to the paraboloid has $n$ hyperplanes. 

By a simple counting argument, one observes that a hyperplane arrangement on $n$ hyperplanes in $\RBB^{d+1}$ has at most $O(n^{d+1})$ vertices. By duality, the number of highest dimensional rhomboids in the Rhomboid tiling is also of the order $O(n^{d+1})$. 
In fact, a more refined analysis gives the same bound on the total number of cells in the Rhomboid tiling of a point set. 
\begin{fact}[{\cite[Proposition 4.8]{OsangThesis},\cite[Section 1.2]{edelsbrunneralgocombgeom}}]\label{Prop:rhombount}
Given a point set $P \subset \RBB^d$ of size $n$, the total number of cells (of all dimensions) in $\rhomb(P)$ is at most $\frac{2^{d+1}}{(d+1)!}(n+1)^{d+1}\leq 2(n+1)^{d+1}$.
\end{fact}
}
\paragraph*{General setup.}
Let $\pointset\subset \mathbb{R}^d$ be an $n$-points set and $\complex$ be a simplicial
complex consisting of linearly embedded simplices with vertices in $P$.
Given $V\subset \pointset$, let $\complex_V$ denote 
the largest subcomplex of $\complex$ with $V$ as its vertex set. We say that
$\complex_V$ is \emph{induced} by $V$.
More generally, we are given a property $\property$,  and  we are interested in determining  subcomplexes $\complex_V$ of $\complex$ that satisfy it.
Specific to our framework, we wish to find a smallest sphere $\ssphere_{c,r}$ that encloses a vertex set $V \subset P$ such that the induced complex $\complex_V$ contains a cycle or a chain satisfying $\property$.
In that case, we say that  $\complex_V$ satisfies property $\property$, else we say that $\complex_V$ does not satisfy property $\property$.  
Note that the smallest sphere containing a cycle/chain is also
the unique smallest sphere that encloses a set $V\subset P$ and thus corresponds
to a base vertex $s_V$ for a rhomboid in $\rhomb(P)$.
Therefore, in order to determine the smallest sphere enclosing a cycle or a chain that satisfies $\property$, one could exhaustively check if there exists a rhomboid $\rho$ in the Rhomboid tiling $\rhomb(\pointset)$ 
with a base vertex $s_V$ so that $\complex_V$ contains such a cycle or a chain. Even this na\"ive approach makes the problem tractable because one has to check only $O(n^{d+1})$ base vertices (Rhomboid tiling as a dual to $\hyperarrange(P)$ has size $O(n^{d+1})$), and checking each case can be achieved in polynomial time. 
Equivalently, the na\"ive approach  consists of checking all possible complexes induced by vertices enclosed by $O(n^{d+1})$ minimum circumspheres  in $\mathbb{R}^d$.

\paragraph*{Hierarchical property $\Pi$.}
We replace the exhaustive search with a generic binary search on the Rhomboid tiling which examines only $O(\log n)$ number of levels of the tiling.
The properties we consider are \emph{hierarchical}, meaning that, if a complex
$\complex_V$ satisfies $\property$, then any complex $\complex_W$ with $W\supseteq V$ necessarily
satisfies $\property$. Symmetrically, if a complex $K_V$ does not satisfy $\property$,
then any complex $\complex_W$ with $W\subseteq V$ does not satisfy $\property$. Since
$V\subseteq W$ if and only if $s_V$ has a level no more than the level of $s_W$, 
this hierarchical property
allows a binary search on the levels of vertices $s_V$ in the Rhomboid tiling. Refer to \Cref{alg:bin-search} for a pseudocode.

\cancel{
if there exists a sphere $S$  enclosing a vertex set $V$ where $\complex_V$ satisfies property $\property$, then  any sphere $S'$  enclosing a vertex set $W \supsetneq V$  induces a complex $\complex_W$ satisfying $\property$ because $\complex_V\subset \complex_W$. Symmetrically, if a sphere $S$  encloses a vertex set $V$ where $\complex_V$ does not satisfy  $\property$,  any sphere $S'$ enclosing a vertex set $W \subsetneq V$   induces a complex  $\complex_W$ that fails to satisfy $\property$. We use this hierarchy on point sets to guide our binary search. Refer to \Cref{alg:bin-search} for a pseudocode.}

\cancel{\begin{center}
\begin{algorithm}[!h] 

\SetKwProg{myproc}{Procedure}{}{}
 \myproc{\textsc{BinSearch}{($\complex,\property,\rhomb(\pointset),start,end$)}}{
{$\low \gets$ start. } \quad {$\high \gets$ end.} \\
{$r_{\min} \gets \infty$;\quad $\vertexset = \emptyset$};\quad $\level = n$.\\
\While{$\low < \high$} { {$\middling=\left\lfloor \frac{\low+\high}{2}\right\rfloor $.}  \\
{Compute $S_{\middling}$ = \{ $s_V\in \rhopoints_{\middling}  \mid$ $\complex_V$ satisfies property $\property$.\}} \\
{Compute $N_{\middling}$ = \{ $s_V\in \rhopoints_{\middling} \mid$ $\complex_V$ does not satisfy property $\property$.\}} \\

\If{$S_{\middling}$ is non-empty}{ 
\If{there exists a  vertex $s_V \in S_{\middling}$ and a rhomboid $\rho$ with $s_V$ as its base vertex such that $r(\rho) < r_{\min}$}{
{$r_{\min} \gets r(\rho)$.}\\
{$\vertexset \gets V$.} \\
{$ \level \gets \middling$.}
}
{Remove all the vertices from $\rhopoints_{k}$ for $k \in [\middling,\high]$.} \\ 
\For{all $s_V\in N_{\middling}$}{
\For{ every $k<\middling$}{
\For{ all $s_W \in \rhopoints_{k}$  with $W \subset V$}{
{Remove $s_W$ from $\rhopoints_k$.}}}
}
{$\high \gets {\middling} - 1$.}
} 

\Else{ 
{Remove all the vertices from $\rhopoints_{k}$ for $k \in [\low,\middling]$.} \\
{$\low \gets \middling + 1$.} 
}
} 
{\textbf{return } $r_{\min}, \,\, \vertexset, \,\,\level$.}
}

\caption{Binary search on the Rhomboid tiling for property $\property$}
\label{alg:bin-search}
\end{algorithm}
\end{center}}

\begin{center}
\begin{algorithm}[!htb] 

\SetKwProg{myproc}{Procedure}{}{}
 \myproc{\textsc{BinSearch}{$(\complex,\property,\rhomb(\pointset),\textnormal{\startshere},\textnormal{\toomuch})$}}{
{$\low \gets \startshere$; } \quad {$\high \gets \toomuch$} \\
{$r_{\min} \gets \infty$;\quad $\vertexset \gets \emptyset$;\quad $\level \gets |P|$;\quad $\zeta_{\mathsf{OPT}} \gets \emptyset$}\\
\While{$\low < \high$} { {$\middling\gets\left\lfloor \frac{\low+\high}{2}\right\rfloor $;
$\found \gets \texttt{false}$} \\
\For{every vertex $s_V \in \rhombvertices{\middling}$ }{
\If{$(\zeta \gets \property(\complex,V,\cdot))$ is non-empty }{
\If{$\exists$ a rhomboid $\rho$ with $s_V$ as its base vertex s.t. $r(\rho) < r_{\min}$}{
{$r_{\min} \gets r(\rho)$; \quad $\vertexset \gets V$; \quad $\zeta_{\mathsf{OPT}} \gets \zeta$;}\\
{$ \level \gets \middling$; \quad $\found \gets \texttt{true}$}}}
\Else{
\For{ every $k \in [\low,\middling-1]$}{
\For{every vertex $s_W \in \rhombvertices{k}$  with $W \subset V$}{
{Remove $s_W$ from $\rhombvertices{k}$}}}
}
}

{\textbf{if} $\found$ is \texttt{true} \,\textbf{then} \, $\high \gets {\middling} - 1$} \\
{\textbf{else}\,\, $\low \gets \middling + 1$}


} 
{\textbf{return } $(r_{\min}, \,\, \vertexset, \,\,\level,\,\,\zeta_{\mathsf{OPT}})$}
}
\caption{Binary search on the Rhomboid tiling for property $\property$}
\label{alg:bin-search}
\end{algorithm}
\end{center}

The procedure \textsc{BinSearch} in \Cref{alg:bin-search} is the centerpiece of our localization of topological constructs like chains and cycles within Euclidean balls of small radii. At its core, \textsc{BinSearch} performs a binary search on the levels between $\mathrm{start}$ and
$\mathrm{end}$ of the Rhomboid tiling of a point set $P \subset \RBB^d$. It takes as input a simplicial complex $\complex$  with vertices in $P$, and the Rhomboid tiling $\rhomb(P)$. The argument $\Pi$ is a property that is checked on complexes $\complex_{V}$ induced by vertex sets $V \subset P$. Importantly, $\Pi$ must be viewed as a pointer to a function with variable number of arguments. That is, for each computational problem, \textsc{BinSearch} is invoked by a main subroutine for optimizing a certain objective function and the function $\Pi$ is specific to that objective function. 
\Cref{alg:bin-search} returns four items, namely, 
\begin{inparaenum}
\item[(1)] $\vertexset$: the vertex set that induces a complex with the 
smallest enclosing sphere satisfying $\Pi$,
\item[(2)]$r_{\min}$: the radius of the smallest sphere  enclosing  $\vertexset$,
\item[(3)]$\zeta_{\mathsf{OPT}}$: the optimal cycle or chain satisfying $\Pi$ in $\complex_V$, and
\item[(4)] $\level = |\vertexset|$.
\end{inparaenum}
Every time a smaller Euclidean  sphere enclosing an induced complex is found, these four items are updated.
The lines~180--182 removes a vertex $s_W$ from the rhomboid tiling, where the vertex set $W$ induces a complex that is a subcomplex of a larger complex which is already found to not satisfy the property $\Pi$ of interest (in the \textbf{if} condition on Line~175). Since $\Pi$ is hierarchical,
the vertex $s_W$ need not be examined. Again, by the virtue of the fact that $\Pi$ is hierarchical, once the level $\middling$ has been fully examined, depending on whether the search was successful (or not), all the levels higher than $\middling$ (lower than $\middling$) can be removed.

\paragraph*{\textbf{Complexity Analysis.}}
The algorithm {\sc BinSearch} visits all vertices in a level in the worst case and checks at most $O(\log n)$ levels. Thus it takes $O(U\log n$) time to visit the Rhomboid tiling where $U$ is an upper bound on the size of any level $\rhomb_k(P)$, 
$k\geq 1$, in the
Rhomboid tiling. As mentioned earlier $U=O(\kappa(d+1))$ where
$\kappa(d+1)$ bounds the number of $k$-sets of $n$ points in $\mathbb{R}^{d+1}$ for any $k\geq 1$. If checking property $\Pi$ takes $O(T)$ time, we get a time complexity of $O(\kappa(d+1)T\log n)$.
}

\section{The $\radmeasure$ metric}
\cancel{
\amritendu{Since the $\radmeasure$ is a contribution, list a few statements on the intuition behind this measure. Illustrate with  a figure.}

\amritendu{We have to argue that the $\radmeasure$ can be computed in polynomial time. Requires arguments from osang's thesis.}
\amritendu{Need to fill complexity results}
}

Given a complex $\complex$, let  $Z_p(\complex)$ denote its $p$-th cycle group, $B_p(\complex)$ its $p$-th  boundary group and $H_p(K)$ its $p$-th homology group with $\ZBB_2$ coefficients.
Given a cycle $\zeta$, our goal is to define a non-negative weight function $w: Z_p(\complex) \rightarrow \mathbb{R}^+$ on the cycles
in $Z_p(\complex)$ and 
compute a minimum-weight (optimal) cycle $\zeta^*$ in its homology class $[\zeta]$, that is,
\begin{equation}
\zeta^*\in \arg\min_{\hat\zeta\in [\zeta]} w(\hat\zeta). \label{eq1}
\end{equation}
We show that the $\radmeasure$ is an alternative natural geometric objective function defined on cycles that guarantees tractability. 
Let $P$ be the vertex set of $\complex$. Then, $\complex_V$ denotes the subcomplex of $\complex$ induced
by a subset $V\subseteq P$. Extending this notation, we say
a complex is \emph{induced} by a sphere $\sphere{c}{r}$ if
it is induced by the subset of vertices of $\pointset$ that are enclosed by $\sphere{c}{r}$ (including on the sphere).
Let the complex induced by $\sphere{c}{r}$ be denoted
as $\complex_{c,r}$. We define a weight function
$r: C_p(\complex)\rightarrow \mathbb{R}^+$, $\xi \mapsto r(\xi)$ where
\begin{equation} \label{eq:randfunc}
    r(\xi)=\min_{c,\delta} \{\delta \,|\,\xi \in C_p(\complex_{c,\delta})\}.  
\end{equation}

In words, $r(\xi)$ is the radius of the smallest Euclidean sphere
whose induced complex in $\complex$ contains $\xi$.

We now define an $\radmeasure$ measure for intervals in a barcode.
For an interval $[b,d) \in \barcode_p(\mathcal{F})$, we define $r([b,d))$  as the radius of the smallest  sphere that encloses a subset of vertices $V$ of $\complex_b$ that induces a subcomplex  $\complex_b^{V} \subset \complex_b$, which supports a representative cycle for $[b,d)$. Equivalently,
\begin{equation} \label{eq:persrad}
   r([b,d)) = \min_{\zeta \in \mathcal{R}([b,d))} \,\, r(\zeta),  
\end{equation}
where in \Cref{eq:persrad}, the radius function $r$ is restricted to the subcomplex $\complex_b \subset \complex$.

\section{Computing optimal homologous cycle} \label{sec:homloc}




Following \Cref{eq1}, we define an optimal
cycle $\zeta^*$ in the class $[\zeta]$ by requiring $\zeta^*\in \arg\min_{\hat\zeta\in [\zeta]} r(\hat\zeta)$.
The cycle $\zeta^*$ represents an optimal localization of the class $[\zeta]$ with respect to the $\radmeasure$.
We consider the following {\sc Optimal Homologous Cycle} problem: 
\begin{framed}
Given an $p$-cycle $\zeta\in Z_p(\complex)$, compute an optimal cycle $\zeta^*$ in $[\zeta]$ and $r(\zeta^*)$.
\end{framed}


\begin{remark}
\label{rem:optimal_l2rad}
To compute the optimal homologous cycle, it is sufficient to look at the minimum circumspheres of all $k$-subsets of points  $P = V(\complex)$, where $k \in \{2,\dots,d+1
\}$, and check if the circumsphere encloses a cycle homologous to the input cycle. When the dimension $d$ of the complex $\complex$ is fixed,
the search terminates in polynomial time. This describes a trivial exact algorithm which was found to be too expensive in our experiments in spite of  polynomial time complexity.    
\end{remark}

\begin{remark}
\label{rem:2-approx_l2rad}
 By restricting the centers of the spheres in \Cref{eq1} to the the sites $P = V(\complex)$ (vertices of $\complex$) yields a $2$-approximation of $\radmeasure$ as follows: let $S$ be a sphere that minimizes $\radmeasure$ of a chain $\xi$ and let $v$ be a vertex on $S$.
Then, a sphere of twice the optimal radius centered at $v$ encloses $S$, and therefore also encloses $\xi$. 
We define $r_c(\xi) = \min\{\delta \,|\, \xi \in C_p(\complex_{c,\delta})\}$ and $r_P(\xi) = \min_{c \in P, \delta}\{\delta \,|\, \xi \in C_p(\complex_{c,\delta})\}$.
\end{remark}


\cancel{
Sometimes we use restrictions of the function $r$. For instance, $r|_{\subcomplex}$ is the function $r$ restricted to a subcomplex $\subcomplex \subset \complex$, defined as follows: 
\[
r|_{\subcomplex}(\xi)=\min_{c,\delta} \{\delta \,|\,\xi \in C_p(\subcomplex_{c,\delta})\}.
\]
\amritendu{The next 2 para. needs to be moved or cancelled. 
Can't where we are using this}
Likewise we also consider restrictions of $r$ to cycle subgroups ($r|_{Z_p(\subcomplex)}$) and boundary subgroups ($r|_{B_p(\subcomplex)}$) of $C_p(\subcomplex)$, defined as follows:
\begin{align*}
    r|_{Z_p(\subcomplex)}(\xi) =\min_{c,\delta} \{\delta \,|\,\xi \in Z_p(\subcomplex_{c,\delta})\}. \\
r|_{B_p(\subcomplex)}(\xi) =\min_{c,\delta} \{\delta \,|\,\xi \in B_p(\subcomplex_{c,\delta})\}.
\end{align*}

\smallskip
}
\paragraph{Notations and Conventions.} 
The notations and conventions described are common to all the problems in the paper.
In our algorithms, a cycle (or a chain) $\zeta$ is represented by a $0$--$1$ vector in the standard chain basis. That is, a $p$-cycle $\zeta$ is represented by a vector $\zeta$ where $\zeta[i] = 1$ ($\zeta[i] = 0$) if a $p$-simplex $\sigma_i$ is (not) in the support of $\zeta$.
We often use cycle vectors of subcomplexes in computations involving cycles and boundaries of larger complexes. To ensure that we are working with vectors/matrices of the right dimensions, we make the following adjustment.
For complexes $\subcomplex \subset \complex$, the inclusion map $\subcomplex \xhookrightarrow{} \complex$ induces maps $Z_p(\subcomplex) \to Z_p(\complex)$ for every $p$. A cycle $\xi$ in $\subcomplex$ is mapped to a cycle $\overline{\xi}$ in $\complex$ with $\overline{\xi}[i] = \xi[i]$ for simplices  $\sigma_i \in \subcomplex$, and  $\overline{\xi}[i] = 0$ for simplices  $\sigma_i \in \complex \setminus \subcomplex$ (using standard chain basis).
Likewise, a matrix $\boldM$ of cycle vectors of $\subcomplex$ can be treated as a matrix of cycle vectors $\overline{\boldM}$ of $\complex$ by padding zeros in the rows corresponding to the simplices in $\complex \setminus \subcomplex$.
We call such cycle vectors  $\overline{\xi}$ and matrices $\overline{\boldM}$, the \emph{extensions} of $\xi$ and $\boldM$ in $\complex$.

\cancel{
Let $\complex$ be a simplicial complex, $\complex \subset \Real^{d}$. Let $v \in \Real^{d}$. For a $u \in \Real^{d}$ we define $r_{v}(u)$ as the Euclidean distance between $v$ and $u$.
If $\sigma \in \complex$ be a simplex  we define $r_{v}(\sigma) = \max_{x \in V(\sigma)}{r_{v}(x)}$ where $V(\sigma)$ is the set of vertices of $\sigma$.
For a chain $\zeta \in C_{p}(\complex)$, we define $r_{v}(\zeta) = \max_{\sigma \in \zeta}{r_{v}(\sigma)}$.
Equivalently we may define $r_v(\zeta) = \min_{}\{\delta \,|\, \zeta \in \complex_{v,\delta}\}$.
Finally $r_{v}([\zeta]) = \min_{\zeta \in [\zeta]}{r_v(\zeta)}$. 
}
Let $\complex$ be a simplicial complex, $\complex \subset \Real^{d}$.
For any $v \in \Real^d$ we can define a total ordering $\prec_{v}$ on the simplices of $\complex$ as follows. 
If $\sigma_{1}$ is a face of $\sigma_{2}$ or $r_{v}(\sigma_1) < r_{v}(\sigma_2)$, then $\sigma_{1} \prec_{v} \sigma_{2}$.
Otherwise (when $r_{v}(\sigma_1) = r_{v}(\sigma_2)$ and $\sigma_1$ is neither a face or coface of $\sigma_2$), ties are arbitrarily broken.
If $\zeta = \sigma_1 + \ldots + \sigma_s$ such that $\sigma_1 \prec_v \ldots \prec_v \sigma_s$, then we define $\maxsmplx(\zeta) = \sigma_s$.
Further, we extend this ordering to chains as follows: If $\zeta_1, \zeta_2 \in C_{p}(\complex)$ such that $\zeta_1 = \sigma_1 + \ldots + \sigma_s, \: \zeta_2 = \sigma'_{1} + \ldots + \sigma'_{s'} \textnormal{with} \: \sigma_1 \prec_{v} \ldots \prec_{v} \sigma_s \textnormal{ and } \sigma'_{1} \prec_{v} \ldots \prec_{v} \sigma'_{s'}$, then $\zeta_1 \prec_{v} \zeta_2$ if $\sigma_s \prec_{v} \sigma'_{s'}$ i.e. $\maxsmplx(\zeta_1) \prec_{v} \maxsmplx(\zeta_2)$. 
Note that $r_v(\zeta) = r_v(\maxsmplx(\zeta))$.
The ordering $\prec_{v}$ induces a simplex-wise filtration on $\complex$ which we denote by $\filtrationv_{v}(\complex)$.
\cancel{
We define $\complex_{\tau} = \{\sigma \,|\, \sigma \prec_v \tau\} \cup \{\tau\}$, that is, $\complex_{\tau}$ is the subcomplex consisting of $\tau$ and  all simplices that precede $\tau$ in the filtration $\filtrationv_v(\complex)$.
}
The standard reduction algorithm ~\cite{bauer2017phat} is used in many of our algorithms as subroutines.
For completeness, we present an outline of the algorithm and recall some facts arising out of it in \Cref{sec:algcorrectness}.
\Cref{alg:approx-optimalhomologouscycle} relies on the following proposition (Proof in \Cref{sec:algcorrectness}).
\cancel{
For complex $\complex$ and the filtration $\filtrationv_v$ let $\zeta_1, \ldots, \zeta_s$  be the essential $p-$cycles computed by the standard reduction algorithm with $\maxsmplx(\zeta_1) \prec_v \ldots \prec_v \maxsmplx(\zeta_s)$.
The essential cycles form a basis of the $p^{th}$ homology of $\complex$ that is, $[\zeta_1], \ldots, [\zeta_s]$ form a basis of $H_p(\complex)$.
}

\begin{proposition}
\label{prop:standardred-minimality}
    Let $\zeta_1 \prec_v \ldots \prec_v \zeta_s$ be the essential $p-$cycles of $\filtrationv_v(\complex)$ computed using standard reduction.
    Let $\zeta \textnormal{ be a p-cycle, } [\zeta] \neq 0 \:\in\: H_p(\complex)$,
    such that $\zeta = \zeta_{i_1} + \ldots + \zeta_{i_m} + \partial c_{p+1}$ where each $\zeta_{i_k} \in \{\zeta_1 ,\ldots, \zeta_s\}$ and $c_{p+1}$ is a $p+1$ chain. If $i_1 <  \ldots < i_m$ then $r_v([\zeta]) = r_v(\zeta_{i_m})$.
    In particular, $\zeta_{i_1} + \ldots + \zeta_{i_m} \in \arg\min_{\xi \in [\zeta]}{r_{v}(\xi)}$.
\end{proposition}

We now describe a  $2$-approximation algorithm for   \textsc{Optimal Homologous Cycle} for an input cycle $\zeta$ by optimizing with respect to $r_P$.
For each site $v$, the algorithm  invokes the subroutine \textsc{optimal-hom-cycle-forsite} (\Cref{proc:optimalhomcycleforsite} of \Cref{alg:approx-optimalhomologouscycle}) which computes $r_v([\zeta]) = \min_{\eta \in [\zeta]}\{ r_v(\eta) \}$ and $\zeta^{*}_v \in \arg\min_{\eta \in [\zeta]}\{ r_v(\eta) \}$.
Finally it reports the minimum among all sites and the corresponding optimal homologous cycle.
Procedure \textsc{optimal-hom-cycle-forsite} is motivated by \Cref{prop:standardred-minimality}.
It first sorts the simplices of $\complex$ based on distance from $v$.
The ordering is monotonic, that is, faces gain precedence over cofaces.
\cancel{
If $r_v(\sigma) = r_v(\tau)$ and $\sigma$ is a face of $\tau$ then $\sigma$ gains precedence over $\tau$, otherwise the tie is arbitrarily broken.
}
In this way the ordering $\prec_v$ and hence the filtration $\filtrationv_v(\complex)$ is defined.
Let $\zeta_1 \prec_v \ldots \prec_v \zeta_m$ be the essential $p-$cycles of $\filtrationv_v(\complex)$ computed using standard reduction.
As noted before we consider cycle vectors to represent the cycles.
To compute the linear combination of cycles $\{\zeta_i\}$ which is homologous to $\zeta$, we solve for the system of equations $[\zeta_1 \ldots \zeta_s \,|\, B_p(\complex)].x = \zeta$. 
(We invoke subroutine \textsc{SolveByReduction} which solves $Ax = b$ over  $\mathbb{Z}_2$, using standard reduction as a subroutine.
Refer to \Cref{sec:algcorrectness}, \Cref{alg:solvebyreduction} for definition of this routine).
If $i_1, \ldots i_s, j_1 \ldots , j_t$ is a solution where indices $i_1, \ldots i_s$ correspond to cycles in $\{\zeta_i\}_{i = 1}^m$) and $j_1 \ldots j_t $ correspond to boundaries in $B_p$, then by \Cref{prop:standardred-minimality} $\zeta_{i_1} + \ldots + \zeta_{i_s} \in \arg\min_{\eta \in [\zeta]\{r_v(\eta)\}}$.

\begin{remark}
    \Cref{alg:approx-optimalhomologouscycle} runs in $O(|P|N^3)$, where $N$ is the number of simplices in $\complex$.
\end{remark}

\begin{algorithm}[h]
\caption{Computing optimal homologous cycle for given set of sites
}\label{alg:approx-optimalhomologouscycle}
\SetKwInOut{Input}{Input}\SetKwInOut{Output}{Output}
\Input{$\complex, \zeta \in C_{p}(\complex)$}
\Output{$r_P([\zeta]), \zeta^*$ (Optimal homologous cycle)}
\SetKwFunction{optHomologouscycle}{OptHomologousCycle}
\SetKwFunction{OptimalHomCycleForSite}{Optimal-Hom-Cycle-ForSite}
\SetKwProg{myproc}{Procedure}{}{}

\myproc{\optHomologouscycle}{
{$r_P([\zeta]) \gets \infty, \zeta^* \gets \emptyset$} \\
\For{$v \in P$}{
{$ r_{v}([\zeta]), \zeta^{*}_{v} \gets \textsc{Optimal-Hom-Cycle-ForSite}(v)$} \\
{If $r_v([\zeta])$ is less than the current value of $r_P([\zeta])$, then update $r_P([\zeta])$ with $r_v([\zeta])$ and $\zeta^{*}$ with $\zeta^{*}_v$}
}
}
\myproc{\OptimalHomCycleForSite{$v$}}
{
\Comment{Description: Computes $r_{v}([\zeta]) = \min_{\eta \in [\zeta]}{r_v(\eta)}, \, \zeta^{*}_v \in \arg\min_{\eta \in [\zeta]}{r_v(\eta)}$}

\BlankLine
{Define $\prec_v$ on $\complex$. Compute $\filtrationv_{v}(\complex)$}\\
{Compute the essential cycles of $\filtrationv_{v}(\complex)$ by standard reduction.
Let $\zeta_{1}, ..., \zeta_{m} $ be essential cycles ordered with respect to $\prec_v$.} \\
{Compute the $p^{th}$boundary matrix of $\complex$, denote it by $B_p$.}\\
{Assemble matrix $\partial = [\zeta_{1},..., \zeta_{m} \,|\, B_{p} ]$} \\
{
Solve $\partial .x  = \zeta$. Invokes(
$\textsc{SolveByReduction}(\partial, \zeta)$).
Let $i_1, \ldots , i_s, j_1, \ldots , j_t$ be the solution where $i_1, \ldots, i_s \leq m$ (indices that correspond to cycles in $\{\zeta_i\}_{i = 1}^m$) and $j_1 \ldots j_t >m$ ( indices correspond to boundaries in $B_p$.)
} \label{procstep:invokesolvebyreduction} \\

{$\zeta^{*}_v \gets \zeta_{i_1} + \ldots + \zeta_{i_s}$}.\\
{$r_v([\zeta]) \gets r_v(\zeta^{*}_v)$}\\

{Return $r_{v}([\zeta]), \zeta^{*}_v$}\\
}\label{proc:optimalhomcycleforsite}

\end{algorithm}

\section{Optimal homology basis}
\label{sec:opt_homologybasis}
\cancel{
Let $\complex$ be a simplicial complex defined on $\pointset$ as defined
before. Let $n=|\pointset|$, $n_p=|\complex^{(p)}|$, and  $\complexsize=|\complex|$. 
Using standard bases for $\mathsf{C}_{p}(\complex)$ and $\mathsf{C}_{p-1}(\complex)$, the $p$-th boundary matrix
 of $\complex$, is denoted by $\partial_{p}(\complex)$ or simply by $\partial_{p}$.
 Let $\beta = \beta_p(K)$.
 }
A set of $p$-cycles $\left\{ \zeta_{1},\dots,\zeta_{\beta}\right\} $ ($\beta_p = \beta_p(\complex)$) is called a \emph{homology cycle basis} if the set of classes $\left\{ [\zeta_{1}],\dots,[\zeta_{\beta}]\right\} $ forms a  basis for $H_p(\complex)$. For simplicity, we use the term \emph{homology basis} to refer to the set of cycles $\left\{ \zeta_{1},\dots,\zeta_{\beta}\right\} $.


\begin{definition}
    A $p-$homology basis $\{\zeta_1, \ldots, \zeta_m\}$ will be called a minimum $p-$homology basis ($p>0$) with respect to a non-negative weight function $w:Z_p(K) \rightarrow \Real$, if for all $p-$homology bases $\zeta'_1 \ldots \zeta'_m$, $w([\zeta_1]) + \ldots + w([\zeta_m]) \leq w([\zeta'_1]) + \ldots + w([\zeta'_m])$ and each $\zeta_i \in \arg\min_{\eta \in [\zeta_i]}{w(\eta)}$
\end{definition}

We consider the following \textsc{Optimal Homology Basis} problem.
\begin{framed}
    \noindent For a given $p > 0$ compute a minimum $p-$homology basis with respect to the weight function $r$.
\end{framed}

\cancel{
A variant is the following  \textsc{Optimal Homology Basis For Sites}
problem.
\begin{framed}
    \noindent For a given $p > 0$ and a finite collection of sites $S$, compute a minimum $p-$homology basis with respect to the weight function $r_P$.
\end{framed}
}

\Cref{alg:minhomologybasis-sites} describes a  $2\beta_p$-approximation algorithm for  \textsc{Optimal Homology Basis} by computing a minimum homology basis with respect to $r_P$ by restricting the centers of minimal spheres to sites.
To compute the minimum homology basis $\minhombasis$ from $\Omega$, (see \Cref{step:selectminhombasis}) standard reduction is performed on $\partial = [B_p(\complex) \, | \, \Omega]$.
We examine the columns of the reduced matrix $\Tilde{\partial}$ from left to right.
For every non-zero column $i$ that is an index from $\Omega$, we add the corresponding cycle in $\Omega$ to $\minhombasis$. $\Cref{alg:minhomologybasis-sites}$ runs in $O(|P|N^3)$. See \Cref{sec:correctness_mhb} for a proof of correctness.

\begin{algorithm}[h]
\caption{Optimal homology basis for sites}
\label{alg:minhomologybasis-sites}
\SetKwInOut{Input}{Input}\SetKwInOut{Output}{Output}
\Input{Complex $\complex \subset \Real^{d}$, $p > 0$}
\Output{A minimum homology basis with respect to $r_P$}
\SetKwFunction{Opthombasis}{Opt-hom-basis-for-sites}
\SetKwProg{myproc}{Procedure}{}{}
\myproc{\Opthombasis{$\complex,p$}}{
{For each $v\in P$, define $\prec_v$.
Using standard reduction compute the essential $p-$cycles of the filtration $\filtrationv_v(\complex)$, denote them by $\zeta_{v,1}, \ldots, \zeta_{v,m}$.}\\
{Let $\Omega = \{\zeta_{v,i}\}_{v\in P,1\leq i \leq m}$.
Sort the cycles in $\Omega$ so that if $r_v(\zeta_{v,i}) < r_{v'}(\zeta_{v',i'})$ then $\zeta_{v,i}$ precedes $\zeta_{v',i'}$ in $\Omega$.
If $\zeta_{v,i} \prec_v \zeta_{v,i'}$, then $\zeta_{v,i}$ precedes $\zeta_{v,i'}$ as well.
Ties are broken arbitrarily.
Denote this ordering on $\Omega$ as $\prec_{\Omega}$.
}\label{step:minhombasis-aggregatedcycles} \\
{$\mathscr{M} \gets \emptyset$.} \\

\For{$\zeta$ \textnormal{ in the ordered list } $\Omega$}
{ \label{step:selectminhombasis}
{ Let $\eta_1, \ldots \eta_k$ be the cycles currently in $\mathscr{M}$.}\\
 \If{$[\zeta] \in span\{[\eta_1], \ldots [\eta_k]\}$}{
{Discard $\zeta$ and continue.}
}   
\Else{
Add $\zeta$ to $\mathscr{M}$.}
}

{Report $\mathscr{M}$ as a minimum homology basis.}
}
\end{algorithm}

\cancel{
The correctness of \Cref{alg:minhomologybasis-sites} follows from the following proposition (see \Cref{sec:correctness_mhb} for a proof).
\begin{proposition}
    Let $\zeta_1, \ldots, \zeta_m$ be the cycles in $\minhombasis$ such that $\zeta_1 \prec_{\Omega} \ldots \prec_{\Omega} \zeta_m$, where $\minhombasis$ and $\prec_\Omega$ are as computed by \Cref{alg:minhomologybasis-sites}. 
    Let $\zeta'_1, \ldots, \zeta'_m$ be an ordered collection of cycles forming a $p-$homology basis.
    Assume further that if $r_P([\zeta'_i]) < r_P([\zeta'_j])$ then $i < j$.
    Then $r_P(\zeta_j) \leq r_P([\zeta'_j]), 1 \leq j \leq m$.
\end{proposition}
}

\section{Optimal persistent homology basis} \label{sec:phbasis}
We now consider a filtration of a simplicial complex $\complex$ with the aim of studying an extension of the problem to persistent homology~\cite{ELZ02}. We introduce the \persMHB problem:
\begin{framed}
Given a filtration $\mathcal F$ of complex $\complex$, compute a persistent
$p$-basis $\Lambda_p = \left\{ \zeta_i \mid i \in [|\barcode_p(\mathcal{F})|] \right \}$
  that minimizes 
    $ r(\Lambda_p) = \sum_{i=1}^{|\barcode_p(\mathcal{F})|} r(\zeta_i)$.  
\end{framed}

\Cref{thm:deypersone} states that for computing an optimal persistent homology basis it suffices to compute the minimum representative of each bar.
Formally, an optimum representative of a bar $[b,d)$ is a cycle $\zeta^* \in \arg\min_{\eta \in \represents([b,d))}\{r_P(\eta)\}$.

\Cref{alg:2-approx_min_persistence} computes an minimum representative of a input bar $[b,d) \in \barcode_p(\filtration)$ for a simplex-wise filtration $\filtration$ of $\complex$ with $V(\complex) = P$ with respect to $r_P$. 
For each site $v \in P$ the subroutine \textsc{Opt-Pers-Cycle-Site} is invoked which computes 
$\zeta^*_v \in \arg\min_{\eta \in \represents([b,d))}\{r_v(\eta)\}$.
Finally the minimum $\zeta^*_P = \arg\min_{v \in P}\{ r_v(\zeta^*_v) \}$ among all sites is reported.
Similar to \Cref{alg:approx-optimalhomologouscycle} a filtration $\filtrationv_v$ is defined on $\complex_b$.
The essential $p-$cycles $Y = \{\zeta_1 \prec_v \ldots \prec_v \zeta_m\}$ of $\filtrationv_v(\complex_b)$ are computed using standard reduction.
We then compute the smallest $i > 0$ such that $\exists \, \xi \in span\{\zeta_1, \ldots, \zeta_i\}, \xi \in \represents([b,d))$.
If $\sigma_b$ was added at index $b$ of $\filtration$ and $\alpha$ is the index of the first cycle in $Y$ containing $\sigma_b$, then update Y by adding $Y_\alpha$ to all other cycles containing $\sigma_b$.
This ensures that only a single cycle now contains $\sigma_b$.
Denoting these cycles of $Y \setminus \{ \zeta_\alpha \}$  by $Y'$ and the first $i$ cycles of $Y'$ by $Y'_{\leq i}$, it suffices to check if $[B_p(\complex_d)\,|\, Y_{\leq i}].x = Y_\alpha$ has a solution.
This is determined in \Cref{step:approxminpers_smallest_i} with a binary search over $i \in [1..m-1]$.

The proof of correctness of \Cref{alg:2-approx_min_persistence} can be found in \Cref{sec:correctness_persmhb}.
It runs in $O(|P|N^3\log{N})$.

\begin{algorithm}[h]
\caption{Computing optimal representative of bar of persistence  wrt $r_P$.}\label{alg:2-approx_min_persistence}
\SetKwInOut{Input}{Input}\SetKwInOut{Output}{Output}
\Input{$\complex, \filtration \textnormal{(simplex-wise filtration)}, [b,d) \in \barcode_{p}(\filtration)$}
\Output{$\zeta^{*}_{P}([b,d)) \in \arg\min_{v \in P, \eta \in \represents([b,d))}\{r_v(\eta)\}$}
\SetKwFunction{optimalpersistenthomologyrepresentative}{Opt-PersHom-Rep}
\SetKwProg{myproc}{Procedure}{}{}
\BlankLine
\myproc{\optimalpersistenthomologyrepresentative{$\complex, \filtration, [b,d)$}}{
{$r_{P}([b,d)) \gets \infty, \zeta^{*}_{P}([b,d)) \gets \emptyset$} \\
\For{$v \in P$}{
{$ r_{v}(\zeta^*_v), \zeta^*_v  \gets \textsc{Opt-Pers-Cycle-Site}([b,d),v)$} \\
\If{$r_{v}(\zeta^*_v) < r_P([b,d)])$ }{
{$r_{P}([b,d)) \gets r_{v}(\zeta^*_v)$}
{$\zeta^{*}_{P}([b,d)) \gets \zeta^*_v $}}
}
{Return $r_{P}([b,d)) , \zeta^{*}_{P}([b,d))$}
}

\SetKwFunction{OptimalPersistentCycleForSite}{Opt-Pers-Cycle-Site}
\BlankLine
\myproc{\OptimalPersistentCycleForSite{$[b,d), v$}}{
Define $\prec_v$ on $\complex_b$. Compute $\filtrationv_{v}(\complex_{b})$ \\
{Compute the essential p-cycles $\zeta_1 \prec_v \ldots \prec_v \zeta_m$ of $\filtrationv_{v}(\complex_{b})$ using standard reduction}\\
{$Y \gets \zeta_{1} ,..., \zeta_{m}$. ($Y$ is a matrix of $m$ columns, the column $Y_i$ is the cycle-vector of $\zeta_i$).
Let $ \alpha$ be index of first cycle in $Y$ containing $\sigma_b$} \\
{Add cycle $Y_{\alpha}$ to all other cycles in $Y$ containing $\sigma_b$, resulting in matrix $\hat{Y}$.}\label{step:approxminpers_addsigmab} \\
{Assemble matrix $Y'$ by dropping the $\alpha^{th}$ column of $\hat{Y}$. Denote by $Y'_{\leq i}$ the first $i$ columns of $Y'$. }\\
{$\partial_{d} \gets  B_p( K_{d})$($\partial_{d}$ is empty if $d = \infty$) }\\
{Compute the smallest $i \in [1 .. m-1]$ such that  $[\partial_{d} | Y'_{\leq i}].x =  Y_\alpha$ has a solution. }\label{step:approxminpers_smallest_i}\\
{Let $b_{1}, ..., b_{t}, {i_1}, ..., {i_s}$ be the solution computed by the previous step where $b_{1}, ..., b_{t}$ are indices in $\partial_d$ and ${i_1}, ..., {i_s} \textnormal{ are in } Y'_{\leq i}$} \\
{$\zeta^*_v \gets Y'_{\leq i,i_1} + ... + Y'_{\leq i,i_s} + Y_\alpha$} \\
{Return $\zeta^*_v, r_{v}(\zeta^*_v)$} \\
}

\end{algorithm}

\section{Experiments}

\label{sec:results}
We report results of experiments on real world datasets with a focus on computing cycle representatives of $H_1$.
These results demonstrate the utility of the $\radmeasure$ towards the identification of meaningful representatives of homology classes.
We consider three applications: localizing individual 1-cycles, optimal 1-homology basis computation, and minimum persistent 1-homology basis computation. 
Computing the exact $\radmeasure$ via an enumeration of all circumspheres is expensive.
So, all experiments were conducted on  implementations of the approximate algorithms described above.

We implement a heuristic to minimize the length of the cycle representative while preserving its radius. Essentially, we replace one of the paths $P$ between two vertices in the cycle with the shortest length path between them if it is homologous to $P$. This heuristic results in smoother and shorter cycles for all datasets. This also addresses the issue with the non-unicity of the $\ell_2$-metric in the sense that in our experiments we always find tighter cycles within a sphere when there are several homologous cycles of varying lengths within a sphere. 

All experiments were performed on an Intel Xeon(R) Gold 6230 CPU @ 2.10 powered workstation with 20 cores and 384GB RAM running Ubuntu Linux. The algorithms were parallelized using Intel Thread Building Blocks (TBB).
The PHAT library~\cite{bauer2017phat} was used in all routines that invoke the standard reduction algorithm.

\begin{figure}[!htb]
\centering
\begin{tabular}{cc}
\includegraphics[height=1.6in]{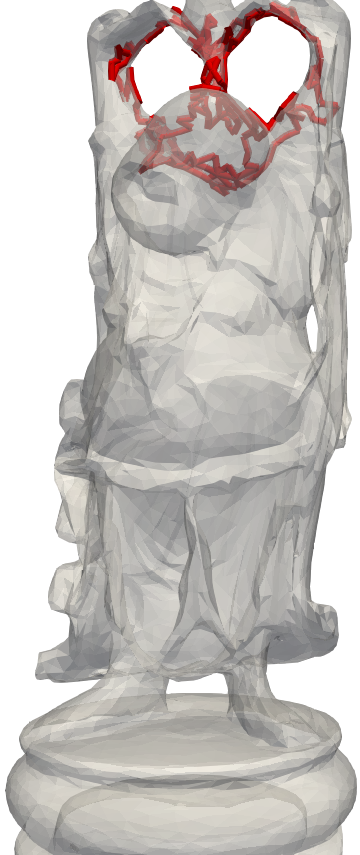} &
\includegraphics[height=1.6in]{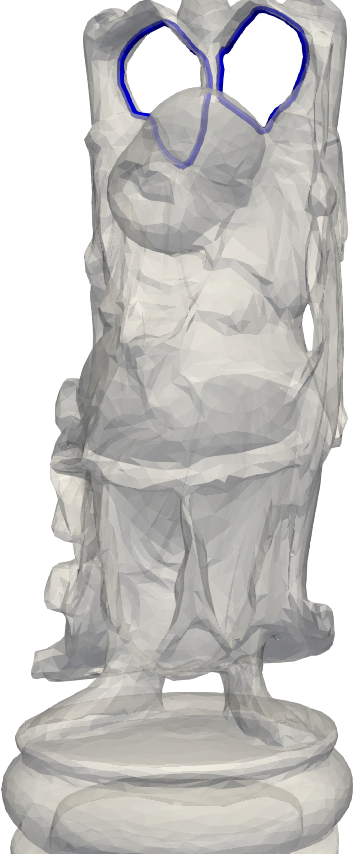} \\
\includegraphics[height=1.4in]{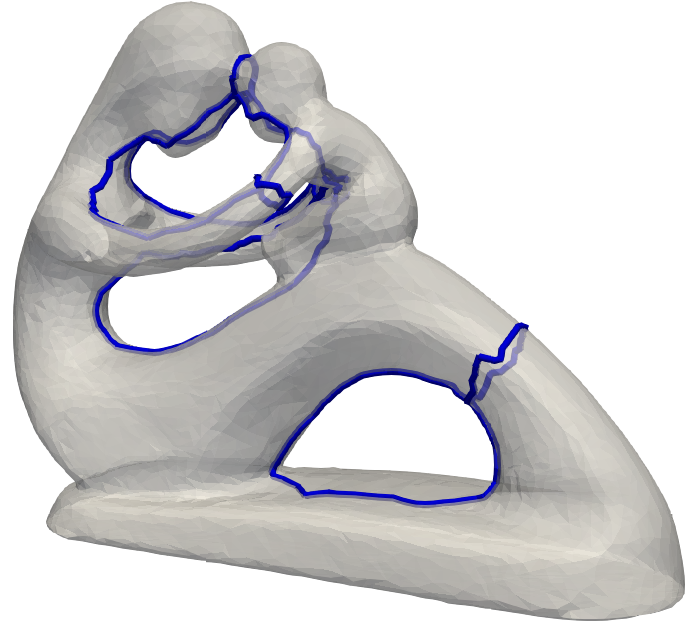} &
\includegraphics[height=1.4in]{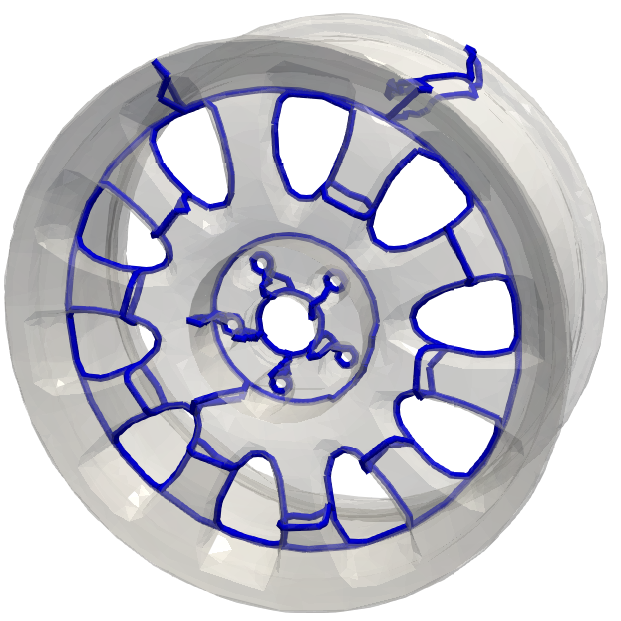}
\end{tabular}
\caption{(\textbf{top})~The localization algorithm computes optimal (blue) 1-cycles that are homologous to input 1-cycles (red). (\textbf{bottom})~Minimum 1-homology basis.}
    \label{fig:homloc-hombasis}
\vspace{-0.2in}
\end{figure}
\begin{figure}[!hbt]
\centering
\begin{tabular}{cc}
\includegraphics[height=1.6in]{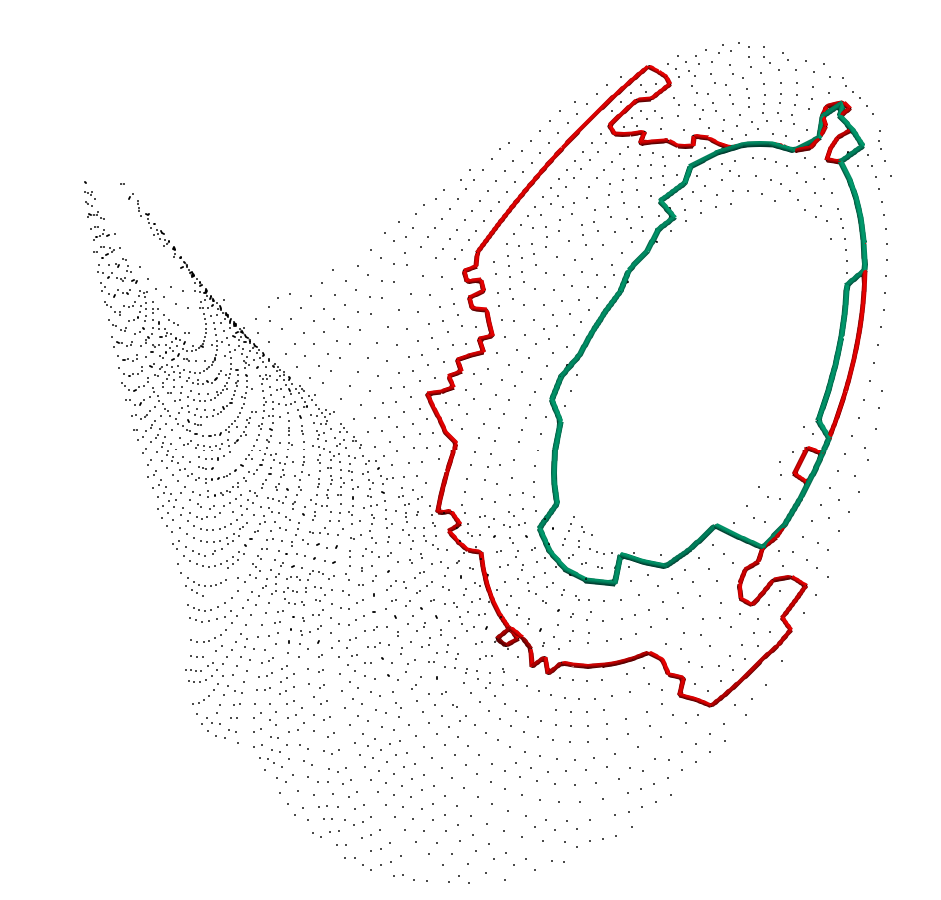}
 \label{fig:lorenz63_60} &
\includegraphics[height=1.6in]{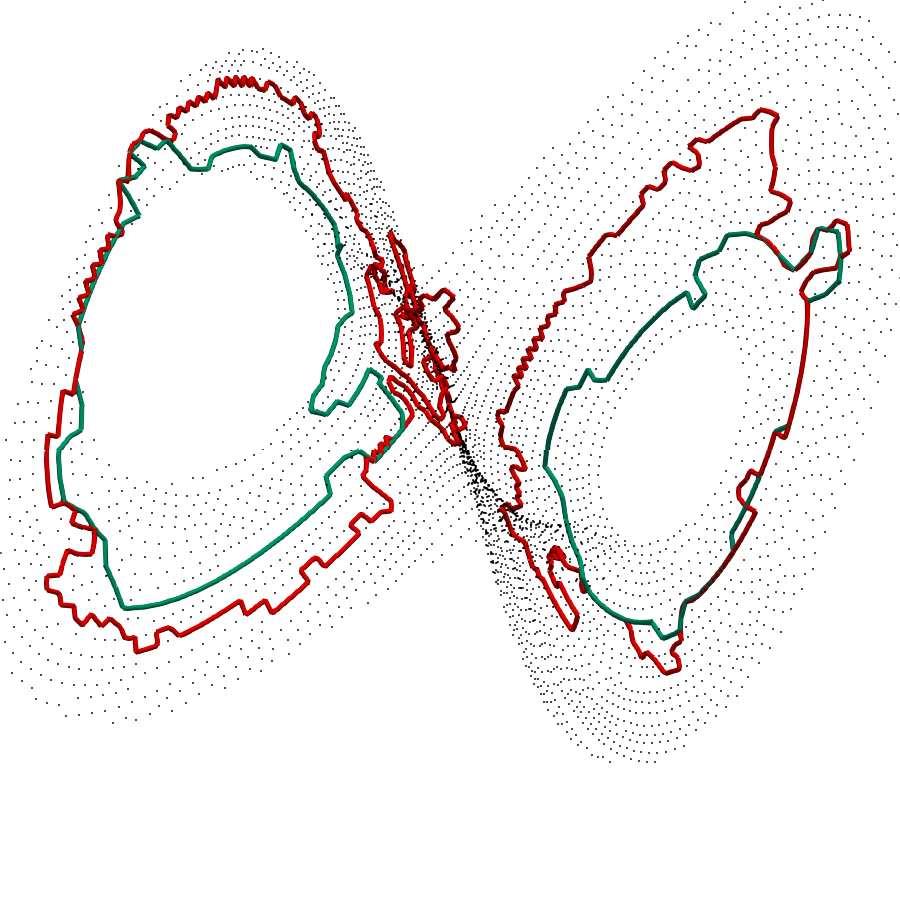} \\
\textbf{(a)} Lorenz'63(60\%) & \textbf{(b)} Lorenz'63(80\%) \\
\includegraphics[height=1.6in]{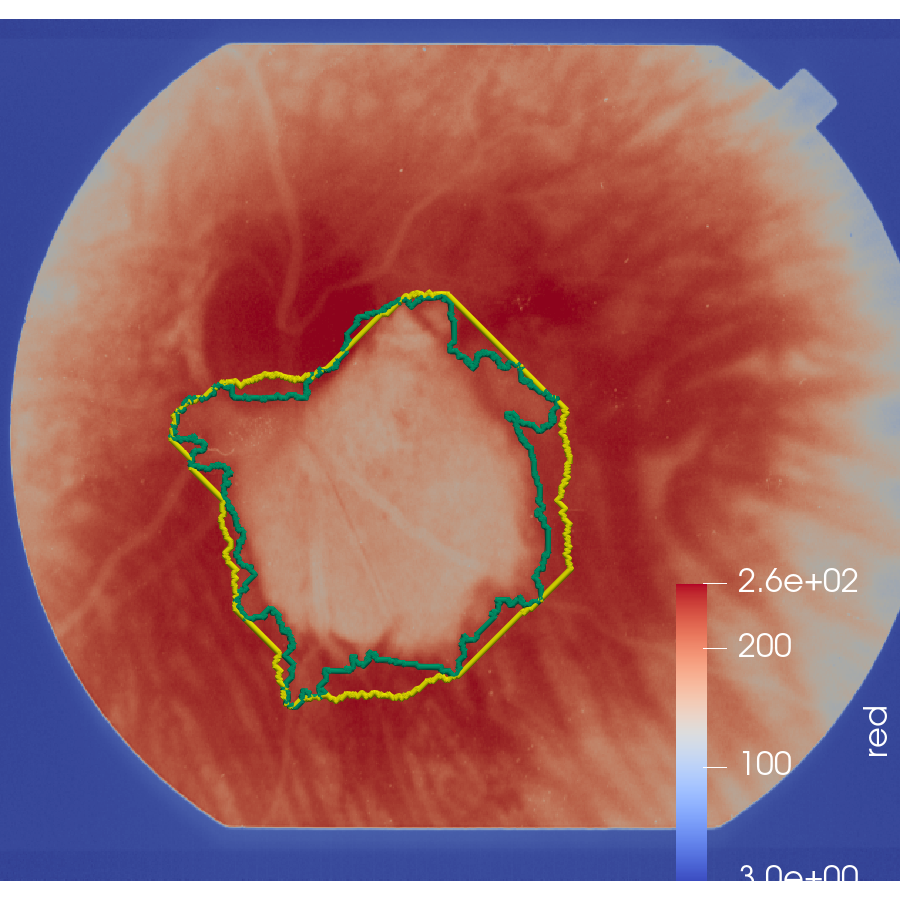} &
\includegraphics[height=1.6in]{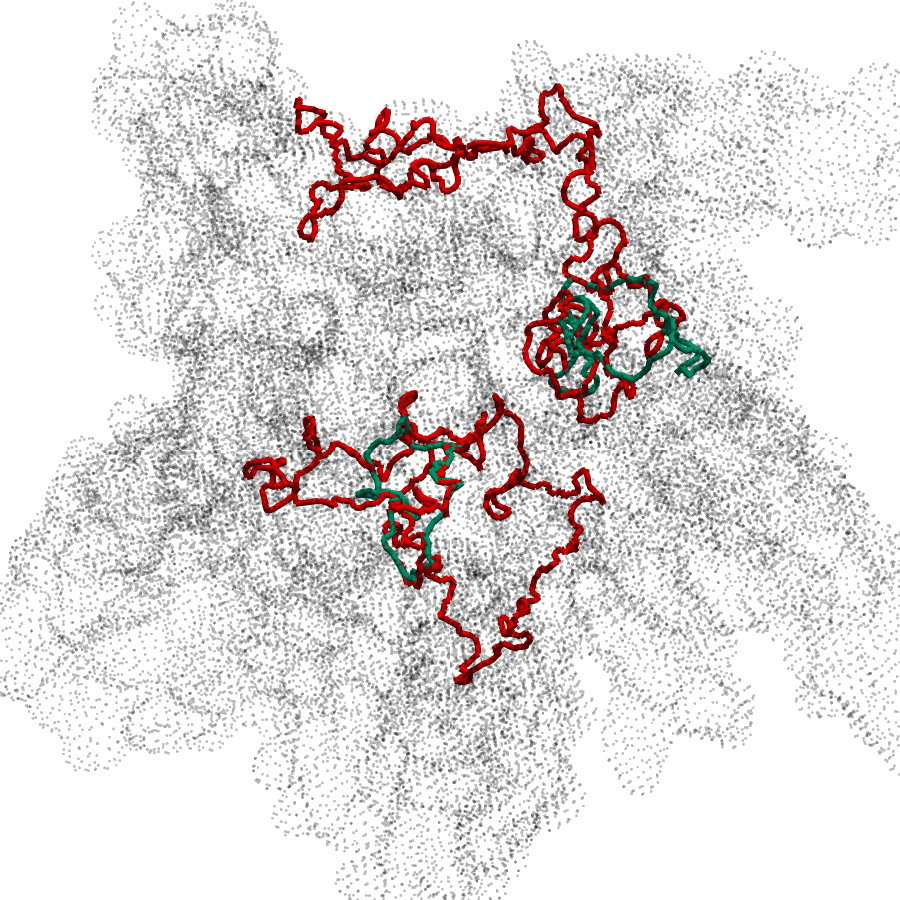} \\
\textbf{(c)} Retina & \textbf{(d)} 1OED
\end{tabular}
\caption{The green persistent 1-homology cycles computed by our algorithm are tighter than the red (or yellow) cycles computed by PersLoop~\cite{deyperstwo}.}
    \label{fig:pers_hom_representative}
\vspace{-0.1in}
\end{figure}

\paragraph{Homology localization.}
Given an input simplicial complex representing a surface and a 1-cycle, we compute a localized cycle that is homologous to the input cycle.
{\Cref{fig:homloc-hombasis}}(top) shows two input cycles (red) and their localized versions (blue) for the Happy Buddha dataset. 
We can visually infer that the localized cycle computed by our approximate algorithm is close to the optimal cycle.

\paragraph{Optimal homology basis.}
{\Cref{fig:homloc-hombasis}}(bottom) shows the optimal homology basis computed for two 3D models. We observe that the cycles are tight and capture all tunnels and loops of the model. Results on additional datasets are available in~\Cref{sec:furtherexperiments}.

\paragraph{Optimal persistent homology cycle.}
We report our results on three classes of filtrations: Rips, lower star, and Delaunay.
\Cref{fig:pers_hom_representative}~\textbf{(a,b)} shows the Lorenz'63 data set for 60\% and 80\% densities and the representatives of the longest and the top two longest bars, respectively, of a Rips filtration.  
These are of relevance in simulating weather phenomena~\cite{StrommenWeather}. 
\Cref{fig:pers_hom_representative}~\textbf{(c)} highlights the representative of the longest lived bar for a lower star filtration on a retinal image~\cite{STARERetina} with a retinal disorder. The cycle represents the region of the disorder. 
\Cref{fig:pers_hom_representative}~\textbf{(d)} highlights representatives of the top two bars of an alpha complex on a protein molecule (PDB-ID: 1OED).
Cycles computed by our algorithm (green) appear ``tighter'' than those computed by PersLoop (red or yellow).

\paragraph{Execution time.}
Our algorithms are parallelizable if the optimization subroutines for each site can be executed independently.
We obtain fast running times ($\sim$~few minutes) for moderate to large-sized datasets ($\sim$~millions of simplices) for all algorithms, see \Cref{table:persmhb_runtime}.
\begin{table}[h!]
\begin{center}
\begin{tabular}{|c|c|c|}
\hline
Data set  &  \#Simplices & Execution time \\
\hline
Lorenz-63(60\%) &   $\sim 2\times 10^6$ & 120s \\
Lorenz-63(80\%) &   $\sim 3\times 10^6$ & 120s \\
Retina &   $\sim 5\times 10^6$ & 140s \\
1OED &   $\sim 2.5\times 10^6$ & 130s\\
\hline
\end{tabular}
\caption{Mean time to compute optimal representative of each of the top-40 bars in the persistence barcode.}
\label{table:persmhb_runtime}
\end{center}
\end{table}

%

\section*{Acknowledgements}
This work is partially supported by the PMRF, MoE Govt. of India, a SERB grant CRG/2021/005278, an NSF grant CCF 2049010 and ERC 101039913. 
VN acknowledges support from the Alexander von Humboldt Foundation, and Berlin MATH+ under the Visiting Scholar program. Part of this work was completed when VN was a guest Professor at the Zuse Institute Berlin. AR would like to thank Tamal K. Dey for several helpful discussions and for his invaluable support.

\bibliographystyle{abbrv}
\bibliography{main}

\newpage
\onecolumn
\appendix

\cancel{
\section{Further algorithms for localization of chains and cycles}

In this section, we consider two additional optimization problems for $\radmeasure$, namely, minimum homology basis and minimum bounding chain.

\subsection{Computing minimum homology basis} \label{sec:minhombasis}

Given a simplicial complex $\complex$ defined on vertices $P \subset \RBB^d$ for some fixed $d$,  we want to compute a basis of $H_p(\complex)$ with the minimum
weight.  Recall that a basis of $H_p(\complex)$ is taken as a set of cycles
$\{\zeta_1,\ldots,\zeta_{\beta_p}\}$, $\beta_p=\mathrm{rank} \,\, H_p(\complex)$, whose classes are independent and generate
$H_p(K)$.
Precisely, we consider the  problem of computing \pMHB  
\begin{framed} \label{box:eqwtfun}
Given $\complex$, compute a basis
$\Lambda_p = \left\{ \zeta_i \mid i \in [\beta_p(\complex)] \right \}$
  that minimizes 
    $ r(\Lambda_p) = \sum_{i=1}^{\beta_p(\complex)} r(\zeta_i)$. 
 \end{framed}

\cancel{
\begin{center}
\begin{algorithm}[!h] 

\SetKwProg{myproc}{Procedure}{}{}
 \myproc{\textsc{MinHomBasis}{$(\complex,p)$}}{
 {$\ell\gets 1$}\\
 {Let $\boldM$ be an empty matrix} \\
 {Compute $\beta_p(\complex)$} \\
 \For{$i=1$ to $\beta_p(\complex)$}{
{$(r(\zeta_i), \vertexset, \level,\zeta_i) \gets \textsc{BinSearch}{(\complex,\textsc{PropertyMHB},\rhomb(\pointset),p,\ell,n)}$} \\
{$\ell \gets \level$} \\
{$\boldM \gets [\boldM \mid \zeta_i]$}
 }
{ \textbf{return } $\boldM$}
 }
 \medskip
  \myproc{\textsc{PropertyMHB}{$(\complex, V, p, \boldM)$}}{
  {Compute a matrix $\boldH$ with a $p$-th homology basis  for complex $\complex_V$ in its columns.} \\
  {Let $\overline{\boldH}$ be the extension of $\boldH$ in $\complex$.}\\
  {Compute the column rank profile of the matrix $\boldW = [\partial_{p+1}(\complex) \mid \boldM \mid \overline{\boldH}]$.} \\
  \If{the rank profile of $W$ returns a nonzero column at some index $i$ in the submatrix $\overline{\boldH}$ of $\boldW$}{\textbf{return } $\zeta = \overline{\boldH}[i]$}
  \Else{\textbf{return } $\emptyset$}
  }
 
\caption{Algorithm for computing minimum $p$-th homology basis}
\label{alg:minhombasis}
\end{algorithm}
\end{center}

  The property $\property$ for \pMHB is described by the procedure \textsc{PropertyMHB} in \Cref{alg:minhombasis}. \textsc{PropertyMHB} takes the matrix $\boldM$ as an (extra) argument. The columns of $\boldM$ constitute a partially computed basis for $H_p(\complex)$. The goal of \textsc{PropertyMHB} is to  check if this basis can  be extended using a $p$-cycle supported by $\complex_V$ that is linearly independent of cycles represented by columns of $\boldM$. This is achieved as follows. We first compute a matrix $\boldH$ whose columns represent an arbitrary homology basis of $H_p(\complex_V)$. We then build the matrix $\overline{\boldH}$, which is the extension of $\boldH$ in $\complex$.
  Next, we assemble the matrix $\boldW = [\partial_{p+1}(\complex) \mid \boldM \mid \overline{\boldH}]$.  By the \emph{column rank profile} of the matrix $\boldW$, we mean the lexicographically smallest sequence of indices of linearly independent columns of $\boldW$.  It is not difficult to check that the basis computed so far in matrix $\boldM$ can be extended by some cycle in $\overline{\boldH}$ if and only if the column rank profile of $\boldW$ has a nonzero column in the submatrix $\overline{\boldH}$ of $\boldW$.
  
  The main subroutine for computing a minimum $p$-th homology basis for a complex $\complex$ is given by procedure \textsc{MinpHomBasis} in \Cref{alg:minhombasis}. The main idea is to iteratively build a matrix $\boldM$ whose columns maintain a partially computed minimum homology basis. For every $i \in [\beta_p(\complex)]$, we invoke \textsc{BinSearch} to extend the basis matrix $\boldM = [\zeta_1 \mid \dots \mid \zeta_{i-1}]$ with an additional basis element. 
  Specifically, \textsc{BinSearch} computes a cycle represented by the vector $\zeta_i$ that is linearly independent of the cycles represented by $\{ \zeta_1,\dots,\zeta_{i-1} \}$.
  The correctness of this greedy algorithm
  follows from the standard argument of
  matroid extension that has been used
  in earlier works~\cite{dey2010approximating,erickson2005greedy}.
 The variable $\ell$ is the level of the Rhomboid tiling at which the current basis element $\zeta_i$ is found. 
We see that {\sc MinHomBasis} calls
{\sc BinSearch} $\beta_p(K)=O(N)$ times and {\sc PropertyMHB} runs in $O(N^\omega)$ time with matrix operations. Thus, {\sc MinHomBasis} takes at most
$O(\kappa(d+1)N^{\omega+1}\log n+n^{d+1})$ time.
}

\subsection{Computing minimum bounding chain.} \label{sec:boundchain}

The algorithm for optimal homologous cycle can easily be adapted to solve the following \MBC problem:
\begin{framed}
Given $\complex$ and a trivial cycle $\zeta \in B_{p-1}(\complex)$, compute a $p$-chain $\xi$ with minimum $r(\xi)$ and $\partial \xi = \zeta$.
\end{framed}

\cancel{
We need to define the property $\property$ appropriately. We say that
for a rhomboid $\rho\in\rhomb(\pointset)$ 
with a base vertex $s_V$, the complex $\complex_V$ satisfies property $\property$ if and only if \Cref{eq:linsolvetwo} has a solution.
\begin{equation} \label{eq:linsolvetwo}
\overline{\partial_p(\complex_V)}\cdot \xi = \zeta   
\end{equation}
Invoking \textsc{BinSearch} in \Cref{alg:bin-search} with $\property$ as defined above with \mbox{start $=1$} and \mbox{end $=n$}, we obtain an algorithm for \MBC. For any complex $\complex_V$ property $\property$ can be checked in $O(N^{\omega})$ time using standard matrix operations.  Hence, as in the previous case, the algorithm runs in 
$O(\kappa(d+1)N^\omega\log n+n^{d+1})$ time.
}


}
\section{Persistent homology}

In this section, we provide the basic definitions for persistent homology, and the usual distances used in this context, namely, the interleaving distance and the matching distance.  

Given a nested sequence of simplicial complexes indexed over $\NBB$, $\ZBB$, or
$\RBB$,
persistent homology of this sequence
captures how a homology class evolves
over the sequence. Formally, let $\PCC$ denote one of the poset categories $\NBB$, $\ZBB$, or $\RBB$. Given a finite complex $\complex$, let
$\mathcal K$ denote the set of all possible subcomplexes of $\complex$ including the empty one. A $\PCC$-indexed \emph{filtration} of 
$\complex$ is a map $\mathcal F: \PCC\rightarrow \mathcal K$ that satisfies
$\mathcal F (a)\subseteq \mathcal F(a')$ for
every pair of indices $a,a'\in \PCC$ with $a\leq a'$. In this case, $\mathcal F$ takes a sequence of real numbers 
\[a_0 \leq a_1 \leq a_2 \leq \dots \] 
to a nested sequence of simplicial complexes $\complex_{a_i}\subseteq \complex$
\[
{\mathcal F}: \dots\hookrightarrow  \complex_{a_0} \hookrightarrow \complex_{a_1} \hookrightarrow  \complex_{a_2} \hookrightarrow \dots \]
Using $p$-th homology groups of the complexes over the field $\ZBB_2$, we get a sequence 
of vector spaces connected by inclusion-induced linear maps:
\[
{H_p\mathcal F}: \dots\rightarrow  H_p(\complex_{a_0}) \rightarrow H_p(\complex_{a_1}) \rightarrow  H_p(\complex_{a_2}) \rightarrow \dots \]
For any pair $a_i,a_j$, let
$\iota_{i,j}: H_p(\complex_{a_i})\rightarrow H_p(\complex_{a_j})$ denote the linear
map (internal morphism) induced by the inclusion $\complex_{a_i}\hookrightarrow \complex_{a_j}$.
The sequence $H_p\mathcal F$ with the linear maps is called a \emph{persistence
module} which satisfies the following two
properties : (i) for any 
triple $a_i\leq a_j \leq a_k$ in $\PCC$, one has
$\iota_{i,k}=\iota_{j,k}\circ\iota_{i,j}$ and (ii) $\iota_{i,i}$ is identity for each $a_i\in \PCC$.

Let $b,d \in \PCC $. 
We define the \emph{interval} $[b,d)$ as 
\[[b,d)=\left\{ a\in \PCC \mid b \leq a < d \right\}. \]
There is a special persistence module
called
the \emph{interval module} ${\interval}^{[b,d)}$ associated to the interval $[b,d)$. Denoting the vector space indexed at $a\in \PCC$ as ${\interval}_a$, this interval module 
is given by 

\begin{equation*} 
{\interval}_a^{[b,d)} = 
\begin{cases} \mathbb{Z}_2 &\text{if $a\in [b,d)$} \\ 
0 &\text{otherwise} 
\end{cases} 
\end{equation*} 

together with identity maps $\id_{a,a'}: {\interval}_a^{[b,d)} \to {\interval}_{a'}^{[b,d)}$
for all $a,a' \in [b,d)$ with $a\leq a'$.

It is known due to a result of Gabriel~\cite{Gabriel72} that
a persistence module defined with finite complexes admits a decomposition
\begin{equation} \label{eq:decomp2}
   H_p{\mathcal F}\cong \bigoplus_\alpha {\interval}^{[b_\alpha,d_\alpha)} 
\end{equation}
 which is unique up to isomorphism and permutation of the intervals. The intervals $[b_\alpha,d_\alpha)$ are called the
\emph{bars}. The multiset of bars forms the
\emph{barcode} of the persistence module
$H_p\mathcal F$. Let $B_p(\mathcal F)$ denote this barcode for the filtration $\mathcal F$.

\subsection{Interleavings and matchings for persistence modules}
We need the following concept of interleavings between persistence modules in order to establish a  stability property for the so called $\radmeasure$ for $\cechfull$ filtrations.
To make the discussion accessible to a computer science audience, we strip the presentation off its original category-theoretic formulation, and instead provide a  more concrete and accessible description. 

\cancel{The definitions for epsilon shifts presented some additional challenges. I have tried hard to write it in a non-category theory language. As expected, it is now very concrete but also somewhat verbose.  Unfortunately, we do need the concept of epsilon-shift of a morphism of persistence modules in the stability proof.}

\begin{definition}[$\epsilon$-shifts of persistence modules]
Let $\epsilon$ be a non-negative real number, and let $\pmodule$ be a persistence module with internal linear maps given by ${\pself}_{a,a'}$.
Then, the \emph{$\epsilon$-shift} of $\pmodule$, denoted by $\pmodule^{\epsilon}$, is given by ${\pmodule}_a^{\epsilon} = {\pmodule}_{a+\epsilon}$ with internal linear maps ${\pself}^{\epsilon}_{a,a'} ={\pself}_{a+\epsilon,a'+\epsilon}$. 
\end{definition}

\begin{definition}[$\eps$-interleaving] \label{defn:interleave}

Let $\mathbb P$ and $\mathbb Q$ be two persistence modules indexed over the real numbers and with the internal linear
maps as ${\pself}_{a,a'}$ and ${\qself}_{a,a'}$ respectively. We say $\mathbb P$ and $\mathbb Q$ are $\eps$-interleaved if there exist two families of maps $F_a: \mathbb{P}_a\rightarrow \mathbb{Q}_{a+\eps}$ and
$G_a: \mathbb{Q}_a\rightarrow \mathbb{P}_{a+\eps}$ satisfying the following two conditions:
\begin{enumerate}
    \item ${\qself}_{a+\eps,a'+\eps}\circ F_a=F_{a'}\circ {\pself}_{a,a'}$ and ${\pself}_{a+\eps, a'+\eps}\circ G_a=G_{a'}\circ {\qself}_{a,a'}$ [rectangular commutativity]\index{rectangular commutativity}
    \item $G_{a+\eps}\circ F_a={\pself}_{a,a+2\eps}$ and $F_{a+\eps}\circ G_{a}={\qself}_{a,a+2\eps}$ [triangular commutativity]\index{triangular commutativity}
\end{enumerate}

\begin{align*} 
\xymatrix
{
\mathbb{P}:\,\ldots\ar[r] & \mathbb{P}_{a} \ar[r] \ar[dr]
& \  \mathbb{P}_{a+\eps} \  \ar[r]\ar[dr]
& \ \mathbb{P}_{a+2\eps} \  \ar[r]\ar[dr]
& \ \ldots \ldots\ 
\\
\mathbb{Q}: \,\ldots\ar[r] & \mathbb{Q}_{a} \ar[r]\ar[ur]
& \  \mathbb{Q}_{a+\eps} \  \ar[r]\ar[ur]
& \ \mathbb{Q}_{a+2\eps} \  \ar[r]\ar[ur]
& \ \ldots \ldots  
}
\end{align*} 

Some of the relevant maps for interleaving between two modules are shown above whereas
the two parallelograms and the two triangles below depict the rectangular and the triangular commutativities respectively.
\label{def:eps-interleaving}
\end{definition}
\begin{align*}
\xymatrix{
\mathbb{P}_{a} \ar[dr]^{F_a} \ar[rr]^{{\pself}_{a,a'}} && \mathbb{P}_{a'}\ar[dr]^{F_{a'}}&\\
&\mathbb{Q}_{a+\eps}\ar[rr]_{{\qself}_{a+\eps,a'+\eps}} &&\mathbb{Q}_{a'+\eps}
 }&
\xymatrix{
&\mathbb{P}_{a+\eps} \ar[rr]^{{\pself}_{a+\eps,a'+\eps}} && \mathbb{Q}_{a'+\eps}\\ \mathbb{Q}_{a}\ar[ur]^{G_{a}}\ar[rr]_{{\qself}_{a,a'}} &&\mathbb{Q}_{a'}\ar[ur]^{G_{a'}}
 }
 \end{align*}
\begin{align*}
\xymatrix{
	\mathbb{P}_{a} \ar[dr]^{F_a} \ar[rr]^{{\pself}_{a,a+2\eps}} && \mathbb{P}_{a+2\eps}\\
	&\mathbb{Q}_{a+\eps} \ar[ur]^{G_{a+\eps}}&
} &
\xymatrix{
	&\mathbb{P}_{a+\eps} \ar[rd]^{F_{a+\eps}}&\\
	\mathbb{Q}_{a}\ar[ur]^{G_a}\ar[rr]^{{\qself}_{a,a+2\eps}} & &\mathbb{Q}_{a+2\eps}
} 
\end{align*}
\begin{definition}[Interleaving distance]\index{interleaving!distance}
Given two persistence modules $\mathbb{P}$ and $\mathbb{Q}$, their interleaving distance is defined as
$$
\dd_I(\mathbb{P},\mathbb{Q})=\inf\{\eps\,|\, \mbox{$\mathbb{P}$ and $\mathbb{Q}$ are $\eps$-interleaved}\}.
$$
\end{definition}

\begin{remark}
In \Cref{defn:interleave}, the family of maps $\{F_a \mid a\in \RBB \}$ assemble to give a map $F: \pmodule \to \qmodule^{\epsilon}$, and the family of maps $\{G_a \mid a\in \RBB \}$ assemble to give a map $G: \qmodule \to \pmodule^{\epsilon}$.
\end{remark}

\begin{definition}[$\epsilon$-shift of a family of maps]
Suppose there exists a family of maps $F: \pmodule \to \qmodule$.
Then, a shift of $F$ by $\epsilon$, denoted by $F^{\epsilon}$, gives a new family maps $\{{F}^{\epsilon}_a = {F}_{a+\epsilon} \mid a \in \RBB \}$ from $\pmodule^{\epsilon}$ to $\qmodule^{\epsilon}$.  
\end{definition}

We now define the notion of $\epsilon$-matchings between two barcodes. 

\begin{definition}[$\epsilon$-matching of barcodes]
Suppose that we are given two barcodes (which are multisets of intervals) $X$ and $Y$.
Then, a matching $\matching$ between $X$ and $Y$ is a collection
of pairs $\matching=\left\{ (\mathbf{i},\mathbf{j})\mid\mathbf{i}\in X,\mathbf{j}\in Y\right\} $
such that each $\mathbf{i}$ and $\mathbf{j}$ occur in at most one
pair. 

Let $\mathcal{U}(\matching)$ be the collection of intervals in $X\cup Y$
that do not appear in any of the pairings of $\matching$.

For pairs $(\mathbf{i},\mathbf{j})\in\matching$ , where $\mathbf{i}=(b,d)$
and $\mathbf{j}=(b',d')$, define $c(\mathbf{i},\mathbf{j})=\max(|b-b'|,|d-d'|)$
and for intervals $\mathbf{i}=[b,d)\in\mathcal{U}$, define $c(\mathbf{i})=\frac{d-b}{2}$. Then, the cost of the matching $\matching$, denoted by $c(\matching)$, is defined as follows:
\[
c(\matching)=\max\left(\sup_{(\mathbf{i},\mathbf{j})\in\matching}c((\mathbf{i},\mathbf{j})),\sup_{\mathbf{i}\in \mathcal{U}(\matching)}c(\mathbf{i})\right).
\]
Finally, we say that $\matching$ is an $\epsilon$-matching if $c(\matching)\leq\epsilon$.

\end{definition}


\cancel{

\section{A structural result for persistent homology basis}

We now provide a new characterization for the set of representative cycles $\represents(\boldi)$ for intervals $\boldi \in \fpbarcode$. 
Suppose that we are given a filtration $\mathcal{F}$ on a complex $\complex$. For simplicity, we assume that $\mathcal{F}$ is simplex-wise. For filtrations that are not simplex-wise, an analogous characterization can be obtained. 
Let $ \fpbarcode$ be the barcode  of the persistence module  $H_p{\mathcal F}$. 
Let $\xi$ be a representative $p$-cycle for the interval $\boldj = [b,d) \in \fpbarcode$. 
Let $\basisset$ be a persistent $p$-basis for $\mathcal{F}$, and let  $ \basisset_b = \{ \zeta_i \mid i \in W \} $ where $W$ indexes the intervals $[s,t)$ of $\fpbarcode$ with $s<b$. Let $ \basisset_b^{\leq d}$ be the restriction of $ \basisset_b$ to intervals $[s,t)$ with $t\leq d$. Finally, $ \basisset_b^{>d}$ be the complement of $ \basisset_b^{\leq d}$ in $ \basisset_b$. 

To begin with, we consider a strict partial order (an anti-reflexive, assymetric and transitive relation)  on the intervals of $\fpbarcode$. 
Let $\boldi = [b_1,d_1)$ and $\boldj = [b_2,d_2)$ be any two intervals in $\fpbarcode$.  Then we write $\boldi <_{B} \boldj$ if and only if $b_1 < b_2$ and $d_1 < d_2$. The relation $<_B$ on the intervals of $\fpbarcode$ descends to a relation $<_B$ on the representative cycles for the intervals in $\fpbarcode$.  Precisely, for any $\zeta \in \represents(\boldi)$ and any $\xi \in \represents(\boldj)$, we say that $\zeta <_{B} \xi$ if and only if $\boldi <_{B} \boldj$. Then, we can describe the representative cycles for intervals in a barcode in terms of the relation $<_{B}$.

\begin{thm} \label{thm:perschar}
Let $\xi$ be any representative cycle for the interval $\boldj =  [b,d)$.
 Then, every representative cycle $z \neq \xi$ of interval $\boldj$ can be written in the form 
$z = \xi + \sum_{i\in S} \zeta_{i} + \delta$, where $S$ indexes a subset of cycles in $ \basisset_b^{\leq d}$ and $\delta$ is a boundary in $\complex_b$. 
\end{thm}
\begin{proof*}
Since $z$ and $\zeta$ are distinct representatives of $\boldj$, $z + \xi$ is only incident on   simplices with index less than $b$.
Thus, $z + \xi$ is a nonzero cycle that is born before $b$.  Also, because $z$ and $\xi$ are both trivial at $d$, it follows that they can be written as $z = \partial \alpha$ and $\xi = \partial \gamma$, respectively, for some chains $\alpha, \gamma \in C_p(\complex_d)$. Then, $z + \xi$ can be written as $z + \xi =\partial (\alpha + \gamma) $ at index $d$. That is, $z + \xi$ is trivial at $d$.  Since $\basisset$ is a persistent homology basis, at index $b$, $z+\xi$ can be written as 
\[ z + \xi = \sum_{i \in S} \zeta_i + \delta \]
where $S$ indexes some subset of $ \basisset_b^{\leq d}$, and $\delta$ is a boundary in $\complex_b$. The claim follows.
\end{proof*}

In other words, \Cref{thm:perschar} says that given a representative $\xi$ for an interval $\boldj$, any other representative $z$  of $\boldj$ can be written as  $z = \xi + \sum_{i \in S} \zeta_{i} + \delta$, where for every $i \in S$, the cycles $\zeta_i$ satisfy $\zeta_i <_{B} \xi$. Equivalently, for every $i\in S$, where $\zeta_i$ represents an interval $\boldi$, we have $\boldi <_{B} \boldj$.

}





\cancel{

\section{Optimal persistent homology basis for non-simplexwise filtrations} \label{sec:nonsimp}

In this section, we expand upon the notion of persistent homology basis for general (non-simplexwise) filtrations. One important way in which non-simplexwise filtrations are different is that unlike in the case of simplexwise filtrations, the barcodes of such filtrations may have multiple instances of certain intervals. That is, in general, the  barcodes of non-simplexwise filtrations are  multisets. Formally, a multiset $M$ can be written as an ordered pair $M = (S,m)$, where $S$ is the \emph{underlying set}, and $m$ is a positive integer valued function on $S$, where for all $a\in S$, $m(a)$  is the multiplicity of $a$ in $M$.

Next, we establish some notation for the remainder of the section. For a non-simplexwise filtration $\mathcal{F}$, and its $p$-th  persistence barcode $B_p(\mathcal{F})$, let $\widetilde{B_p(\mathcal{F})}$ denote the underlying set of $B_p(\mathcal{F})$. We write $B_p(\mathcal{F}) = (\widetilde{B_p(\mathcal{F})},m)$ where $m$ is the multiplicity function for the intervals of $B_p(\mathcal{F})$.
For an interval $\boldi \in \widetilde{B_p(\mathcal{F})}$, we call the ordered pairs $(\boldi,j)$ for $j \in [m(\boldi)]$ the \emph{instances of $\boldi$} in $B_p(\mathcal{F})$.

\begin{definition}
For an interval $\boldi = [b,d) \in B_p(\mathcal{F})$, we say that the cycles $\{\zeta_1,\dots,\zeta_{m(\boldi)}\}$ are the \emph{representative cycles} for the instances of $\boldi$, or  simply  $\{\zeta_j \mid j \in [m(\boldi)]\}$ \emph{represent} $\boldi$, if the following  conditions are satisfied:
\begin{itemize}
\item The cycles in the set $\{\zeta_j \mid j \in [m(\boldi)]\}$ are linearly independent. 
\item One of the following two conditions hold for every $j \in m(\boldi)$.
\begin{itemize}
\item[$d\neq +\infty$] \quad  $\zeta_j$  is a cycle in $\complex_{b}$ containing a simplex that was inserted at index $b$, and $\zeta_j$ is not a boundary in $\complex_{d-1}$, but becomes a boundary in $\complex_{d}$.
\item[$d= +\infty$] \quad $\zeta_i$ is a cycle in $\complex_{b}$ containing a simplex inserted at index $b$.
\end{itemize}
\end{itemize}
\end{definition}

\begin{definition}[Persistent cycles]
\label{defn:pers_cyc_stronger}
For an interval $\boldi \in \widetilde{B_p(\mathcal{F})}$, the $p$-cycles  $\{\zeta_j \mid j \in [m(\boldi)]\}$ that represent  an interval $\boldi = [b,d)\in B_p(\mathcal{F})$ are called   \emph{persistent $p$-cycles} for $\boldi$.
\end{definition}

\begin{definition}[Persistent basis] \label{defn:pers_basisgen}
Let $I$ be the indexing set for the intervals in the set $\widetilde{B_p(\mathcal{F})}$ associated to the filtration $\mathcal{F}$. 
That is, for every $i\in I$, $\boldi = [b_i,d_i)$ is an interval in $\widetilde{B_p(\mathcal{F})}$.
Then a set of $p$-cycles $\{\zeta_i^j \mid i\in I, j\in [m(\boldi)]\}$ is called a \emph{persistent $p$-basis} for $\mathcal{F}$ if 
\[H_p{\mathcal F} \,\, = \,\, \bigoplus_{i\in I, j\in [m(\boldi)]} {\interval}^{\zeta_i^j}
\]
where ${\interval}^{\zeta_i^j}$ is defined by 
\begin{equation*} 
{\interval}_a^{\zeta_i^j} = 
\begin{cases} [\zeta_i^j] &\text{if $a\in [b_i,d_i)$} \\ 
0 &\text{otherwise.} 
\end{cases} 
\end{equation*} 
Here, for every $j\in J$ and  every $a,a' \in [b_j,d_j)$ with $a\leq a'$ the maps ${\interval}_a^{\zeta_i^j} \to {\interval}_{a'}^{\zeta_i^j}$
 are the induced maps on homology restricted to $[\zeta_i^j]$, respectively.
\end{definition}


The following theorem that relates persistence cycles to persistent bases for  non-simplexwise filtrations.

\begin{thm}
Let $I$ be an indexing set for the  set  $\widetilde{B_p(\mathcal{F})}$ associated to filtration $\mathcal{F}$.
Then, the set of $p$-cycles $\{\zeta_i^j \mid i\in I, j\in [m(\boldi)]\}$ is a persistent $p$-basis for a filtration $\mathcal{F}$ if and only if the cycles $\{\zeta_i^j \mid j \in [m(\boldi)]\}$ represent $\boldi$ for every $i \in I$.
\end{thm}
\begin{proof*}
The proof uses the same argument as that of  Theorem 1~\cite{deypersone}.
\end{proof*} 

\begin{remark}
Recall that we use the notation 
$\overline{\xi}$ and matrices $\overline{\boldM}$, for \emph{extensions} of $\xi$ and $\boldM$ in $\complex$. The inverse operation of  extension of a cycle is called a \emph{contraction}.
That is, if a cycle $\overline{\xi}$ in $\complex$ is an extension of  $\xi$ in $\subcomplex \subset \complex$, then  $\xi$ is the contraction of $\overline{\xi}$ in $\subcomplex$.

\end{remark}

\cancel{

\begin{center}
\begin{algorithm}[!htb] 

\SetKwProg{myproc}{Procedure}{}{}
 \myproc{\textsc{PropertyPMHBGen}{$(\complex,V,p,b,d,\boldM_{[b,d)})$}}{
 
 	\KwIn{$[b,d)$ is a $p$-th interval. $s_V$ is a vertex in the Rhomboid tiling}
 	\Comment{The subroutine \textsc{PropertyPMHBGen} checks if $V$ induces a complex that supports a representative for the interval  $[b,d)$}\\
 {Let $\complex_{b}^{V}$ be the complex induced by vertices $V$ in $\complex_b$} \\
 {Compute a basis for $H_p(\complex_{b}^{V})$, and let $B_{V}$ be the matrix with basis vectors in its columns}\\
 {Let $\overline{B_V}$ be the extension of $B_V$ in $\complex_b$} \\ 
   {Compute the column rank profile of the matrix $\boldW = [\partial_{p+1}(\complex_b) \mid \boldM_{[b,d)} \mid \overline{B_V}]$} \\  
   {Let $I$ be the set of indices of columns in $\overline{B_V}$ that are zeroed out in the column rank profile computation} \\
   {Remove the columns with indices in $I$ from $\overline{B_V}$} \\
 \If{none of the columns of $\overline{B_V}$ are incident on a simplex inserted at $b$}{\textbf{return} \quad \texttt{false}} 
 {Let $z$ be a column of $\overline{B_V}$  that is incident on some simplex $\sigma$ inserted at $b$.}\\
 {Modify $\overline{B_V}$ so that there is a unique column $z$ incident on $\sigma$ (by adding $z$ to any other column of $\overline{B_V}$ that is also incident on $\sigma$).} \\ 
 {Let $B$ be the matrix formed by removing $z$ from $\overline{B_V}$} \\
   {Let $\overline{B}$ be the extension of $B$ in $\complex_d$, and let $\overline{z}$ be the extension of $z$ in $\complex_d$} \\
 {Let $A_d^p$ be the $p$-th annotation matrix for $\complex_d$} \\
 \uIf{$A_d^p \cdot \overline{z} = 0$}{
 \textbf{return} the contraction of $\overline{z}$ in $\complex_b$}
 \uElseIf{the linear system $A_d^p\cdot \overline{B}\cdot x = A_d^p\cdot \overline{z}$ has a solution $x$}{
 \textbf{return} the contraction of $\overline{B}\cdot x + \overline{z}$ in $\complex_b$}
 {\textbf{return }$\emptyset$}
 }
 \medskip
\myproc{\textsc{MinPersHomBasisGen}{$({\mathcal F},\complex,p)$}}{
	\KwIn{Filtration $\mathcal F$ of complex $\complex$ \quad \textbf{Output:} Min. persistent homology basis for $H_p(\complex)$}

 {Compute $B_p(\mathcal{F})$} \\
  {Let $\boldM$ be an empty matrix} \\
 \For{every interval $[b,d)\in B_p(\mathcal{F})$ with multiplicity $m$}{
  {Let $\boldM_{[b,d)}$ be an empty matrix} \\
 \For{$i=1$ to $m$}{
 {$(r_{\min}, \vertexset, \level,\zeta) \gets \textsc{BinSearch}{(\complex,\textsc{PropertyPMHBGen},\rhomb(\pointset),1,n)}$} \\
{$\boldM_{[b,d)} = [\boldM_{[b,d)} \mid \zeta]$}
 }
 {$\boldM = [\boldM  \mid \overline{\boldM_{[b,d)}}$], where $\overline{\boldM_{[b,d)}}$ is the extension of $\boldM_{[b,d)}$ in $\complex$} 
 }
{ \textbf{return }$\boldM$}
   }

\caption{Algorithm for computing minimum persistent homology basis}
\label{alg:persbasismodgen}
\end{algorithm}
\end{center}
}

\cancel{

\Cref{alg:persbasismodgen} is a minor variation of \Cref{alg:persbasismodgen} wherein we address the case when the barcode  $B_p(\mathcal{F})$ of a filtration $\mathcal{F}$ is a multiset. 
In particular, for every interval $\boldi = [b_i,d_i)$ we compute the representative cycles $ \Omega_i = 
\{\zeta_i^j \mid \boldi \in B_p(\mathcal{F}), j\in m(\boldi)\}$.  

Another notable change from \Cref{alg:persbasis} occurs in Lines~765~and~767 of \Cref{alg:persbasismodgen} where the contraction of column vectors is carried out to ensure that all vectors in the matrix $\boldM_{[b,d)}$ have the same size.

The property $\property$ for \persMHB is described by the procedure \textsc{PropertyPMHBGen} in \Cref{alg:persbasismodgen}. The only difference from \textsc{PropertyPMHB} in \Cref{alg:persbasis} is that for every interval $\boldi = [b_i,d_i)$, we augment an entire matrix of representative cycles $\overline{\boldM_{[b,d)}}$ instead of augmenting a single cycle vector. 

}

\cancel{

\section{Missing proofs}
}
\cancel{
\begin{corollary} \label{cor:maincoreapx}
There is an matching of representative cycles of $\Cech(\pointsetone)$ with representative cycles of $\Cech(\pointsettwo)$ such that 
\begin{itemize}
    \item All persistent cycles of $\Cech(\pointsetone)$  and $\Cech(\pointsettwo)$ representing intervals of length greater than $2\epsilon$ are matched.
    \item If a representative $\zeta_i$ of  $\Cech(\pointsetone)$ is matched to a representative $z_j$ of $\Cech(\pointsettwo)$, then $\iinterval$ is $\epsilon$-interleaved with $\jinterval$, and $F_{\boldi,\boldj}$ and $G_{\boldj,\boldi}$ are nonzero.
\end{itemize}
\end{corollary}
\begin{proof*}

\end{proof*}
}

}


\section{Approximate stability for  \texorpdfstring{$\ell_2$}--radius} \label{sec:repmatch}

We now study notions of stability for weight functions that serve as objective functions for intervals in a barcode. We limit the discussion to $\cechfull$ filtrations, while noting that similar definitions can be derived for other commonly encountered filtrations like Rips or lower star. 

\subsection{Stability of weight functions on intervals}

We recall the notion of $\cechfull$ complexes before proceeding to introduce some new definitions of our own concerning stability of weight functions on intervals.

\begin{definition}[$\cechfull$ complexes]
Let $\pointset$ be a finite point set in $\RBB^d$. Let $D_{r,x}$ denote a Euclidean ball of radius $r$ centered at $x$. The $\cechfull$ complex of $P$ for radius $r$ is the abstract simplicial complex given by 
\[
\Cech_r(P) = \{ X \subset P \mid \bigcap_{x\in X}  D_{r,x} \neq \emptyset\}.
\]

The $\cechfull$ filtration of $P$, denoted by $ \Cech(P)$, is the nested sequence of complexes $\{\Cech_r(P)\}_{r\geq 0}$, where $\Cech_s(P) \subseteq \Cech_t(P) $ for $s \leq t$. 
We use the notation $B(\Cech(P))$ to denote the barcode of $\Cech(P)$. 
\end{definition}

\cancel{
\begin{definition}[Vietoris-Rips complexes]
Let $\pointset$ be a finite point set in $\RBB^d$. 
The Vietoris-Rips complex of $\pointset$ at scale $t$ consists of all simplices with diameter at most $t$, where the diameter of a simplex is the maximum distance between any two points in the simplex. In other words, 
\[ \Rips_t(P) = \{ S \subset X \mid \diam S \leq t  \}. \]
The Rips filtration of $P$, denoted by $\Rips(P)$, is the nested sequence of complexes $\{\Rips_t(P)\}_{t\geq 0}$, where $\Rips_s(P) \subseteq \Rips_t(P) $ for $s \leq t$. 
We use the notation $B(\Rips(P))$ to denote the barcode of $\Rips(P)$. 
\end{definition}
}

\begin{definition}[$\epsilon$-perturbations of point sets]
We say that $\pointsettwo$ is an \emph{$\epsilon$-perturbation} of $\pointsetone$ \emph{realized through} a bijective map $f$, if $f$ maps  points in $\pointsetone$ to points in $\pointsettwo$ such that $\|p-f(p)\|_2\leq \epsilon$. 
\end{definition}

\begin{definition}[Weight functions on persistent homology bases]
Given a filtration $\mathcal{F}$,
and a $p$-th persistent homology basis $\basisset_p(\mathcal{F})$ of  $\mathcal{F}$,
a function that assigns a positive real number to every  cycle in $\basisset_p(\mathcal{F})$ is called a \emph{weight function on the $p$-th persistent homology basis $\basisset_p(\mathcal{F})$} of $\mathcal{F}$. In this case, the (total) weight on $\basisset_p(\mathcal{F})$ is simply the sum of weights of basis cycles. 
\end{definition}

 \begin{definition}[Weight functions on persistent barcodes]
For a filtration $\mathcal{F}$, its $p$-th barcode $B_p(\mathcal{F})$, a persistent $p$-homology basis $\basisset_p(\mathcal{F})  = \{\zeta_1\dots\zeta_{|\basisset_p(\mathcal{F})|}\}$, and a weight function $s$ on $\basisset_p(\mathcal{F})$ defined as 
 $ s(\basisset^p(\mathcal{F})) = \sum_{i=1}^{|\basisset_p(\mathcal{F}) |} s(\zeta_i)$, a weight function $w$ on $B_p(\mathcal{F})$ is defined as: 
\begin{equation} \label{eq:perswt}
   w(B_p(\mathcal{F})) = \min_{\basisset_p(\mathcal{F})} \,\, s(\basisset_p(\mathcal{F})) \quad \text{ where } \basisset_p(\mathcal{F})\text{ is a persistent $p$-homology basis for $\mathcal{F}$.} 
\end{equation}
If $\basisset^{\star}_{p}(\mathcal{F})$ is a persistent $p$-homology basis of minimum weight in \Cref{eq:perswt}, then the weight $w([b,d))$ for an interval $[b,d)\in B_p(\mathcal{F})$ is defined as $w([b,d)) = s(\zeta^{\star})$ where $\zeta^{\star}$ is the representative for $[b,d)$ in  $\basisset^{\star}_{p}(\mathcal{F})$.
\end{definition}

\begin{definition}[Stability of weight functions for filtration families] \label{defn:stabmain}
Suppose that we are given a weight function $w$  on barcodes of a family of filtrations.
Then, for a point set  $P$, an associated filtration of complexes $\mathcal{F}_P$ and for real numbers $\epsilon, \delta\geq 0$, we say  $w$ is \emph{$(\epsilon,\delta)$-stable} for an interval $[b,d) \in B(\mathcal{F}_P)$ with $(d - b) > 2\epsilon$, if
for every $\epsilon$-perturbation  $Q$ of $P$ with an associated filtration of complexes $\mathcal{F}_Q$,
there exists an $\epsilon$-matching $\matching$ between the intervals of $B(\mathcal{F}_P)$ and the intervals of $B(\mathcal{F}_Q)$    such that if the length of the interval $\matching([b,d)) > 2\epsilon$ then
\[
|w([b,d)) - w(\matching([b,d)))| \leq 2\delta.
\] 
Furthermore,  if $w$ is $(\epsilon,\delta)$-stable for all intervals of $B(\mathcal{F}_P)$, then we say that $w$ is \emph{$(\epsilon,\delta)$-stable for the point set} $P$. 
Finally, if $w$ is $(\epsilon,\delta)$-stable for every point set embedded in a Euclidean space, then we say that $w$ is \emph{$(\epsilon,\delta)$-stable for the  filtration family}. 
\end{definition}

\begin{definition}[Instability of weight functions]
Given a family of filtrations $\filfam$, and a weight function $w$ defined on the persistent barcodes of filtrations belonging to $\filfam$, we say that the pair $(\filfam,w)$ is unstable, if for every $\epsilon, \alpha >0$, there exists filtrations $\mathcal{F}_1,\mathcal{F}_2 \in \filfam$ such that the barcodes of $\mathcal{F}_1,\mathcal{F}_2 \in \filfam$ admit an $\epsilon$-matching while satisfying $|w(\mathcal{F}_1) - w(\mathcal{F}_2)| > \alpha$. 
\end{definition}

\begin{figure*}[htbp]
\centering   
        \includegraphics[width=0.6\textwidth]{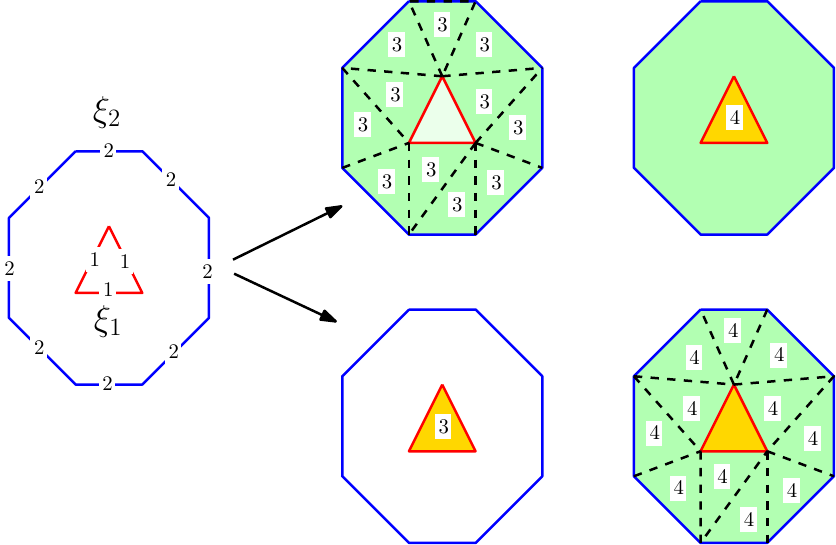}
\caption{Two $1$-cycles $\xi_1$ and $\xi_2$ appear at time (index) $1$ and $2$ respectively. In the upper filtration, the green triangles appear at time $3$ filling up the annulus and the golden triangle appears at time $4$ giving rise to the bars $[2,3)$ and $[1,4)$ with the minimal (lengthwise) persistent
cycles $\xi_1+\xi_2$ and $\xi_1$ resp. In the lower filtration the golden triangle comes first and green triangles next giving rise to slightly perturbed bars $[1,3)$ and $[2,4)$. 
Their minimal persistent cycles change to $\xi_1$ and $\xi_2$, resp. suggesting that the cycles may change considerably with respect to the length function.}
\label{fig:perscycle}
\end{figure*}

\cancel{
\begin{figure}[!h]
    \centering
\scalebox{0.6}{    
\begin{tikzpicture}[line cap=round,line join=round,>=triangle 45,x=1cm,y=1cm]
\clip(-3.653333333333338,-10.216666666666686) rectangle (8.38666666666667,4.523333333333344);
\fill[line width=2pt,color=rvwvcq,fill=rvwvcq,fill opacity=0.10000000149011612] (-2,3) -- (-2,1) -- (0,3) -- cycle;
\fill[line width=2pt,color=rvwvcq,fill=rvwvcq,fill opacity=0.10000000149011612] (-2,3) -- (-2,1) -- (-3,2) -- cycle;
\fill[line width=2pt,color=rvwvcq,fill=rvwvcq,fill opacity=0.10000000149011612] (-2,1) -- (-3,0) -- (-3,2) -- cycle;
\fill[line width=2pt,color=rvwvcq,fill=rvwvcq,fill opacity=0.10000000149011612] (-2,1) -- (-2,-1) -- (-3,0) -- cycle;
\fill[line width=2pt,color=rvwvcq,fill=rvwvcq,fill opacity=0.10000000149011612] (-2,1) -- (0,-1) -- (-2,-1) -- cycle;
\fill[line width=2pt,color=rvwvcq,fill=rvwvcq,fill opacity=0.10000000149011612] (-2,1) -- (8,4) -- (0,3) -- cycle;
\fill[line width=2pt,color=rvwvcq,fill=rvwvcq,fill opacity=0.10000000149011612] (-2,1) -- (8,-2) -- (8,4) -- cycle;
\fill[line width=2pt,color=rvwvcq,fill=rvwvcq,fill opacity=0.10000000149011612] (-2,1) -- (0,-1) -- (8,-2) -- cycle;
\fill[line width=2pt,color=dtsfsf,fill=dtsfsf,fill opacity=0.1] (0,3) -- (0,-1) -- (8,-2) -- (8,4) -- cycle;
\fill[line width=2pt,color=rvwvcq,fill=rvwvcq,fill opacity=0.10000000149011612] (-2,-5) -- (-2,-7) -- (0,-5) -- cycle;
\fill[line width=2pt,color=rvwvcq,fill=rvwvcq,fill opacity=0.10000000149011612] (-2,-5) -- (-2,-7) -- (-3,-6) -- cycle;
\fill[line width=2pt,color=rvwvcq,fill=rvwvcq,fill opacity=0.10000000149011612] (-2,-7) -- (-3,-8) -- (-3,-6) -- cycle;
\fill[line width=2pt,color=rvwvcq,fill=rvwvcq,fill opacity=0.10000000149011612] (-2,-7) -- (-2,-9) -- (-3,-8) -- cycle;
\fill[line width=2pt,color=rvwvcq,fill=rvwvcq,fill opacity=0.10000000149011612] (-2,-7) -- (0,-9) -- (-2,-9) -- cycle;
\fill[line width=2pt,color=rvwvcq,fill=rvwvcq,fill opacity=0.10000000149011612] (-2,-7) -- (8,-4) -- (0,-5) -- cycle;
\fill[line width=2pt,color=rvwvcq,fill=rvwvcq,fill opacity=0.10000000149011612] (-2,-7) -- (8,-10) -- (8,-4) -- cycle;
\fill[line width=2pt,color=rvwvcq,fill=rvwvcq,fill opacity=0.10000000149011612] (-2,-7) -- (0,-9) -- (8,-10) -- cycle;
\fill[line width=2pt,color=dtsfsf,fill=dtsfsf,fill opacity=0.1] (0,-5) -- (0,-9) -- (8,-10) -- (8,-4) -- cycle;
\draw [line width=2pt,color=wrwrwr] (8,4)-- (8,-2);
\draw [line width=2pt,color=wrwrwr] (8,4)-- (0,3);
\draw [line width=2pt,color=wrwrwr] (0,3)-- (-2,3);
\draw [line width=2pt,color=wrwrwr] (-2,3)-- (-3,2);
\draw [line width=2pt,color=wrwrwr] (-3,2)-- (-3,0);
\draw [line width=2pt,color=wrwrwr] (-3,0)-- (-2,-1);
\draw [line width=2pt,color=wrwrwr] (-2,-1)-- (0,-1);
\draw [line width=2pt,color=wrwrwr] (0,-1)-- (8,-2);
\draw [line width=2pt,color=rvwvcq] (-2,3)-- (-2,1);
\draw [line width=2pt,color=rvwvcq] (-2,1)-- (0,3);
\draw [line width=2pt,color=rvwvcq] (0,3)-- (-2,3);
\draw [line width=2pt,color=rvwvcq] (-2,3)-- (-2,1);
\draw [line width=2pt,color=rvwvcq] (-2,1)-- (-3,2);
\draw [line width=2pt,color=rvwvcq] (-3,2)-- (-2,3);
\draw [line width=2pt,color=rvwvcq] (-2,1)-- (-3,0);
\draw [line width=2pt,color=rvwvcq] (-3,0)-- (-3,2);
\draw [line width=2pt,color=rvwvcq] (-3,2)-- (-2,1);
\draw [line width=2pt,color=rvwvcq] (-2,1)-- (-2,-1);
\draw [line width=2pt,color=rvwvcq] (-2,-1)-- (-3,0);
\draw [line width=2pt,color=rvwvcq] (-3,0)-- (-2,1);
\draw [line width=2pt,color=rvwvcq] (-2,1)-- (0,-1);
\draw [line width=2pt,color=rvwvcq] (0,-1)-- (-2,-1);
\draw [line width=2pt,color=rvwvcq] (-2,-1)-- (-2,1);
\draw [line width=2pt,color=rvwvcq] (-2,1)-- (8,4);
\draw [line width=2pt,color=rvwvcq] (8,4)-- (0,3);
\draw [line width=2pt,color=rvwvcq] (0,3)-- (-2,1);
\draw [line width=2pt,color=rvwvcq] (-2,1)-- (8,-2);
\draw [line width=2pt,color=rvwvcq] (8,-2)-- (8,4);
\draw [line width=2pt,color=rvwvcq] (8,4)-- (-2,1);
\draw [line width=2pt,color=rvwvcq] (-2,1)-- (0,-1);
\draw [line width=2pt,color=rvwvcq] (0,-1)-- (8,-2);
\draw [line width=2pt,color=rvwvcq] (8,-2)-- (-2,1);
\draw [line width=2pt,color=dtsfsf] (0,3)-- (0,-1);
\draw [line width=2pt,color=dtsfsf] (0,-1)-- (8,-2);
\draw [line width=2pt,color=dtsfsf] (8,-2)-- (8,4);
\draw [line width=2pt,color=dtsfsf] (8,4)-- (0,3);
\draw (-0.5933333333333362,1.223333333333337) node[anchor=north west] {3};
\draw [line width=2pt,color=rvwvcq] (-2,-5)-- (-2,-7);
\draw [line width=2pt,color=rvwvcq] (-2,-7)-- (0,-5);
\draw [line width=2pt,color=rvwvcq] (0,-5)-- (-2,-5);
\draw [line width=2pt,color=rvwvcq] (-2,-5)-- (-2,-7);
\draw [line width=2pt,color=rvwvcq] (-2,-7)-- (-3,-6);
\draw [line width=2pt,color=rvwvcq] (-3,-6)-- (-2,-5);
\draw [line width=2pt,color=rvwvcq] (-2,-7)-- (-3,-8);
\draw [line width=2pt,color=rvwvcq] (-3,-8)-- (-3,-6);
\draw [line width=2pt,color=rvwvcq] (-3,-6)-- (-2,-7);
\draw [line width=2pt,color=rvwvcq] (-2,-7)-- (-2,-9);
\draw [line width=2pt,color=rvwvcq] (-2,-9)-- (-3,-8);
\draw [line width=2pt,color=rvwvcq] (-3,-8)-- (-2,-7);
\draw [line width=2pt,color=rvwvcq] (-2,-7)-- (0,-9);
\draw [line width=2pt,color=rvwvcq] (0,-9)-- (-2,-9);
\draw [line width=2pt,color=rvwvcq] (-2,-9)-- (-2,-7);
\draw [line width=2pt,color=rvwvcq] (-2,-7)-- (8,-4);
\draw [line width=2pt,color=rvwvcq] (8,-4)-- (0,-5);
\draw [line width=2pt,color=rvwvcq] (0,-5)-- (-2,-7);
\draw [line width=2pt,color=rvwvcq] (-2,-7)-- (8,-10);
\draw [line width=2pt,color=rvwvcq] (8,-10)-- (8,-4);
\draw [line width=2pt,color=rvwvcq] (8,-4)-- (-2,-7);
\draw [line width=2pt,color=rvwvcq] (-2,-7)-- (0,-9);
\draw [line width=2pt,color=rvwvcq] (0,-9)-- (8,-10);
\draw [line width=2pt,color=rvwvcq] (8,-10)-- (-2,-7);
\draw [line width=2pt,color=dtsfsf] (0,-5)-- (0,-9);
\draw [line width=2pt,color=dtsfsf] (0,-9)-- (8,-10);
\draw [line width=2pt,color=dtsfsf] (8,-10)-- (8,-4);
\draw [line width=2pt,color=dtsfsf] (8,-4)-- (0,-5);
\draw (-0.5733333333333362,-6.756666666666679) node[anchor=north west] {4};
\begin{scriptsize}
\draw [fill=rvwvcq] (8,4) circle (2.5pt);
\draw[color=rvwvcq] (8.226666666666668,4.333333333333344) node {$v_8$};
\draw [fill=rvwvcq] (8,-2) circle (2.5pt);
\draw[color=rvwvcq] (8.306666666666668,-1.6666666666666692) node {$v_7$};
\draw[color=wrwrwr] (8.146666666666667,1.1533333333333367) node {1};
\draw [fill=rvwvcq] (0,3) circle (2.5pt);
\draw[color=rvwvcq] (-0.013333333333336,3.713333333333342) node {$v_1$};
\draw[color=wrwrwr] (3.666666666666665,3.7533333333333423) node {1};
\draw [fill=rvwvcq] (-2,3) circle (2.5pt);
\draw[color=rvwvcq] (-1.9533333333333367,3.6333333333333417) node {$v_2$};
\draw[color=wrwrwr] (-0.9333333333333362,3.4333333333333416) node {2};
\draw [fill=rvwvcq] (-3,2) circle (2.5pt);
\draw[color=rvwvcq] (-3.473333333333337,2.5333333333333394) node {$v_3$};
\draw[color=wrwrwr] (-2.653333333333337,2.9733333333333403) node {2};
\draw [fill=rvwvcq] (-3,0) circle (2.5pt);
\draw[color=rvwvcq] (-3.453333333333337,0.09333333333333446) node {$v_4$};
\draw[color=wrwrwr] (-3.253333333333337,1.2333333333333367) node {2};
\draw [fill=rvwvcq] (-2,-1) circle (2.5pt);
\draw[color=rvwvcq] (-1.9933333333333367,-1.2266666666666683) node {$v_5$};
\draw[color=wrwrwr] (-2.653333333333337,-0.48666666666666675) node {2};
\draw [fill=rvwvcq] (0,-1) circle (2.5pt);
\draw[color=rvwvcq] (-0.03333333333333602,-1.2066666666666683) node {$v_6$};
\draw[color=wrwrwr] (-0.9333333333333362,-1.086666666666668) node {2};
\draw[color=wrwrwr] (2.966666666666665,-1.046666666666668) node {1};
\draw [fill=rvwvcq] (-2,1) circle (2.5pt);
\draw[color=rvwvcq] (-1.3933333333333364,1.1533333333333367) node {w};
\draw[color=rvwvcq] (-1.2533333333333363,2.5733333333333395) node {3};
\draw[color=rvwvcq] (-0.5733333333333361,1.9333333333333382) node {3};
\draw[color=rvwvcq] (-2.253333333333337,2.2333333333333387) node {3};
\draw[color=rvwvcq] (-2.793333333333337,0.8333333333333359) node {3};
\draw[color=rvwvcq] (-2.253333333333337,0.23333333333333478) node {3};
\draw[color=rvwvcq] (-1.5333333333333365,-0.20666666666666617) node {3};
\draw[color=rvwvcq] (-0.5333333333333361,0.37333333333333507) node {3};
\draw[color=dtsfsf] (4.0666666666666655,1.2333333333333367) node {4};
\draw[color=dtsfsf] (0.14666666666666406,1.1533333333333367) node {1};
\draw [fill=rvwvcq] (8,-4) circle (2.5pt);
\draw[color=rvwvcq] (8.226666666666668,-3.6466666666666736) node {$v_8$};
\draw [fill=rvwvcq] (8,-10) circle (2.5pt);
\draw[color=rvwvcq] (8.326666666666668,-9.666666666666686) node {$v_7$};
\draw [fill=rvwvcq] (0,-5) circle (2.5pt);
\draw[color=rvwvcq] (-0.013333333333336,-4.306666666666675) node {$v_1$};
\draw [fill=rvwvcq] (-2,-5) circle (2.5pt);
\draw[color=rvwvcq] (-1.9933333333333367,-4.326666666666674) node {$v_2$};
\draw [fill=rvwvcq] (-3,-6) circle (2.5pt);
\draw[color=rvwvcq] (-3.473333333333337,-5.4666666666666766) node {$v_3$};
\draw [fill=rvwvcq] (-3,-8) circle (2.5pt);
\draw[color=rvwvcq] (-3.453333333333337,-7.906666666666682) node {$v_4$};
\draw [fill=rvwvcq] (-2,-9) circle (2.5pt);
\draw[color=rvwvcq] (-1.9933333333333367,-9.226666666666684) node {$v_5$};
\draw [fill=rvwvcq] (0,-9) circle (2.5pt);
\draw[color=rvwvcq] (-0.03333333333333602,-9.206666666666685) node {$v_6$};
\draw [fill=rvwvcq] (-2,-7) circle (2.5pt);
\draw[color=rvwvcq] (-1.3133333333333364,-6.84666666666668) node {w};
\draw[color=rvwvcq] (-1.2533333333333363,-5.4266666666666765) node {4};
\draw[color=rvwvcq] (-0.6933333333333362,-6.0466666666666775) node {4};
\draw[color=rvwvcq] (-0.9733333333333363,-4.606666666666675) node {2};
\draw[color=rvwvcq] (-2.253333333333337,-5.766666666666677) node {4};
\draw[color=rvwvcq] (-2.753333333333337,-5.146666666666676) node {2};
\draw[color=rvwvcq] (-2.653333333333337,-6.8266666666666795) node {4};
\draw[color=rvwvcq] (-3.293333333333337,-6.94666666666668) node {2};
\draw[color=rvwvcq] (-2.253333333333337,-7.766666666666682) node {4};
\draw[color=rvwvcq] (-2.753333333333337,-8.506666666666684) node {2};
\draw[color=rvwvcq] (-1.4333333333333365,-8.246666666666682) node {4};
\draw[color=rvwvcq] (-0.5333333333333361,-7.626666666666681) node {4};
\draw[color=rvwvcq] (-1.1333333333333364,-9.146666666666684) node {2};
\draw[color=rvwvcq] (2.966666666666665,-4.206666666666675) node {1};
\draw[color=rvwvcq] (8.166666666666668,-6.84666666666668) node {1};
\draw[color=rvwvcq] (2.966666666666665,-9.526666666666685) node {1};
\draw[color=dtsfsf] (4.0666666666666655,-6.766666666666679) node {3};
\draw[color=dtsfsf] (0.14666666666666406,-6.86666666666668) node {1};
\end{scriptsize}
\end{tikzpicture} 
}
    \caption{Caption}
    \label{fig:my_label}
\end{figure}

}

In \Cref{fig:perscycle}, we provide an example depicting
instability for the weight function given by length of $1$-cycles.
While this is not the same as the $\radmeasure$, we maintain that similar counterexamples can be designed for $\radmeasure$ as well.
We leave out the details here. 
In particular, it is easily seen that $\radmeasure$ is an unstable function for $\cechfull$ filtrations.

\subsection{Approximate representatives and bases}

\cancel{
\abhishek{I think there is a general instability theorem lurking around here. We should be able to prove that if the weight function satisfies yada yada yada conditions, then it is $(\epsilon,\delta)$-unstable for all $\epsilon, \delta$.}

\info[inline]{\Large{I now have a full understanding of how the instability theorem goes! I would like to rewrite this section.}}
}
\cancel{
\begin{definition}[Vertices belonging to a cycle]
Let $\zeta$ be a $p$-cycle. 
Viewing $\zeta$  as a subset of $\complex^{(p)}$, we say that the vertex $v$ \emph{belongs to} $\zeta$ if  $v \in \sigma^{p}$ for some $\sigma^{p} \in \zeta$. 
\end{definition}
}

\cancel{
More generally, one can define interval modules associated to cycles that need not arise as interval summands of the persistence homology of a filtration.
\begin{definition}[Intervals modules associated to cycles]
Given a filtration $\mathcal{F}$ on a complex $\complex$, suppose that $[b,d)$ is an interval for which a $p$-cycle $\zeta$ is non-bounding.
Then, the interval module associated to a cycle $\zeta$ is defined by
\begin{equation*} 
{\interval}_a^{\zeta} = 
\begin{cases} [\zeta] &\text{if $a\in [b,d)$} \\ 
0 &\text{otherwise.} 
\end{cases} 
\end{equation*} 
  Here, for every $a,a' \in [b,d)$ with $a\leq a'$ the maps ${\interval}_a^{\zeta} \to {\interval}_{a'}^{\zeta}$
 are the induced maps on homology restricted to $[\zeta]$.
 Additionally, if $[b,d)\in B_p(\mathcal{F})$, then ${\interval}_a^{\zeta}$ is a summand in $H_p{\mathcal F}$
\end{definition}
}


Given the inherent instability associated to the $\radmeasure$, we formulate a  slightly weaker notion of stability for the $\radmeasure$. To state the result, we need a few definitions.

\cancel{

\begin{definition}[$\epsilon$-approximate persistent basis] \label{defn:apx_pers_basis}
Let $J$ be the indexing set for the intervals in the barcode  $B_p(\mathcal{F})$ of filtration $\mathcal{F}$, and let $J_{A^{\epsilon}} \subset J$ be such that if $j\in J$, where $[b_j,d_j)$ is an interval of $B_p(\mathcal{F})$  with $(d_j - b_j) > \epsilon$, then $j\in J_{A^{\epsilon}}$.

Then a set of $p$-cycles ${A^{\epsilon}} = \{\zeta_j \mid j\in J_{A^{\epsilon}}\}$ is called an \emph{$\epsilon$-approximate persistent $p$-basis} for $\mathcal{F}$ if 
\[\mathbb{A}^{\epsilon} \,\, = \,\, \bigoplus_{j\in J_{A^{\epsilon}}} {\interval}^{\zeta_j}
\textnormal{ is $\epsilon$-interleaved with $H_p \mathcal{F}$,}
\]
where ${\interval}^{\zeta_j}$ is defined by 
\begin{equation*} 
{\interval}_a^{\zeta_j} = 
\begin{cases} [\zeta_j] &\text{if $a\in [b'_j,d'_j)$, where $b'_j \in [b_j-\epsilon,b_j +\epsilon]$ and $d'_j \in [d_j-\epsilon,d_j +\epsilon]$} \\ 
0 &\text{otherwise.} 
\end{cases} 
\end{equation*} 
Moreover, for every $j\in J_{A^{\epsilon}}$ and  every $a,a' \in [b_j,d_j)$ with $a\leq a'$ the maps ${\interval}_a^{\zeta_j} \to {\interval}_{a'}^{\zeta_j}$
are the induced maps on homology restricted to $[\zeta_j]$, respectively. In this case, there is a canonical $\epsilon$-matching between the respective intervals of $B_p(\mathcal{F})$ and ${A^{\epsilon}}$. In particular, for every $j\in J_{A^{\epsilon}}$, $[b_j,d_j)$ is matched to $[b'_j,d'_j)$.
\end{definition}
}

\cancel{
\info[inline]{\large{As defined here the notion of $\epsilon$-approximate representatives works for the stability proof. But there's a sharper notion that makes more sense. For simplicity, I am eschewing the sharper idea for the one above which is simpler.}}
}

\cancel{
Next we define $A^{\epsilon}_{\min}$ and $r^{\epsilon}_{\min}$  as follows.
\[\mathcal{A}^{\epsilon}_{\min}(\mathcal{F}) = \arg\min_{A} \left(\sum_{j\in J_A} r(\zeta_j)\right) \textnormal{ where $A$  constitutes an $\epsilon$-approximate persistent basis for $\mathcal{F}$.} \]
and
\[r^{\epsilon}_{\min}(\mathcal{F}) =    \sum_{j} r(\zeta_j) \textnormal{ where $j\in J_{A_{\min}^{\epsilon}}$.} \]
Note that for a filtration $\mathcal{F}$ even when $|\mathcal{A}^{\epsilon}_{\min}(\mathcal{F})| > 1$,   $r^{\epsilon}_{\min}(\mathcal{F})$ is indeed unique. 
}


\begin{definition}[$\epsilon$-approximate representatives]
For an interval $[b,d)$ with $(d-b) > 2\epsilon$, a cycle $\zeta$ is said to be an \emph{$\epsilon$-approximate representative} for $[b,d)$, if $\zeta$ is a nontrivial cycle that is born at $b' \in [b-\epsilon,b+\epsilon]$ and dies at $d' \in [d-\epsilon,d+\epsilon]$, and the class $[\zeta]$ is nontrivial in $\complex_s$ for every $s\in [b', d')$. 
\end{definition}

Recall that in \Cref{sec:homloc}, we provided a definition for the $\radmeasure$ function $r$ for intervals defined on chains and cycles of all subcomplexes of a complex $\complex$, whose vertex set is embedded in a Euclidean space. We now provide a definition for an approximate variant of this function.

\begin{definition}[$\epsilon$-approximate radii of intervals]
As an approximation of the radius function defined in \Cref{eq:persrad}, we now define $r^{\epsilon}([b,d))$ for an interval $[b,d) \in B(\Cech(P))$ with $d-b>2\epsilon$ to be the radius of the smallest Euclidean sphere that encloses all the vertices of some $\epsilon$-approximate representative cycle of $[b,d)$. In other words,  $r^{\epsilon}([b,d)) = \min_{\zeta} r(\zeta)$, where $\zeta$ is an $\epsilon$-approximate representative cycle for $[b,d)$. 
\end{definition}
In \Cref{sec:repmatch}, we prove the following stability theorem (\Cref{thm:mainstabilitygo} restated as \Cref{thm:mainstability} with a proof) by building on techniques developed by Bjerkevik~\cite{bjerkevik}.

\begin{thm}[Approximate stability for $\radmeasure$] \label{thm:mainstabilitygo}
Let $P$ be a point set embedded in a Euclidean space and $Q$ be an $\epsilon$-perturbation of $P$.
Then, there exists an $\epsilon$-matching $\matching$ between $B(\Cech(\pointsetone))$ and $B(\Cech(\pointsettwo))$ such that 
if $[b',d')=\matching([b,d))$ and both
$[b,d)$, $[b',d')$ have lengths greater than $2\epsilon$, then
\begin{align}  \label{eqn:mainresrestated}
  r^{2\epsilon}([b,d)) &\leq   r([b',d')) + \epsilon, \nonumber\\
    r^{2\epsilon}([b',d')) &\leq   r([b,d)) + \epsilon. 
\end{align}
\end{thm}

Note that, as per \Cref{defn:stabmain}, we would have $(\epsilon,\epsilon)$ stability for $\radmeasure$,  if there exists an $\epsilon$-matching $\matching$ that matches  intervals $[b,d)\in B(\Cech(P))$ of length greater than $2\epsilon$ to  intervals $[b',d') \in B(\Cech(Q))$ of length greater than $2\epsilon$ satisfying
\begin{align}  \label{eqn:mainresrestatedagain}
  r([b,d)) &\leq   r([b',d')) + \epsilon, \nonumber\\
    r([b',d')) &\leq   r([b,d)) + \epsilon. 
\end{align}

In this sense, \Cref{thm:mainstabilitygo} is an approximate version of $\radmeasure$ stability.
\cancel{
Interestingly, the function $r^{\epsilon}$ for intervals can also be computed, and \Cref{sec:apxoptimal} provides an algorithm for doing so.
Finally, since we do not have unconditional stability for the $\radmeasure$, in \Cref{sec:algstab}, we provide a recipe to determine specific values of $(\epsilon,\delta)$ with  $(\epsilon,\delta)$-stability guarantees for $\epsilon$--perturbations of inputs.
}

\bigskip


\cancel{

\section{Counterexample for a stronger version of stability}

Consider a regular polygon $P_1$ of $ n = 100$ sides, where each side $b$ is of length $10$. We shall call $P_1$ the \emph{outer polygon}. 
Then, the radius of the circle is approximately $R = 9118.906$ by the formula:

\[ R \quad = \quad \frac{b}{2 \cdot \sin{\frac{\pi}{n}}}  \quad = \quad \frac{10}{2 \cdot \sin{\frac{\pi}{100}}}
 \quad \approx \quad 9118.906.
\] 

We will now construct another polygon with $n = 100$ sides that is inscribed by a circle with the same center as the outer polygon, but with a smaller radius. First, we will describe how to select the vertices of this polygon.  For every pair of adjacent vertices on the outer regular polygon, fix a vertex that is at length $b+ \epsilon$ from each vertex in the pair, and is inside the circle inscribing the outer polygon. Since an adjacent pair contributes to a single vertex, and since every vertex of the outer polygon belongs to two pairs, in total, $n=100$ vertices are introduced. By symmetry, every newly introduced vertex has the same distance to two vertices that are closest to it. So, these connecting these vertices with two of the closest vertices to it forms a regular polygon, that we call the \emph{inner polygon}. 

The radius of the inner polygon is given by $\frac{R-c}{\cos{(\frac{\pi}{100}})}$ where $c$ is given by:
 
\[ c \quad =\quad \cos{\left(\frac{\pi}{2} - \frac{\pi}{n} - \arccos{ \left(\frac{0.5 \cdot b}{b+\epsilon}\right)}\right)} \cdot (b + 
\epsilon). \]

The length $a$ of the side of the inner polygon is given by 

\[ \frac{a}{2} \quad =\quad \sin{\left(\frac{\pi}{2} - \frac{\pi}{n} - \arccos{\left(\frac{0.5 \cdot b}{b + \epsilon}\right)}\right)} \cdot (b + 
\epsilon). \] 

For $\epsilon = 0.1$, $a \approx 9.443$ and $c \approx 8.928$. For $\epsilon = 0$, $a \approx 9.453$ and $c \approx 8.81$.

\subsection{Stability to do}

\begin{itemize}
    \item Write down 1. Locality: Counterexample.
    \item Why do you need approximate representatives for answering stability questions?
Explain why the definitions force you to use an approximate version? In general, it is unreasonable to ask for a strong version of stability for any measure because of a fundamental asymmetry between the definitions of matching of bars in stability, and persistence basis. For stability, the bars can move left as well as right. For optimal persistence basis to be maintained the optimal bars should move only left.
    \item Complete the stability counterexample. Explain why there's no inequality in the other direction.
    \item Does there exist a perturbation that realizes the approximate version?
    \item Statistical question: Can you discretize the perturbation, and under what condition can you draw inference from sampled perturbations?
\end{itemize}

}


\cancel{
From the above two equations we obtain

\begin{align} \label{eqn:mainder}
F^{\epsilon}(G({\interval}^{z_{j}}_{c})) & =F^{\epsilon}(\sum_{i\in I}\kappa_{j,i}\cdot {\interval}^{\zeta_{i}}_{c+\epsilon}) \nonumber\\
 & =\sum_{i\in I}\kappa_{j,i}\cdot F^{\epsilon}({\interval}^{\zeta_{i}}_{c+\epsilon}) \nonumber\\
 & =\sum_{i\in I}\kappa_{j,i}\cdot(\sum_{\ell\in J}\eta_{i,\ell}\cdot {\interval}^{z_{\ell}}_{c+2\epsilon}) \nonumber\\
 & =\sum_{\substack{i\in I\\
\ell\in J
}
}\kappa_{j,i}\cdot\eta_{i,\ell}\cdot {\interval}^{z_{\ell}}_{c+2\epsilon}
\end{align}

The first equality in \Cref{eqn:mainder} follows from the fact that  using \Cref{eqn:zj,eqn:geq}, a cycle in the class of  $G([z_j])$ in $\Cech_{c+\epsilon}(\pointsetone)$ can be expressed as follows.

\begin{equation} 
   \zeta  = \sum_{i \in I} \kappa_{j,i} \cdot \zeta_i  + \delta_{c + \epsilon}.
\end{equation}

The third equality in \Cref{eqn:mainder} follows from the fact that using \Cref{eqn:zetai,eqn:feq}, a cycle in the class of  $F^{\epsilon}(G([z_{j}]))$ in $\Cech_{c+2\epsilon}(\pointsettwo)$ can be expressed as follows.

\begin{equation} 
   z  = \sum_{\substack{i\in I\\
\ell\in J
}
}\kappa_{j,i}\cdot\eta_{i,\ell}\cdot z_{\ell} + \partial_{c + 2\epsilon}.
\end{equation}

Since we use map $h_{\ast}$ followed by ${h_{\ast}}^{-1}$, the cycle $z$ is homologous to $z_j$. Hence \Cref{eqn:mainder} can be written as

\begin{align} \label{eqn:suppone}
    F^{\epsilon}(G({\interval}^{z_{j}}_{c}))  &= \sum_{\substack{i\in I\\
\ell\in J
}
}\kappa_{j,i}\cdot\eta_{i,\ell}\cdot {\interval}^{z_{\ell}}_{c+2\epsilon} \nonumber\\
&= \sum_{\substack{i\in I\\
\ell\in J
} 
}\kappa_{j,i}\cdot\eta_{i,j}\cdot {\interval}^{z_{j}}_{c+2\epsilon} \nonumber\\
&= {\interval}^{z_j}_{c+2\epsilon}.
\end{align}

\vspace{0.5cm}

Symmetrically, $G^{\epsilon}(F({\interval}^{\zeta_{i}}_{c}))$ can be written as

\begin{align} \label{eqn:supptwo}
    G^{\epsilon}(F({\interval}^{\zeta_{i}}_{c}))  &= \sum_{\substack{i\in I\\
\ell\in J
}
}\eta_{i,j}\cdot\kappa_{j,k}\cdot {\interval}^{\zeta_{k}}_{c+2\epsilon} \nonumber \\
&= \sum_{\substack{i\in I\\
\ell\in J
}
}\eta_{i,j}\cdot\kappa_{j,i}\cdot {\interval}^{\zeta_{i}}_{c+2\epsilon} \nonumber \\
&= {\interval}^{\zeta_i}_{c+2\epsilon}.
\end{align}

\medskip

}


We now build towards a proof for \Cref{thm:mainstabilitygo}.
To begin with, let $P$ be a point set embedded in a Euclidean space, and let $\pointsettwo$ be an $\epsilon$-perturbation of $\pointsetone$ realized through a bijective map between point sets denoted by $f$. We denote the inverse of $f$ by $g$.
Let ${\basisset}^P = \{\zeta_1, \zeta_2, \dots, \zeta_n\}$ be a persistent homology basis for $\Cech(\pointsetone)$, and let ${\basisset}^Q = \{z_1, z_2, \dots, z_m\}$ be a persistent homology basis for $\Cech(\pointsettwo)$.

\smallskip

Then, it is easy to check that for every $s\in \RBB$, the map $f$ induces a simplicial map $f_{s}: \Cech_s(\pointsetone) \to \Cech_{s+\epsilon}(\pointsettwo)$.
Furthermore, the map $f_{s}$ induces a map
on the respective cycle groups ${f}_{s}^{\#}: Z_{\ast}(\Cech_s(\pointsetone)) \to  Z_{\ast}(\Cech_{s+\epsilon}(\pointsettwo))$ as well as a map 
$\hat{f}_{s}: H_{\ast}(\Cech_s(\pointsetone)) \to  H_{\ast}(\Cech_{s+\epsilon}(\pointsettwo))$ on homology groups.

Similarly, for every $s \in \RBB$, the map $g$ induces a simplicial map $g_{s}: \Cech_{s}(\pointsettwo) \to \Cech_{s+\epsilon}(\pointsetone)$, which in turn, induces a map  
on the respective cycle groups ${g}_{s}^{\#}: Z_{\ast}(\Cech_s(\pointsettwo)) \to  Z_{\ast}(\Cech_{s+\epsilon}(\pointsetone))$ as well as  a map
on homology groups $\hat{g}_{s}: H_{\ast}(\Cech_s(\pointsettwo)) \to  H_{\ast}(\Cech_{s+\epsilon}(\pointsetone))$. Moreover, $g_{s+\epsilon}\circ f_{s} = \id$, and $f_{s+\epsilon}\circ g_{s} = \id$ for every $s\in \RBB$. 


\smallskip

We define persistent modules $\pmodule$ and $\qmodule$ as follows.
\begin{alignat*}{2}
    \pmodule \quad &=\quad \bigoplus_{\substack{\boldi\in B(\Cech(\pointsetone))\\ \zeta_i \in \mathcal{R}(\boldi) }}   {\interval}^{\zeta_i}  &&\cong\quad \bigoplus_{\boldi\in B(\Cech(\pointsetone))} \iinterval, \\
    \qmodule \quad &=\quad \bigoplus_{\substack{\boldj\in B(\Cech(\pointsettwo))\\ z_j \in \mathcal{R}(\boldj) }}  {\interval}^{z_j} &&\cong\quad \bigoplus_{\boldj\in B(\Cech(\pointsettwo))} \jinterval .
\end{alignat*}

For some $\boldj = [b_j,d_j) \in B(\Cech(Q))$, let $z_j \in \represents(\boldj)$.  Then, $\hat{g}_{\vphantom{1^{H}}b_j}([z_j])$ is a class in $\Cech_{b_j + \epsilon}(\pointsetone)$. 
Since ${\basisset}^P$ is a persistent homology basis for $\Cech(P)$, $\hat{g}_{\vphantom{1^{H}}b_j}([z_j])$ can be written as follows.
\begin{equation} \label{eqn:zj}
    \hat{g}_{\vphantom{1^{H}}b_j}([z_j]) = \sum_{i \in I} \kappa_{j,i} \cdot [\zeta_i].
\end{equation}
In \Cref{eqn:zj}, \mbox{$I  \subset [n]$} indexes a subset of the representative cycles for the intervals in $ B(\Cech(\pointsetone))$, and the coefficients $\kappa_{j,i} \in \mathbb{Z}_2$. 

We define $ G|_{\boldj}:{\interval}^{z_j} \to \pmodule^{\epsilon}$ as follows.
\begin{equation} \label{eqn:geq}
     G|_{\boldj}({\interval}^{z_j}_s) \nonumber =  \sum_{i \in I} \kappa_{j,i} \cdot {{\interval}^{\zeta_i}_{s+\epsilon}}\quad \text{ for $s\in [b_j,d_j)$}.  
\end{equation}

By linear extension over all intervals $\boldj \in   B(\Cech(\pointsettwo))$, we obtain a map $G: \qmodule \to \pmodule^{\epsilon}$.

\smallskip

Symmetrically, let $\zeta_i \in {\basisset}^P$ be a representative cycle for some $\boldi = [b_i,d_i)  \in B(\Cech(P))$. Then, $\hat{f}_{\vphantom{1^{H}}b_i}([\zeta_i])$ is a class in $\Cech_{b_i + \epsilon}(\pointsettwo)$ 
Thus,  $\hat{f}_{\vphantom{1^{H}}b_i}([\zeta_i])$  can be written as follows.
\begin{equation} \label{eqn:zetai}
     \hat{f}_{\vphantom{1^{H}}b_i}([\zeta_i])  = \sum_{j \in J} \eta_{i,j} \cdot [z_j]. 
\end{equation}

In \Cref{eqn:zetai}, \mbox{$J  \subset [m]$} indexes a subset of the representative cycles for intervals in $B(\Cech(\pointsettwo))$, and the coefficients $\eta_{i,j} \in \mathbb{Z}_2$. 

We define $F|_{\boldi}: {\interval}^{\zeta_i} \to \qmodule^{\epsilon}$ as follows.
\begin{equation} \label{eqn:feq}
     F|_{\boldi}({\interval}^{\zeta_i}_{s}) = \sum_{j \in J} \eta_{i,j} \cdot {{\interval}^{z_j}_{s+\epsilon}} \quad \text{ for $s\in [b_i,d_i)$}. \nonumber
\end{equation}
By linear extension over all intervals $\boldi \in   B(\Cech(\pointsetone))$, we obtain a map $F: \pmodule \to \qmodule^{\epsilon}$.

It is easy to check that the maps $F$ and $G$ consitute an $\epsilon$-interleaving between modules $\pmodule$ and $\qmodule$. This is a simple consequence of the fact that $g_{s+\epsilon}\circ f_{s} = \id$, and $f_{s+\epsilon}\circ g_{s} = \id$ for every $s\in \RBB$.


\smallskip

    Let $\imath^{\pmodule}_{\boldi}: {\interval}^{\zeta_i} \injects \pmodule$
and $\imath^{\qmodule}_{\boldj}: {\interval}^{z_j} \injects \qmodule$ denote the canonical inclusion maps, and let $\pi^{\pmodule}_{\boldi}: \pmodule  \surjects  {\interval}^{\zeta_i}$
and $\pi^{\qmodule}_{\boldj}: \qmodule  \surjects
 {\interval}^{z_j}  $ denote the canonical projection maps.
 
 For the morphism $F: \pmodule \to \qmodule^{\epsilon}$, we have  $F|_{\boldi} = F \circ \imath^{\pmodule}_{\boldi}$. Now, let $F_{\boldi,\boldj} = {(\pi^{\qmodule}_{\boldj})}^{\epsilon} \circ F \circ \imath^{\pmodule}_{\boldi}$. Likewise, for the morphism 
 $G: \qmodule \to \pmodule^{\epsilon}$, we have $G|_{\boldj} = G \circ \imath^{\qmodule}_{\boldj}$. Now, let $G_{\boldj,\boldi} = {(\pi^{\pmodule}_{\boldi})}^{\epsilon} \circ G \circ \imath^{\qmodule}_{\boldj}$.
 
 Let $\phi^{\epsilon}: \pmodule \to \pmodule^{\epsilon}$ denote the collection of maps whose restriction to  $\pmodule_t$ gives the internal linear map $\phi_{t,t+\epsilon}: \pmodule_t \to \pmodule_{t+\epsilon}$.
 For an interval summand ${\interval}^{\boldi}$ of $\pmodule$, let $\phi_{\boldi}^{\epsilon}$ denote  the collection of maps whose restriction to ${\interval}^{\boldi}_t$ gives the internal linear map  $\phi_{t,t+\epsilon}|_{{\interval}^{\boldi}}: {\interval}_t \to {\interval}_{t+\epsilon}$.
 
 Using the fact that $\phi_{\boldi}^{2\epsilon} =  {(\pi^{\pmodule}_{\boldi})}^{2 \epsilon} \circ \phi^{2\epsilon} \circ \imath^{\pmodule}_{\boldi}$, we can write 
 
 \begin{equation} \label{eq:etaex}
     \phi_{\boldi}^{2\epsilon} \quad = \quad \sum_{\boldj \in B(\Cech(\pointsettwo))} G_{\boldj,\boldi}^{\epsilon} \circ F_{\boldi,\boldj}.
 \end{equation}
 
Moreover, because $\pmodule$ is a direct sum of interval modules, for $\boldi \neq \boldi'$, we obtain

 \begin{alignat}{1} \label{eq:zeroex}
   {(\pi^{\pmodule}_{\boldi'})}^{2 \epsilon} \circ \phi^{2\epsilon} \circ \imath^{\pmodule}_{\boldi} \quad &= \quad  \sum_{\boldj \in B(\Cech(\pointsettwo))} G_{\boldj,\boldi'}^{\epsilon} \circ F_{\boldi,\boldj}  \nonumber \\
   &= \quad  0.
 \end{alignat}

\begin{definition}[$\Lambda(\boldi),\Lambda(A) $] \label{defn:candedge}
For intervals $\boldi \in  B(\Cech(P))$, we define
\begin{align*}
    \widetilde{\Lambda}(\boldi) &= \{ \boldj \in B(\Cech(Q)): {\interval}^\boldi\text{ and }{\interval}^\boldj\text{ are } \epsilon\text{-interleaved}\}\text{, and} \\
    \Lambda(\boldi) &= \{ \boldj \in B(\Cech(Q)): {\interval}^\boldi\text{ and }{\interval}^\boldj\text{ are } \epsilon\text{-interleaved and either } \eta_{i,j}\neq 0 \text{ or }\kappa_{j,i}\neq 0\}.
\end{align*}
For a collection of intervals $A  \subset  B(\Cech(P)) $, we define 
\begin{equation*}
\widetilde{\Lambda}(A) = \bigcup_{ \boldi \in  A} \widetilde{\Lambda}(\boldi)\quad \text{ and } \quad
\Lambda(A) = \bigcup_{\boldi \in  A} \Lambda(\boldi).
\end{equation*}
\end{definition} 

\begin{definition}[$\Upsilon(\boldj),\Upsilon(A)$] \label{defn:candedgetwo}
For intervals $\boldj \in  B(\Cech(Q))$, we define
\begin{align*}
    \widetilde{\Upsilon}(\boldj) &= \{ \boldi \in B(\Cech(P)): {\interval}^\boldi\text{ and }{\interval}^\boldj\text{ are } \epsilon\text{-interleaved}\}\text{, and} \\
    \Upsilon(\boldj) &= \{ \boldi \in B(\Cech(P)): {\interval}^\boldi\text{ and }{\interval}^\boldj\text{ are } \epsilon\text{-interleaved and either } \eta_{i,j}\neq 0 \text{ or }\kappa_{j,i}\neq 0\}.
\end{align*}
For a collection of intervals $A  \subset  B(\Cech(Q)) $, we define 
\begin{equation*}
\widetilde{\Upsilon}(A) = \bigcup_{ \boldj \in  A} \widetilde{\Upsilon}(\boldj) \quad \text{ and }  \quad
\Upsilon(A)  = \bigcup_{\boldj \in  A} \Upsilon(\boldj).
\end{equation*}
\end{definition}

\begin{remark}\label{rem:diffsets}
From \Cref{defn:candedge,defn:candedgetwo} we make the following observations.
\[\boldj \in \widetilde{\Lambda}(A) \setminus \Lambda(A) \quad  \Longrightarrow \quad \eta_{i,j} =0
\text{ and } \kappa_{j,i} =0 
\text{ for all } \boldi \in A.\]
\[\boldi \in \widetilde{\Upsilon}(A) \setminus \Upsilon(A) \quad \Longrightarrow \quad  \eta_{i,j} =0
\text{ and } \kappa_{j,i} =0 
\text{ for all }\boldj \in A. \]
\end{remark}

\begin{remark}
The sets $\widetilde{\Lambda}(A)$ is equivalent to $\mu(A)$  as defined in Botnan's lecture notes~\cite{botnan}[Section 13.2].
The rationale behind defining $\Lambda(A)$ and $\Upsilon(A)$ is to ensure that the representative cycles of perturbed sets can be linearly related. Refer to \Cref{rem:howdifferent},\Cref{thm:mainstability} for additional details.
\end{remark}

 We now recall some elementary propositions from Botnan's lecture notes\cite{botnan}. The proofs for \Cref{lem:helperone,lem:helpertwo,lem:helperthree} can be found in Section 13.2 of~\cite{botnan}.

 \medskip

 \begin{proposition}
 \label{lem:helperone}
     For a morphism $\fakef: \intervalone 
     \to \intervaltwo$, if $b_2>b_1$ or $d_2>d_1$, then $\fakef$ is zero. 
     On the other hand, if  $b_2\leq b_1$ and $d_2\leq d_1$, then $\fakef_t$ is determined by $\fakef_{b_1}$ for $t \in [b_1,d_2)$, in the sense that, $\fakef_t$  is nonzero if and only if  $\fakef_{b_1}$ is nonzero.
 \end{proposition}

\Cref{lem:helperone} above says that the morphism $\fakef: \intervalone 
     \to \intervaltwo$ between interval modules is completely determined by the function value at $b_1$. 

For an interval $\boldi =[b,d)$, define $\alpha(\boldi) = b + d$.
Specializing Bjerkevik's arguments~\cite{bjerkevik} to  single parameter persistence, Botnan~\cite{botnan} defines the following preorder on single parameter interval modules: $\boldi \leq_{\alpha} \boldj$ if and only if $\alpha_(\boldi) \leq \alpha_(\boldj)$.

\begin{proposition}
\label{lem:helpertwo}
    Let $\boldi_1 = \oneinterval$, $\boldi_2 = \twointerval$ and $\boldi_3 = \threeinterval$ be intervals such that $\boldi_1 \leq_{\alpha} \boldi_3$. If there exist nonzero maps $\fakef: \Intervalone \to {(\Intervaltwo)}^{\epsilon}$ and $\fakeg: \Intervaltwo \to {(\Intervalthree)}^{\epsilon}$, then $\Intervaltwo$ is $\epsilon$-interleaved with either $\Intervalone$ or $\Intervalthree$. 
\end{proposition}

\begin{remark}\label{rem:contra}
When $\boldi_3 = \boldi_1$, the contrapositive of \Cref{lem:helpertwo} reads as follows: If $\Intervaltwo$ is not $\epsilon$-interleaved with  $\Intervalone$, then either  $\fakef: \Intervalone \to {(\Intervaltwo)}^{\epsilon}$ or $\fakeg: \Intervaltwo \to {(\Intervalthree)}^{\epsilon}$ is a zero map.
\end{remark}

\begin{proposition}
\label{lem:helperthree}
    Let $\boldi_1 = [b_1,d_1)$, $\boldi_2 = [b_2,d_2)$ and $\boldi_3 = [b_3,d_3)$ be such that $(d_1 - b_1) > 2\epsilon$, $(d_3-b_3) > 2\epsilon$ and $\boldi_1 \leq_{\alpha} \boldi_3$. If there exist nonzero maps $\fakef: \Intervalone \to {(\Intervaltwo)}^{\epsilon}$ and $\fakeg: \Intervaltwo \to {(\Intervalthree)}^{\epsilon}$, then $\fakeg^{\epsilon} \circ \fakef \neq 0$.
\end{proposition}

\begin{remark}
 It is worth noting that \Cref{lem:helperthree} applies only when $\boldi_1 {\leq}_{\alpha} \boldi_3$. When $\boldi_1 >_{\alpha} \boldi_3$, 
there is no guarantee that $\fakeg^{\epsilon} \circ \fakef$ is nonzero even when  $\fakef$ and $\fakeg$ are both nonzero. Later, in \Cref{lem:mainhall}, this has implications in how \Cref{eqn:matrixeqn} is set up. 
\end{remark}

\begin{proposition} \label{lem:mainhall}
    Suppose $\pmodule$ and $\qmodule$ are $\epsilon$-interleaved. Let $A$ be a collection of intervals in $\pmodule$ with length greater than $2 \epsilon$. Then, $|A| \leq |\Lambda(A)|$.
\end{proposition}
\begin{proof*}
Let $\mu = |A|$ and $\nu = |\Lambda(A)|$. 
To begin with, order the elements of $A = \{\boldi_1,\boldi_2,\dots,\boldi_{\mu}\}$   in non-decreasing order of $\leq_{\alpha}$.
Also, for $\boldi=[b_i,d_i)$ and $\boldi'=[b_{i'},d_{i'})$,
we write $i  < i'$ if and only if $\boldi \leq_{\alpha} \boldi'$.
Let  $\Lambda(A) = \{ \boldj_1,\boldj_2,\dots,\boldj_{\nu} \}$. 

Next, we obtain an expression for  $\phi_{\boldi}^{2\epsilon}$ as follows.
\begin{alignat}{2} \label{eq:maineqone}
 \phi_{\boldi}^{2\epsilon} \quad &= \quad \sum_{\boldj \in B(\Cech(\pointsettwo))} G_{\boldj,\boldi}^{\epsilon} \circ F_{\boldi,\boldj} \quad & \text{ using \Cref{eq:etaex}} \nonumber\\
            &= \quad \sum_{\boldj \in \widetilde{\Lambda}(A)} G_{\boldj,\boldi}^{\epsilon} \circ F_{\boldi,\boldj} \quad &  \text{ using \Cref{rem:contra}} \nonumber\\
            &= \quad \sum_{\boldj \in \Lambda(A)} G_{\boldj,\boldi}^{\epsilon} \circ F_{\boldi,\boldj}  &  \text{ using \Cref{rem:diffsets}.}
\end{alignat}

Using the same observation, for $\boldi \leq_{\alpha} \boldi'$, we obtain the following expression:

\begin{alignat}{2} \label{eq:maineqtwo}
      \sum_{\boldj \in \Lambda(A)} G_{\boldj,\boldi'}^{\epsilon} \circ F_{\boldi,\boldj} \quad &= \quad \sum_{\boldj \in \widetilde{\Lambda}(A)} G_{\boldj,\boldi'}^{\epsilon} \circ F_{\boldi,\boldj} &  \text{ using \Cref{rem:diffsets}} \nonumber\\
      \quad &= \quad \sum_{\boldj \in B(\Cech(\pointsettwo))} G_{\boldj,\boldi'}^{\epsilon} \circ F_{\boldi,\boldj} \quad &  \text{ using \Cref{lem:helpertwo}} \nonumber\\
      \quad & = \quad 0  \quad &  \text{ using \Cref{eq:zeroex}}.
\end{alignat}

Using \Cref{lem:helperone,lem:helperthree}, \Cref{eq:maineqone,eq:maineqtwo} can be written as 
\begin{alignat}{1}
 1 \quad &= \quad \sum_{\boldj \in \Lambda(\boldi)} \kappa_{j,i} \cdot \eta_{i,j}\quad \text{and} \label{eq:firstone}\\
 0 \quad &= \quad \sum_{\boldj \in \Lambda(\boldi)} \kappa_{j,i'} \cdot \eta_{i,j}\text{ for $i<i'$, respectively.} \label{eq:secondone}  
\end{alignat}

Putting \Cref{eq:firstone,eq:secondone} in matrix form yields the following matrix equation, where the entries above the diagonal in the right hand side matrix are unknown. 

\begin{equation} \label{eqn:matrixeqn}
    \left[
\begin{matrix}
\kappa_{j_1,i_1} &  \dots  & \kappa_{j_{\nu},i_1} \\
\vdots & \ddots & \vdots \\
\kappa_{j_1,i_{\mu}} &  \dots  &\kappa_{j_{\nu},i_{\mu}}
\end{matrix}
\right]
\left[
\begin{matrix}
\eta_{i_1,j_1} &  \dots  & \eta_{i_{\mu},j_1} \\
\vdots & \ddots & \vdots \\
\eta_{i_1,j_{\nu}} &  \dots  & \eta_{i_{\mu},j_{\nu}}
\end{matrix}
\right]
=
\left[
\begin{matrix}
1 & \ast & \dots  & \ast \\
0 & 1 & \dots  & \ast \\
0 & 0 & \ddots & \vdots \\
0 & 0 & \dots  & 1
\end{matrix}
\right].
\end{equation}

Since the matrix on the right hand side is upper triangular with ones on diagonal, it has rank $|A|$. On the other hand, the two matrices on the left have rank upper bounded by $\nu = |\Lambda(A)|$. It follows immediately that $|A| \leq |\Lambda(A)|$.
\end{proof*}

\begin{proposition} \label{lem:mainhalltwo}
    Suppose $\pmodule$ and $\qmodule$ are $\epsilon$-interleaved. Let $A$ be a collection of intervals in $\qmodule$ with length greater than $2 \epsilon$. Then, $|A| \leq |\Upsilon(A)|$.
\end{proposition}
\begin{proof*}
The statement of the theorem is symmetric to \Cref{lem:mainhalltwo}. Hence, we omit the details of the proof.
\end{proof*}

\begin{figure}
    \centering
        \includegraphics[scale=0.5]{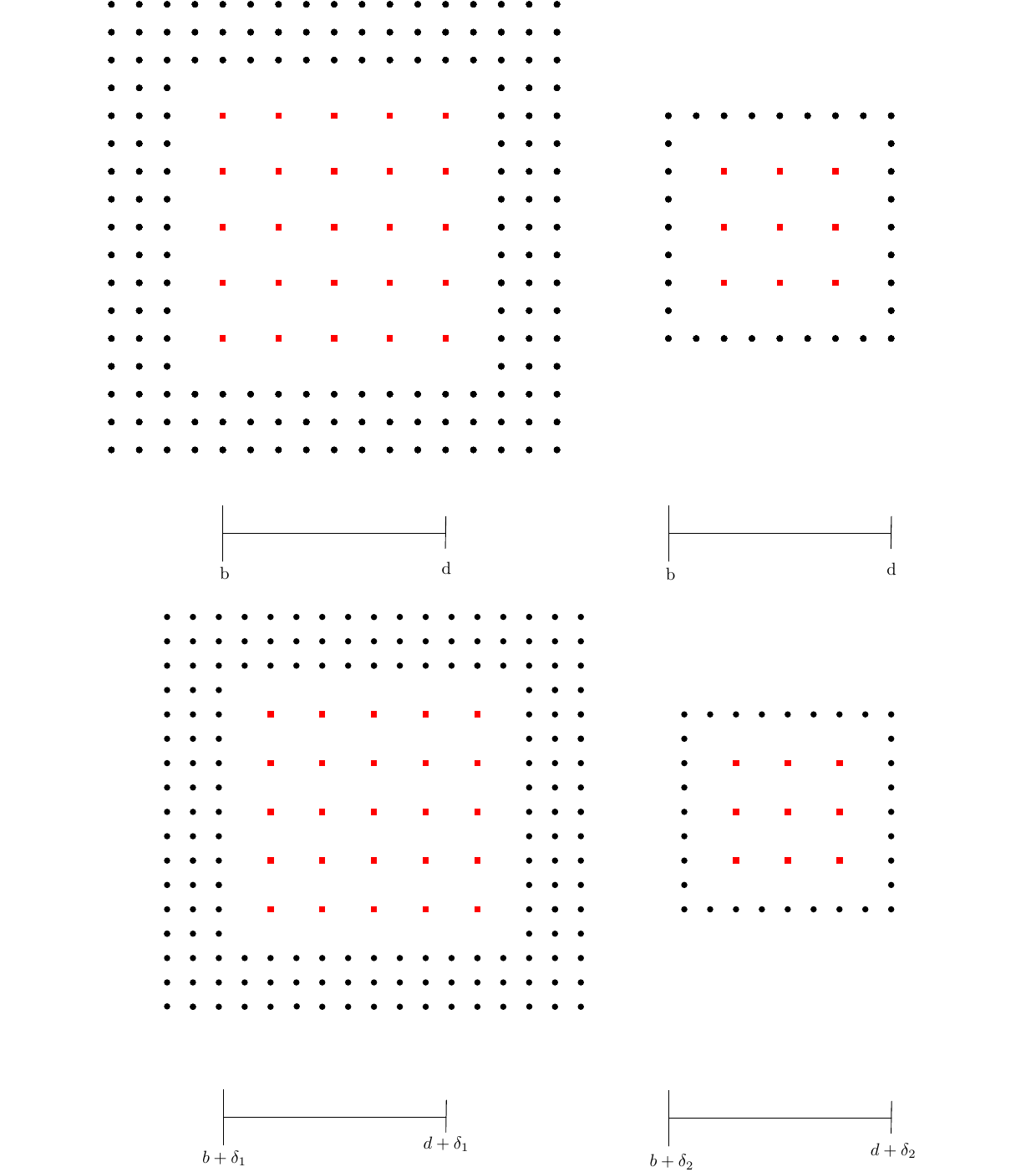}
    \caption{Consider the $\cechfull$ filtrations on the point set shown in the top figure, denoted by $P$, which is $\delta$-perturbed to obtain another point set $Q$ shown in the bottom figure. In $\Cech(P)$, the cycle on the left supported by the inner rim of black grid points, and the cycle on the right supported by black points are born at $b$ and die at $d$.  Owing to a $\delta$-perturbation, where $\delta_1, \delta_2 < \delta$,  in  $\Cech(Q)$, the first cycle is born at ${b+\delta_1} $ and dies at ${d+\delta_1}$, whereas the second cycle is born at   ${b+\delta_2}$ and dies at ${d+\delta_2}$. If one uses Bjerkevik's approach to obtain a $\delta$-matching, the  bar $[b,d)$ (top left) may be matched to either the bar $[b+\delta_1,d+\delta_1) $ (bottom left) or $[b+\delta_2,d+\delta_2) $ (bottom right), whereas the  bar $[b,d)$ (top right) may also be matched to either the bar $[b+\delta_1,d+\delta_1) $ (bottom left) or $[b+\delta_2,d+\delta_2) $ (bottom right). Thus, a naive approach to $\delta$-matching  is not approximately stable as the radius values of the enclosing spheres of matched representatives  can be arbitrarily far apart.
    Our version of matching ensures that for this example, $[b,d)$ (top left) is matched to $[b+\delta_1,d+\delta_1) $ (bottom left) and $[b,d)$ (top right) is matched to the bar $[b+\delta_2,d+\delta_2) $ (bottom right).
    }
    \label{fig:needcoeff}
\end{figure}

\begin{remark} \label{rem:howdifferent}
Note that the proof of \Cref{lem:mainhall} closely follows Botnan's exposition of Bjerkevik's ideas for constructing an $\epsilon$-matching given an $\epsilon$-interleaving between persistence modules using Hall's theorem. However, there are three important differences. 
\begin{compactitem}
    \item In our proof, it is vital to use the canonical $\epsilon$-interleavings that are induced by the simplicial maps $f_s$ and $g_s$ for $s\in \RBB$ as described in \Cref{sec:repmatch}. In Bjerkevik's approach an arbitrary $\epsilon$-interleaving can be used to derive an $\epsilon$-matching. See \Cref{fig:needcoeff} for an example.
    \item While for Bjerkevik's result it suffices to establish the inequality $|A| \leq |\widetilde{\Lambda}(A)|$, for our purposes it is necessary to establish the stricter inequality $|A| \leq |\Lambda(A)|$. In particular, we require that an interval $\boldi$ is matched only to one of the intervals in $\Lambda(\boldi)$. 
    \item While Bjerkevik uses arbitary interval decompositions of  persistent modules, we are required to use the decompositions that come from  fixed choices of persistence homology bases. This has the following consequence: even when the $\epsilon$-interleaving maps $F$ and $G$ are canonical, the maps $F_{\boldi,\boldj}$ and $G_{\boldj,\boldi}$ depend on how we choose to represent the interval summands of $\pmodule$ and $\qmodule$.
    Since $F_{\boldi,\boldj}$ and $G_{\boldj,\boldi}$ determine $\Lambda(\boldi)$ for every $\boldi$, the underlying bipartite graph to be matched is determined by the choice of representative cycles.
    This in turn has a bearing on what kind of $\epsilon$-interleaved interval summands of $\pmodule$ and $\qmodule$ get matched. 
\end{compactitem}
\end{remark}

\begin{thm}[Hall's theorem]
\label{thm:hall}
Let $G$ be a finite bipartite graph on  sets $U$ and $V$. For a subset of vertices $U' \subset U$, let $N_G(U')$ denote the subset of $V$ adjacent to $U'$.
Then, the following are equivalent:
\begin{itemize}
\item for all $U' \subset U$, $|U'| \leq |N_G(U')|$
\item there exists an injective map $\imath: U \xhookrightarrow{} V$ such that $\imath$ maps every vertex $a$ of $U$ to a vertex $b$ of $V$ only if there is an edge $(a,b)$ in $G$. 
\end{itemize}
\end{thm}

Let ${\basisset}^P_{2\epsilon}   \subseteq {\basisset}^P$ be the cycles of ${\basisset}^P$ that represent intervals of length greater than $2\epsilon$. Likewise, let ${\basisset}^Q_{2\epsilon} \subseteq {\basisset}^Q$ be the cycles of ${\basisset}^Q$ that represent intervals of length greater than $2\epsilon$. 
Since $\pmodule$ and $\qmodule$ are $\epsilon$-interleaved, combining \Cref{lem:mainhall} and \Cref{thm:hall}, we obtain two injections $\imath: {\basisset}^P_{2\epsilon} \xhookrightarrow{ } {\basisset}^Q$ and 
$\jmath: {\basisset}^Q_{2\epsilon} \xhookrightarrow{ } {\basisset}^P$ such that 
\begin{align}
    \imath({\zeta_i}) &= {z_j} \quad \Longrightarrow \quad {\interval}^{\zeta_i}\text{ and }{\interval}^{z_j}  \text{ are } \epsilon\text{-interleaved and either } F_{\boldi,\boldj} \text{ or } G_{\boldj,\boldi} \text{ is nonzero.} \\
    \jmath({z_j}) &= {\zeta_i}  \quad \Longrightarrow \quad {\interval}^{\zeta_i} \text{ and }{\interval}^{z_j}  \text{ are } \epsilon\text{-interleaved and either } F_{\boldi,\boldj} \text{ or } G_{\boldj,\boldi} \text{ is nonzero.}
\end{align}

\begin{corollary}    
 \label{cor:maincore}
There is a matching of representative cycles of $\Cech(\pointsetone)$ with representative cycles of $\Cech(\pointsettwo)$ such that 
\begin{itemize}
    \item All persistent cycles of $\Cech(\pointsetone)$  and $\Cech(\pointsettwo)$ representing intervals of length greater than $2\epsilon$ are matched.
    \item If a representative $\zeta_i$ of  $\Cech(\pointsetone)$ is matched to a representative $z_j$ of $\Cech(\pointsettwo)$, then $\iinterval$ is $\epsilon$-interleaved with $\jinterval$, and either $F_{\boldi,\boldj}$ or $G_{\boldj,\boldi}$ is nonzero.
\end{itemize}
\end{corollary}
\begin{proof*}
The proof is essentially a paraphrase of the proof of Theorem~13.14 in Botnan's notes~\cite{botnan}. 

 Construct a bipartite graph $G=(V,E)$ with vertex set $V  = {\basisset}^P\bigcup {\basisset}^Q$. The vertices of ${\basisset}^P$ are colored red and the vertices of ${\basisset}^Q$ are colored blue. 
 The edges of $G$ are built from the two injections $\imath: {\basisset}^P_{2\epsilon} \xhookrightarrow{ } {\basisset}^Q$ and 
$\jmath: {\basisset}^Q_{2\epsilon} \xhookrightarrow{ } {\basisset}^P$. In particular, we have a directed edge $\zeta \de z$ if and only if $\imath(\zeta) = z$, and a directed edge $z \de \zeta$ if and only if $\jmath(z) = \zeta$. It is easy to check that every connected component in $G$ is either a directed cycle or a directed path. 

The matching $\matching$ is constructed as follows. 
For every cycle, pick alternate edges and include them in $\matching$. For every directed path, pick the odd numbered edges and include them in $\matching$. As a consequence, all vertices incident on some directed cycle are matched.
Also, all vertices on directed paths of odd length are matched. The only vertices that are not matched are the terminal vertices of paths with even length, and these terminal vertices are representative cycles for intervals of length smaller than $2\epsilon$. This shows that $\matching$ is an $\epsilon$-matching.

 Also, by construction of $\Lambda(A)$ and $\Upsilon(A)$ as described in \Cref{defn:candedge,defn:candedge}, we have a directed edge from interval $\boldi$ to $\boldj$ in $G$ only if  they are $\epsilon$-interleaved and either $F_{\boldi,\boldj}$ or $G_{\boldj,\boldi}$ is nonzero, which proves the second claim.
\end{proof*}

\begin{proposition} \label{lem:dieswhen}
For an interval $[b',d') \in B(\Cech(\pointsettwo))$ represented by  $z_j$, let $ \hat{g}_{\vphantom{1^{H}}b_j}([z_j]) = \sum_{i \in I} \kappa_{j,i} \cdot [\zeta_i]$,
where $I$ indexes a subset of representative cycles of $\Cech(\pointsetone)$.
Let $\Omega = \{\zeta_i \mid i\in I\}$.
Then, every cycle in the set  $\Omega$ dies at or before $d'+\epsilon$.
\end{proposition}
\begin{proof*}
Targeting a contradiction, suppose that there exists a partition of cycles  $\Omega = \Omega_1 \sqcup \Omega_2$, where the cycles in $\Omega_1$ die at or before $d'+\epsilon$, whereas the cycles in $\Omega_2$ die after $d'+\epsilon$. Then, Eqn.~~\ref{eqn:zj} can be written as 

\begin{equation}
   \sum_{\omega \in \Omega_2} [\omega] \, = \, \sum_{\gamma \in \Omega_1} [\gamma] \, + \,   \hat{g}_{\vphantom{1^{H}}b_j}([z_j]). \, 
\end{equation}

First, note that $\hat{g}_{\vphantom{1^{H}}b_j}([z_j])$ is trivial at $d'+\epsilon$ because $g$ is an $\epsilon$-perturbation and $z_{j}$ dies at  $d'$.
Then, at $d'+\epsilon$, the class on the left hand side, namely, $ \sum_{\omega \in \Omega_2} [\omega]$ is nontrivial because the cycles in $\Omega_2$ persist beyond $d'+\epsilon$, whereas the class on the right hand side, namely, $\sum_{\gamma \in \Omega_1} [\gamma] \, + \,   \hat{g}_{\vphantom{1^{H}}b_j}([z_j]) \,$ 
is trivial at $d'+\epsilon$ because it is a sum of trivial classes. This gives  the required contradiction.
Hence, every cycle in the set  $\Omega$ dies at or before $d'+\epsilon$.
\end{proof*}

\begin{thm}[Approximate stability for $\radmeasure$ for $\cechfull$ filtrations] \label{thm:mainstability}
Let $P$ be a point set embedded in a Euclidean space and $Q$ be an $\epsilon$-perturbation of $P$.
Then, the $\epsilon$-matching described in \Cref{cor:maincore} matches intervals of $B(\Cech(\pointsetone))$ to intervals of $B(\Cech(\pointsettwo))$ such that for every interval $[b,d) \in B(\Cech(\pointsetone))$ with length greater than $2\epsilon$, if the interval $[b',d') = \matching([b,d))$ has length greater than $2\epsilon$, then
\begin{align} 
  r^{2\epsilon}([b,d)) &\leq   r([b',d')) +  \epsilon, \label{eq:firststab} \\
    r^{2\epsilon}([b',d')) &\leq   r([b,d)) +  \epsilon. \label{eq:nextstab}
\end{align}
\end{thm}
\begin{proof*}
We will only prove \Cref{eq:firststab}. \Cref{eq:nextstab} follows from the symmetry of the argument. 
Let ${\basisset}^Q_{\star} = \{z_k \mid k\in [m] \}$ be an optimal persistent homology basis for $\Cech(\pointsettwo)$ and let ${\basisset}^P = \{\zeta_{\ell} \mid \ell \in [n] \}$ be an arbitrary persistent homology basis for $\Cech(\pointsetone)$. 
Suppose that $z_j$, a (optimal) representative cycle for the interval $[b_j,d_j) \in B(\Cech(\pointsettwo))$, is matched to  $\zeta_{i}$, a representative cycle for the interval $[b_i,d_i) \in B(\Cech(\pointsetone))$.

Recall that the matching in \Cref{cor:maincore} guarantees that ${\interval}^{z_j}$ and ${\interval}^{\zeta_{i}}$ are $\epsilon$-interleaved  and either $\eta_{i,j}\neq 0$ or $\kappa_{j,i}\neq 0$. 

If  $\kappa_{j,i} \neq 0$, then we write
$\hat{g}_{\vphantom{1^{H}}b_j}([z_j])$ as 
\begin{equation*}
     \hat{g}_{\vphantom{1^{H}}b_j}([z_j]) = \sum_{\ell \in I} \kappa_{j,\ell} \cdot [\zeta_{\ell}].  
\end{equation*}
where as before, $I$ indexes a subset of $[n]$ and for $\ell=i$, $\kappa_{j,\ell}\neq 0$.
Since  ${\interval}^{z_j}$ is  $\epsilon$-interleaved with ${\interval}^{\zeta_{i}}$, $b_i \in [b_j-\epsilon, b_j+\epsilon]$ and $d_i \in [d_j-\epsilon, d_j+\epsilon]$.
Then, using \Cref{lem:dieswhen}, every cycle $\zeta_{\ell}$ for $\ell \in I$ dies at or before $d_i+2\epsilon$.
Furthermore, every cycle $\zeta_{\ell}$ for $\ell \in I$ is born at or before $b_j+\epsilon$, and hence also at or before $b_i +2\epsilon$.  Therefore, $g_{\vphantom{1^{H}}b_j}^{\#}(z_j)$ is a valid (and possibly optimal) $2\epsilon$-approximate representative for $\zeta_{i}$. Since $g$ is an $\epsilon$-perturbation, by triangle inequality, $r( g_{\vphantom{1^{H}}b_j}^{\#}(z_j)) \leq r(z_j) + \epsilon$. Since the $\radmeasure$ of an optimal choice of $2\epsilon$-representative for $\zeta_{i}$ is upper bounded by  $r( g_{\vphantom{1^{H}}b_j}^{\#}(z_j))$, and is possibly even smaller than $r(g_{\vphantom{1^{H}}b_j}^{\#}(z_j))$, 
we have $r^{2\epsilon}([b,d)) \leq   r(g_{\vphantom{1^{H}}b_j}^{\#}(z_j)) $. Also, by definition, $r(z_j) = r([b',d'))$.
This proves the claim when $\kappa_{j,i}\neq 0$.

On the other hand, if  $\eta_{i,j} \neq 0$, we write
$\hat{f}_{\vphantom{1^{H}}b_i}([\zeta_i])$ as 
\begin{equation} \label{eqn:revmatch}
     \hat{f}_{\vphantom{1^{H}}b_i}([\zeta_i]) = \sum_{k \in J} \eta_{i,k} \cdot [z_{k}].  
\end{equation}
where as before, $J$ indexes a subset of $[m]$ and for $k=j$, $\eta_{i,j}\neq 0$.
Clearly, the cycles $z_{k}$ for $k\in J$ are born before $b_i + \epsilon$. Furthermore, it is easy to prove along the lines of \Cref{lem:dieswhen} that the cycles $z_{k}$ for $k\in J$ also dies before $d_i + \epsilon$.
Rewriting \Cref{eqn:revmatch} we obtain
\begin{equation*} \label{eqn:revmatchmod}
    [z_j] = \eta_{i,k} \cdot \eta_{i,j}^{-1}\sum_{k\in J'} [z_k] + \eta_{i,j}^{-1} \cdot \hat{f}_{\vphantom{1^{H}}b_i}([\zeta_i]) \quad  \text{ where } J' = J\setminus \{j\}.
\end{equation*}

Applying $\hat{g}_{\vphantom{1^{H}}b_i+\epsilon}$ to \Cref{eqn:revmatch} gives
\begin{align*}
    \hat{g}_{\vphantom{1^{H}}b_i + \epsilon}([z_j]) &= \eta_{i,k} \cdot \eta_{i,j}^{-1}\sum_{k\in J''}  \hat{g}_{\vphantom{1^{H}}b_i+\epsilon}([z_k]) + \eta_{i,j}^{-1}  \hat{g}_{\vphantom{1^{H}}b_i+\epsilon}\circ \hat{f}_{\vphantom{1^{H}}b_i}([\zeta_i]).\\
    &= \eta_{i,k} \cdot \eta_{i,j}^{-1}\sum_{k\in J''}  \hat{g}_{\vphantom{1^{H}}b_i+\epsilon}([z_k]) + \eta_{i,j}^{-1} ([\zeta_i]).
\end{align*}

Here, $J'' \subset J'$ indexes cycles $z_k$ with $k\in J'$ for which $\hat{g}_{\vphantom{1^{H}}b_i+\epsilon}([z_k])$ is not trivial.  
As before, $g_{\vphantom{1^{H}}b_{i}+\epsilon}^{\#}(z_j)$ is a valid (and possibly optimal) $2\epsilon$-approximate representative for $\zeta_{i}$. Since $g$ is an $\epsilon$-perturbation, by triangle inequality, $r( g_{\vphantom{1^{H}}b_{i}+\epsilon}^{\#}(z_j)) \leq r(z_j) + \epsilon$. Since the $\radmeasure$ of an optimal choice of $2\epsilon$-representative for $\zeta_{i}$ is bounded from above by $r( g_{\vphantom{1^{H}}b_{i}+\epsilon}^{\#}(z_j))$, 
we have $r^{2\epsilon}([b,d)) \leq   r(g_{\vphantom{1^{H}}b_{i+\epsilon}}^{\#}(z_j)) $. 
Therefore, for $\eta_{i,j}\neq 0$, we have
\begin{align*}
r^{2\epsilon}([b,d)) &\leq   r(g_{\vphantom{1^{H}}b_{i+\epsilon}}^{\#}(z_j)) \\
& \leq  r(z_j) + \epsilon \\
& = r([b',d') + \epsilon. \qedhere
\end{align*}
\end{proof*}

\paragraph*{Approximate stability for $\radmeasure$ for $\ripsfull$ complexes} 

As before, let $P$ be a point set embedded in a Euclidean space, and let $\pointsettwo$ be an $\epsilon$-perturbation of $\pointsetone$ realized through a bijective map between point sets, denoted by $f$. Also, let $g$ be the inverse of $f$.
Then, it is easy to check that for every $s\in \RBB$, the map $f$ (resp. $g$) induces an inclusion induced simplicial map $f_{s}: \Rips_s(\pointsetone) \to \Rips_{s+2\epsilon}(\pointsettwo)$ (resp. $g_{s}: \Rips_s(\pointsettwo) \to \Rips_{s+2\epsilon}(\pointsetone)$).
Moreover, the map $f_{s}$ (resp. $g_{s}$)  induces a map
on the respective cycle groups ${f}_{s}^{\#}: Z_{\ast}(\Rips_s(\pointsetone)) \to  Z_{\ast}(\Rips_{s+2\epsilon}(\pointsettwo))$ (resp. ${g}_{s}^{\#}: Z_{\ast}(\Rips_s(\pointsettwo)) \to  Z_{\ast}(\Rips_{s+2\epsilon}(\pointsetone))$) as well as a map 
$\hat{f}_{s}: H_{\ast}(\Rips_s(\pointsetone)) \to  H_{\ast}(\Rips_{s+2\epsilon}(\pointsettwo))$ (resp. $\hat{g}_{s}: H_{\ast}(\Rips_s(\pointsettwo)) \to  H_{\ast}(\Rips_{s+2\epsilon}(\pointsetone))$) on homology groups.

The proof strategy in \Cref{sec:repmatch} can be repeated more or less vertbatim to obtain the following result for $\ripsfull$ complexes. We leave out the details.

\begin{thm}[Approximate stability for $\radmeasure$  for $\ripsfull$ complexes] \label{thm:mainstabilityrips}
Let $P$ be a point set embedded in a Euclidean space and $Q$ be an $\epsilon$-perturbation of $P$.
Then, there exists an $2\epsilon$-matching that matches intervals of $B(\Rips(\pointsetone))$ to intervals of $B(\Rips(\pointsettwo))$ such that for every interval $[b,d) \in B(\Rips(\pointsetone))$ with length greater than $4\epsilon$, if the interval $[b',d') = \matching([b,d))$ has length greater than $4\epsilon$, then
\begin{align*} 
  r^{4\epsilon}([b,d)) &\leq   r([b',d')) +  \epsilon,   \\
    r^{4\epsilon}([b',d')) &\leq   r([b,d)) +  \epsilon. 
\end{align*}
\end{thm}

\begin{remark}
The astute reader may have noticed that statements analogous to \Cref{thm:mainstability} can be obtained for many commonly encountered filtrations including Delaunay and lower star. To obtain the respective statements for each of these filtrations, the changes required to the proof of \Cref{thm:mainstability} are routine, and hence we do not discuss the topic of extensions of \Cref{thm:mainstability} any further. 
\end{remark}

\cancel{
In the above proof, let $Z_1 =  \{g^{\#}_{\vphantom{1^{H}}b_j}(z_j) \mid \kappa_{j,i} \neq 0\}$ and  $Z_2 = \{ g^{\#}_{\vphantom{1^{H}}b_i + \epsilon}(z_j) \mid \eta_{i,j}\neq 0\}$ and 
$Z = Z_1 \cup Z_2$. Then, it is easy to check that $Z$ is a $2\epsilon$-approximate persistent basis for $\Cech(P)$. 

}



\cancel{
\subsection{Computing optimal {$2\epsilon$}--approximate representatives for intervals} \label{sec:apxoptimal}
Let $\zeta$ be a $p$-cycle that is a representative for some interval $\ibd$. Next, we describe a method to compute an optimal approximate representative for $\ibd$. In this section, we use  slightly different notation for extensions and boundary matrices for the sake of readability.
Let $\hombasis$ be the extension of an arbitrary $p$-th homology basis for $\complex_{b+2\epsilon}^{V}$ in $\complex_{b+2\epsilon}$, and let $\partial_{b+2\epsilon}^V$ be the extension of the $(p+1)$-th boundary matrix for $\complex_{b+2\epsilon}^{V}$ in $\complex_{b+2\epsilon}$. 
Any extension of $p$-cycle of $\complex_{b+2\epsilon}^{V}$ in $\complex_{b+2\epsilon}$ can be written as $\hombasis \cdot x + \partial_{b+2\epsilon}^V\cdot y$.  
Let $\basisset$ be the cycles in a persistent homology basis of $\Cech(P)$ that are born at or before $b+2\epsilon$ and die at or before $d+2\epsilon$. 
Let $\partial_{b+2\epsilon}$ be the  $(p+1)$-th boundary matrix for $\complex_{b+2\epsilon}$. 
Then, there exists a approximate representative for $\ibd$ in the complex induced by $V$ if and only if the following linear equation has a solution:

\begin{equation} \label{eq:apxopt}
    \basisset \cdot w +  \partial_{b+2\epsilon} \cdot z =  \hombasis \cdot x + \partial_{b+2\epsilon}^V\cdot y.
\end{equation}

\cancel{
\abhishek{A more refined notion of approximate representatives would give sharper bounds. We need to discuss this.}
}
Given an interval $[b,d)$, if \Cref{eq:apxopt} is satisfied for some vertex set $V$, then the  $\radmeasure$ of $\complex_V$ that supports the desired cycle is determined by the sphere enclosing $V$. The optimal $2\epsilon$-approximate representative can be found by searching over vertex sets $V$ across various levels of the Rhomboid tiling by employing \Cref{alg:bin-search}. 




~


Let $r$ denote the radius for an optimal representative of $\ibd$, and let $r'$ denote the radius for a $2\epsilon$--approximate representative of  $\ibd$. As elaborated upon in \Cref{sec:algstab}, if $r - r'$ is small, then radius computation for $\ibd$ is stable. 

\cancel{
\subsection{Algorithmic determination of exact (one-way) stability for $\radmeasure$ }
\label{sec:algstab}
We note that \Cref{thm:mainstability} can be employed to obtain some stability guarantees for $r([b,d))$ for small perturbations of an input point set $P \subset \RBB^{d}$.
This can be achieved as follows: In \Cref{sec:phbasis}, we describe a method to compute $r([b,d))$ for an interval $[b,d)$ of $B(\Cech(P))$. In \Cref{sec:apxoptimal}, we describe an algorithm to compute $r^{2\epsilon}([b,d))$. Now, if $r([b,d)) - r^{2\epsilon}([b,d)) = \delta'$ for some $\delta'$, then \Cref{eq:firststab} can be written as 
\begin{equation} \label{eq:secondstab}
  r([b,d)) \leq   r([b',d')) + \epsilon + \delta'. 
\end{equation}
In other words, if $\epsilon + \delta'$ is sufficiently small and within an acceptable tolerance limit, then for any $\epsilon$-perturbation $\pointsettwo$ of  of $\pointsetone$, $[b,d)$ is matched to some interval $[b',d')$ whose radius is not substantially smaller than $[b,d)$ (and possibly larger). If one prescribes a tolerance $\delta$ of instability for $r([b,d))$, then one can do a binary search on $\epsilon$ until one finds a  $\delta'$ small enough to satisfy $ \epsilon + \delta' < \delta$ and hence
$ r([b,d)) \leq   r([b',d')) + \delta$ (as desired).
In other words, for an $(\epsilon,\delta)$ pair that satisfies $ \epsilon + \delta' < \delta$,  irrespective of which $\epsilon$--perturbation function $h$ one uses, an interval $[b,d)$ of length greater than $2\epsilon$ is guaranteed to be matched to an interval of the barcode of the perturbed set of points whose $\radmeasure$ almost upper bounds $r([b,d))$.

On the other hand, at this point, we do not know yet how to choose an $\epsilon$ to obtain an algorithmic guarantee for the equation 
\begin{equation} \label{eq:thirdstab}
  r([b',d')) \leq   r([b,d)) +  \delta.
\end{equation}
Hence, algorithmically we  can only ascertain one-way $(\epsilon,\delta)$-stability.

}

}
\section{Correctness of Algorithms}\label{sec:algcorrectness}
In this section we give proofs for the correctness of algorithms stated and other associated results.
The standard reduction algorithm ~\cite{bauer2017phat} will be used in many of our algorithms as subroutines.
We begin by outlining the standard reduction algorithm (\Cref{alg:standardReduction}) and recalling some facts that arise out of it. For any matrix $\partial$ with entries in $\mathbb{Z}_2$ we define $low(j)$ to the row index of the lowest 1 in column $j$ of $\partial$. It is undefined if column $j$ is 0.

\begin{remark}
    For a $n \times m$ matrix standard reduction runs in $O(n^2m)$ time.
\end{remark}

\begin{algorithm}[h]
\caption{Standard reduction algorithm for matrix reduction~\cite{bauer2017phat}}\label{alg:standardReduction}

\SetKwInOut{Input}{Input}\SetKwInOut{Output}{Output}
\vspace{1.5mm}
\Input{Matrix $\partial$ with $m$ columns and entries in $\mathbb{Z}_2$}
\Output{
$\Tilde{\partial}$( the reduced matrix), $V$ (a collection of cycle-vectors).
$\Tilde{\partial}$ has the property that $low(j)$ is unique for every non-zero column of $\Tilde{\partial}$.
If $\partial$ is the boundary matrix of a filtration and if $i_1, \ldots i_\beta$ are the unpaired column indices then cycle vectors represented by $V_{i_1}, \ldots, V_{i_\beta}$ are the essential cycles of the filtration}

\SetKwFunction{standardreduction}{StandardReduction}
\SetKwProg{myproc}{Procedure}{}{}
\myproc{\standardreduction{$\partial$}}{
{$V \gets I_m$}\\
\For{$j = 1$ to $m$ }{
\While{there exists $j_0 < j$ with $low(j_0) = low(j)$}
{{Add column $\partial_{j_0}$ to $\partial_{j}$}\\
{Add column $V_{j_0}$ to $V_{j}$}\\
}
}
\For{$j = 1$ to $m$ }{
\If{$\partial_j \neq 0$}{
{\textit{pair} $(low(j),j)$}
}
}
{Return $V$}
}
\end{algorithm}

For complex $\complex$ and the filtration $\filtrationv_v$ let $\zeta_1, \ldots, \zeta_s$  be the essential $p-$cycles computed by the standard reduction algorithm with $\maxsmplx(\zeta_1) \prec_v \ldots \prec_v \maxsmplx(\zeta_s)$.
The essential cycles form a basis of the $p^{th}$ homology of $\complex$ that is, $[\zeta_1], \ldots, [\zeta_s]$ form a basis of $H_p(\complex)$.
Let $\partial$ be the boundary matrix corresponding to $\filtrationv_v$ and $\Tilde{\partial}$ be the reduced matrix.
If $\Tilde{\partial}_i$ is $0$ and $i$ is a paired index then the cycle $V_i$ is a $p-$boundary in $\complex$.
If $j$ is an index in the filtration where the simplex $\tau$ appears and $\complex_j$ the subcomplex consisting of all simplices till index $j$ of the filtration, then the cycles $Z_{j} =  \{V_i \,|\, \textnormal{i is an index that corresponds to a p-simplex, } \Tilde{\partial}_i = 0, \, i \leq j  \}$ form a basis of $Z_p(\complex_{j})$.

We now state a few propositions that follow from the standard reduction algorithm.

\begin{proposition}\label{prop:standardred0}
Let $V_i \in Z_j$ be a p-cycle.
Assume that $i$ is a paired index, that is, $V_i$ is not an essential cycle.
Let $V_{i_1}, \ldots, V_{i_s}$ be the essential $p-$cycles in $Z_j$.
Then $[V_i] \in span\{[V_{i_1}], \ldots, [V_{i_s}]\}$ in $H_p(\complex)$.
\end{proposition}
\begin{proof*}
Let $j_1, \ldots, j_t = i$ be the paired indices $\leq i$.
Assume $j_1 < \ldots < j_t$.
We proceed by induction on $t$.
Since $j_1$ is paired there exists a  $p-$boundary $\Tilde{\partial}_{j'_1}$ such that $low(\Tilde{\partial}_{j'_1}) = j_1$.
Therefore $low(\Tilde{\partial}_{j'_1} + V_{j_1}) < j_1$ and so $\Tilde{\partial}_{j'_1} + V_{j_1} \in Z_p(\complex_{j_1-1})$.
But $Z_p(\complex_{j_1-1}) \subset span\{V_{i_1}, \ldots, V_{i_s}\}$ and so $[V_{j_1}] \in span\{[V_{i_1}], \ldots, [V_{i_s}]\}$.
For $t > 1$ argueing as before we have $\Tilde{\partial}_{j'_t} + V_{j_t} \in Z_p(\complex_{j_t-1})$.
But $Z_p(\complex_{j_t-1}) \subset span\{V_{i_1}, \ldots, V_{i_s}, V_{j_1}, \ldots, V_{j_{t-1}}\}$ and so by the inductive hypothesis $[V_{j_t}] \in span\{[V_{i_1}], \ldots, [V_{i_s}]\}$.
\end{proof*}

\begin{proposition}\label{prop:standardred1}
    Let $\zeta_1 \prec_v \ldots \prec_v \zeta_s$ be the essential $p-$cycles  computed by the standard reduction algorithm for $\filtrationv_v(\complex)$. If $\zeta \in Z_p(\complex)$ be such that $\zeta_i \prec_v \zeta \prec_v \zeta_{i+1}$ then  $[\zeta] \in span\{[\zeta_1] \ldots [\zeta_i]\}$.    
\end{proposition}
\begin{proof*}
    $\zeta_i \prec_v \zeta \prec_v \zeta_{i+1} \implies \maxsmplx(\zeta_i) \prec_v \maxsmplx(\zeta) \prec_v \maxsmplx(\zeta_{i+1})$.
    Let $\tau = \maxsmplx(\zeta)$ and $j$ be the index where $\tau$ appears.
    $Z_{j} \supset \{\zeta_1, \ldots \zeta_i\}$ forms a basis of $ Z_p(\complex_j)$.
    Write $\zeta = a_1\zeta_1 + \ldots + a_i\zeta_i + b$ where $b$ is a cycle spanned by $Z_{j} \setminus \{\zeta_1, \ldots \zeta_i\}$.
    By \cref{prop:standardred0} for each $\xi \in Z_{j} \setminus \{\zeta_1, \ldots \zeta_i\}, [\xi] \in span\{[\zeta_1] \ldots [\zeta_i]\}$ in $H_p(\complex)$.
\end{proof*}

\begin{proposition}
\label{prop:standardred2}
    Let $\zeta_1 \prec_v \ldots \prec_v \zeta_s$ be as in \Cref{prop:standardred1}. 
    Let $\zeta \textnormal{ be a p-cycle, } [\zeta] \neq 0 \:in\: H_p(\complex)$.
    such that $\zeta = \zeta_{i_1} + \ldots + \zeta_{i_m} + \partial c_{p+1}$ where each $\zeta_{i_k} \in \{\zeta_1 ,\ldots, \zeta_s\}$ and $c_{p+1}$ is a $p+1$ chain. If $i_1 <  \ldots < i_m$ then $r_v([\zeta]) = r_v(\zeta_{i_m})$.
    In particular, $\zeta_{i_1} + \ldots + \zeta_{i_m} \in \arg\min_{\xi \in [\zeta]}{r_{v}(\xi)}$.
\end{proposition}
\begin{proof*}
    First note that $\maxsmplx(\zeta_{i_1}) \prec_v \ldots \prec_v \maxsmplx(\zeta_{i_m})$.
    So $r_v(\zeta_{i_1} + \ldots + \zeta_{i_m}) = r_v(\maxsmplx(\zeta_{i_m})) = r_v(\zeta_{i_m})$.
    We prove by contradiction.
    Let $\hat{\zeta} \in \arg\min_{\eta \in [\zeta]}r_v{ \eta}$.
   If $r_v(\hat{\zeta}) < r_v(\zeta_{i_m}) \implies \maxsmplx(\hat{\zeta}) \prec_v \maxsmplx(\zeta_{i_m}) \implies \hat{\zeta} \prec_v \zeta_{i_m}$.
    By the \Cref{prop:standardred1} we would have $[\hat{\zeta}] \in span\{[\zeta_1] \ldots [\zeta_{i_{m}-1}] \}$. 
    But by assumption $[\hat{\zeta}] = [\zeta_{i_1}] + \ldots + [\zeta_{i_m}]$.
\end{proof*}

\cancel{\begin{algorithm}[h]
\caption{Solve with Reduction}\label{alg:solvebyreduction}
\begin{algorithmic}[1]
\State{Description: Finds a solution of $A.x = b$. Reports if the system has a solution and if under-determined computes the lexicographically smallest solution}
\Procedure{SolveByReduction}{$A$, $b$}
\State{$C \gets [A | b]$. Let $s$ be the index of the last column of $C$}
\State{$V \gets \textsc{StandardReduction}(C)$}\\
\If{$C_s \neq 0$}{
\State{Let $j_1, \ldots , j_q$ be the row indices of $V_s$ that are $1$.
$soln \gets \{j_1, \ldots , j_{q-1}\}$}
\State{Return true, $soln$}
}\\
\Else{
\State{Return false}.
}
\EndProcedure
\end{algorithmic}
\end{algorithm} 
}
\begin{algorithm}[h]
    \caption{Solve with Reduction}\label{alg:solvebyreduction}
    \SetKwInOut{Input}{Input}\SetKwInOut{Output}{Output}
    \Input {$A(\mathbb{Z}_2 \textnormal{ matrix}), b(\mathbb{Z}_2 \textnormal{ vector})$}
    \Comment{computes a solution to $Ax=b$ if one exists}\\
    \Output {If $x$ is the solution computed, returns the indices of $x$ equal to $1$, returns \textit{false} if no solution exists. }
    \SetKwFunction{solvebyreduction}{SolveByReduction}
    \SetKwProg{myproc}{Procedure}{}{}
    \myproc{\solvebyreduction{$A,b$}}{
    {$C \gets [A | b]$. Let $s$ be the index of the last column of $C$} \\
    $V \gets \textsc{StandardReduction}(C)$. Continue to denote the reduced matrix by $C$.\\
    \If{$C_s = 0$}
    {
     Let $j_1, \ldots , j_q$ be the row indices of $V_s$ that are $1$.
    $soln \gets \{j_1, \ldots , j_{q-1}\}$ \\
    Return true, $soln$
    }
    \Else{
    Return false.}
    }
\end{algorithm}

\cancel{
For a cycle $\zeta, \, [\zeta] \in H_p(\complex) \neq 0$ and a set of \textit{sites} $S$ (a collection of points in $\mathbb{R}^d$), we define $r_P([\zeta]) = \min_{\eta \in [\zeta], v \in S}{r_v(\eta)}$.
Note that when $S = \mathbb{R}^d,\, r_P([\zeta] = r([\zeta]))$. 
Therefore $r_P([\zeta])$ is a \textit{restricted} $\radmeasure$ with the centers of the spheres $S_{c,\delta}$ constrained to lie in $S$.
}

\subsection{Optimal Homology Basis}
\label{sec:correctness_mhb}
\begin{proposition}
    Let $\zeta_1, \ldots, \zeta_m$ be the cycles in $\minhombasis$ such that $\zeta_1 \prec_{\Omega} \ldots \prec_{\Omega} \zeta_m$. 
    Let $\zeta'_1, \ldots, \zeta'_m$ be any collection of cycles such that $span\{[\zeta'_1], \ldots, [\zeta'_m]\} = H_p(\complex)$.
    Assume further that if $r_P([\zeta'_i]) < r_P([\zeta'_j])$ then $i < j$.
    Then $r_P(\zeta_j) \leq r_P([\zeta'_j]), 1 \leq j \leq m$.
\end{proposition}

\begin{proof*}
    Let $\zeta_j$ correspond to cycle $\zeta_{u,\Tilde{j}}, u \in S$ in the list $\Omega$. 
    Let $\Omega_{-} = \{\eta \in \Omega| \, \eta \prec_{\Omega} \zeta_j\}$ and $\Omega_{+} = \{\eta \in \Omega| \, \zeta_j \prec_{\Omega} \eta \} \cup \{\zeta_j\}$.
    Let $[\Omega_{-}] = span\{[\eta], \eta \in \Omega_{-} \}$.
    First note that $r_P([\zeta_j]) = r_u(\zeta_j)$, for if $u^* \in \arg\min_{x \in S}{r_x([\zeta_j])}$ and say $\zeta_j = \zeta_{u^*,k_1} + \ldots + \zeta_{u^*,k_q} + b', \zeta_{u^*,k_1} \prec_{u^*} \ldots \prec_{u^*} \zeta_{u^*,k_q}, b' \in B_p(\complex)$, then $\zeta_{u,\Tilde{j}} \prec_{\Omega}  \zeta_{u^*,k_q}$.
    (This is because if $\zeta_{u^*,k_q} \in \Omega_{-} \implies [\zeta_j] \in [\Omega_{-}]$. But $[\zeta_j] \notin [\Omega_{-}]$).
    Thus $r_u(\zeta_{u,\Tilde{j}}) \leq r_{u^*}(\zeta_{u^*,k_q}) = r_P([\zeta_j])$.
    
    \noindent Now assume the contrary, that is, $r_P([\zeta'_j]) < r_P([\zeta_j]$.
    Let $v \in \arg\min_{v \in S}{r_v([\zeta'_j])}$.
    Let $\zeta'_j = \zeta_{v,i_1} + \ldots + \zeta_{v,i_s} + b$, where $b \in B_p(\complex)$ and $\zeta_{v,i_1} \prec_v \ldots \prec_v \zeta_{v,i_s}$
    By \Cref{prop:standardred2}, $r_v([\zeta'_j]) = r_v(\zeta_{v,i_s}) = r_P([\zeta'_j])$.
    $r_P([\zeta'_j]) < r_P([\zeta_j] \implies r_v(\zeta_{v,i_l}) < r_P(\zeta_j), \, 1\leq l \leq s$.
    We have $\zeta_{v,i_1} ,\ldots , \zeta_{v,i_s} \in \Omega_{-}$.
    Clearly $dim([\Omega_{-}]) = j-1$.
    The classes $\{[\zeta'_1], \ldots [\zeta'_{j-1}]\}$ cannot all be in $[\Omega_{-}]$ as $[\zeta'_j] \in [\Omega_{-}]$ and the classes $\{[\zeta'_1], \ldots [\zeta'_{j}]\}$ are assumed to be linearly independent.
    Let $\zeta'_k, k < j$ be such that $[\zeta'_k] \notin [\Omega_{-}]$.
    Let $w \in \arg\min_{x \in S}{r_x([\zeta'_k])}$.
    Let $\zeta'_k = \zeta_{w,j_1} + \ldots + \zeta_{w,j_t} + b'$ where $b' \in B_p(\complex)$ and $\zeta_{w,j_1} \prec_w \ldots \prec_w \zeta_{w,j_t}$.
    Then $[\zeta'_k] \notin \Omega_{-} \implies \zeta_{w,j_t} \in \Omega_{+}$.
    In particular, $r_P([\zeta'_k]) = r_w(\zeta_{w, j_t}) \geq r_u(\zeta_j) = r_P([\zeta_j]) > r_P([\zeta'_j])$.
    This violates the assumption that the cycles $\{\zeta'_i\}$ are ordered by $r_P([\zeta'_i])$.
\end{proof*}

\subsection{Minimal Persistent Homology Basis}
\label{sec:correctness_persmhb}
In the following propositions we assume a simplex-wise filtration $\filtration$ on $\complex$, $\partial_{\filtration}$ is the corresponding boundary matrix, $[b,d) \in \barcode_p(\filtration)$ and $\sigma_b$ is the simplex added at index $b$ of $\filtration$.
\cancel{
\begin{proposition}
    If $\omega$ is a $p-$boundary in $\complex_b$ then $\sigma_b$ is not incident on $\omega$.
\end{proposition}
\begin{proof*}  
    Let $V_{\filtration}$ be the output representative cycles of the standard reduction algorithm on $\partial_{\filtration}$.
    Let $\widetilde{\partial_{\filtration}}$ be the reduced boundary matrix.
    If $b$ be the index in $\filtration$ where $\sigma_b$ appears then the cycle $\xi = V_{\filtration,b} \in \represents([b,d))$.
    In particular $\xi$ is not a $p-$boundary in $\complex_b$.
    The set $Z_{\leq b} = \{V_{\filtration,j} \,|\, \Tilde{\partial_{\filtration,j}} = 0, j\leq b \}$ is a basis for $Z_p(\complex_b)$.
    $\xi$ not being a boundary, $Z_{\leq b} \setminus \{\xi\}$ contains a basis of $B_p(\complex_b)$.
    But $\sigma_b$ is not incident on any cycle in $Z_{\leq b} \setminus \{\xi\}$.
\end{proof*}
}
\begin{proposition}
    If $\omega$ is a cycle in $\complex_b$ such that $\sigma_b$ is incident on $\omega$ then $\omega$ is not a boundary in $\complex_{d-1}$.
\end{proposition}
\label{prop:persistence_notboundary}
\begin{proof*}
    Let $\partial_\filtration$ be the boundary matrix corresponding to $\filtration$, $\widetilde{\partial_\filtration}$ be the reduced boundary matrix after applying standard reduction on $\partial_\filtration$ and $V$ be the output cycle vectors.
    For any $j$ that corresponds to an index in $\filtration$ the non-zero columns of $\widetilde{\partial_\filtration}$ with index $\leq j$ form a basis for the boundaries of $\complex_j$.
    Let $\rho(\sigma)$ be the index in $\filtration$ where the simplex $\sigma$ appears.
    For a cycle $\zeta$, let $\rho(\zeta) = \max_{\sigma \in \zeta}{\rho(\sigma)}$.
    Since $\omega \in Z_p(\complex_b)$ and $\sigma_b \in \omega \implies \rho(\omega) = b$.
    Since $\filtration$ is simplex-wise a death index $d$ can occur in at most 1 bar.
    Therefore $ \nexists \, \textnormal{a boundary } \widetilde{\partial_{\filtration,i}}, i < d$ with $\rho(\widetilde{\partial_{\filtration,i}}) = b $.
    Since all non-zero columns $\widetilde{\partial_{\filtration,i}}$ have unique values of $\rho$, if $\eta$ is a boundary that is a linear combination of cycles $\{ \widetilde{\partial_{\filtration,i}} \}, i < d$, then $\rho(\eta) \neq b$.
    It follows that $\omega$ is a not a boundary in $\complex_{d-1}$.    
\end{proof*}

\begin{proposition}
\label{prop:persistence_first_i}
    Let ${\zeta_1} \prec_v \ldots \prec_v {\zeta_m}$ be the essential $p-$cycles of $\filtrationv_v(\complex_b)$ computed using standard reduction. Let $\xi^* \in \arg\min_{\eta \in \represents([b,d))}\{r_v(\eta)\}$. If $[\xi^*] \in span\{[\zeta_1], \ldots , [\zeta_i]\}$ in $H_p(\complex_b)$, then $\exists \, \zeta^* \in span\{\zeta_1, \ldots , \zeta_i\}, \zeta^* \in  \arg\min_{\eta \in \represents([b,d))}\{r_v(\eta)\}$.
\end{proposition}
\begin{proof*}
    Let $\xi^* = a_1\zeta_1 + \ldots + a_i\zeta_i + b$ where $a_i = 1$ and $b \in B_p(\complex_b)$. Then $\kappa(\xi^*) \nprec_v \kappa(\zeta_i)$ for otherwise by \Cref{prop:standardred1} we would have $[\xi^*] \in span\{[\zeta_1], \ldots, [\zeta_{i-1}]\}$ in $H_p(\complex_b)$. It follows that $\zeta^* = \xi^* + b  = a_1\zeta_1 + \ldots + a_i\zeta_i \in \arg\min_{\eta \in \represents([b,d))}\{r_v(\eta)\}$.
\end{proof*}

\paragraph{Correctness of \cref{alg:2-approx_min_persistence}}
\Cref{alg:2-approx_min_persistence} correctly computes a representative of the input bar $[b,d)$ as the cycle contains $\sigma_b$ and is a boundary at $\complex_d$. 
By \cref{prop:persistence_notboundary} it is not a boundary in $\complex_{d-1}$.
By \cref{prop:persistence_first_i} the subroutine \textsc{Opt-Pers-Cycle-Site} correctly computes $\zeta^*_v$ for each site $v$.

    
\section{Further Experimental Results}
\label{sec:furtherexperiments}
\subsection{Homology Localization}
 Our localization algorithm is robust to these inputs and produces cycles that correspond to the relevant geometric features in the 3D models. We can visually infer for the Wheel and Happy Buddha datasets that the localized cycle computed by our approximate algorithm is nearly equal to the optimal cycle. Note that even though the input cycle spans both tunnels in the Happy Buddha dataset, the localized cycle is a sum of the two disjoint cycles. These experimental results demonstrate how the localization algorithm may be utilized as a supplement to existing methods that compute cycle representatives.

 \begin{figure*}[!htb]
\centering
\begin{tabular}{ccc}
\includegraphics[height=1.6in]{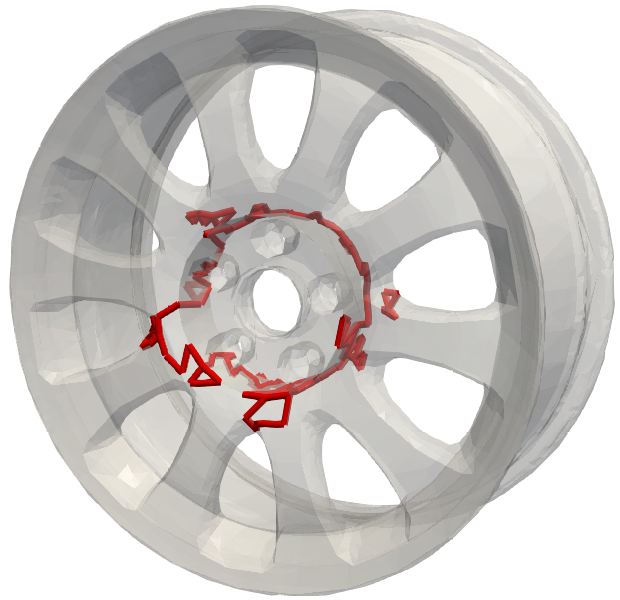} &
\includegraphics[height=1.7in]{Images/HomLoc/buddha_rand_cyc.png} &
\includegraphics[height=1.7in]{Images/HomLoc/buddha_local_cyc.png}\\ 
\includegraphics[height=1.6in]{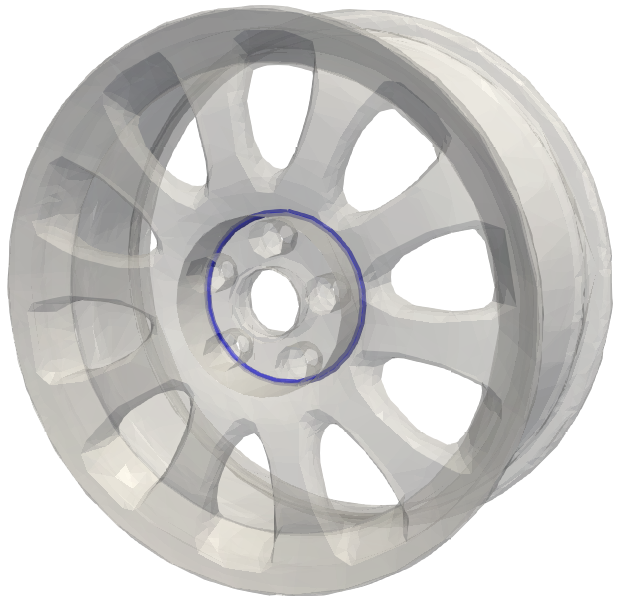} &
\includegraphics[height=1.7in]{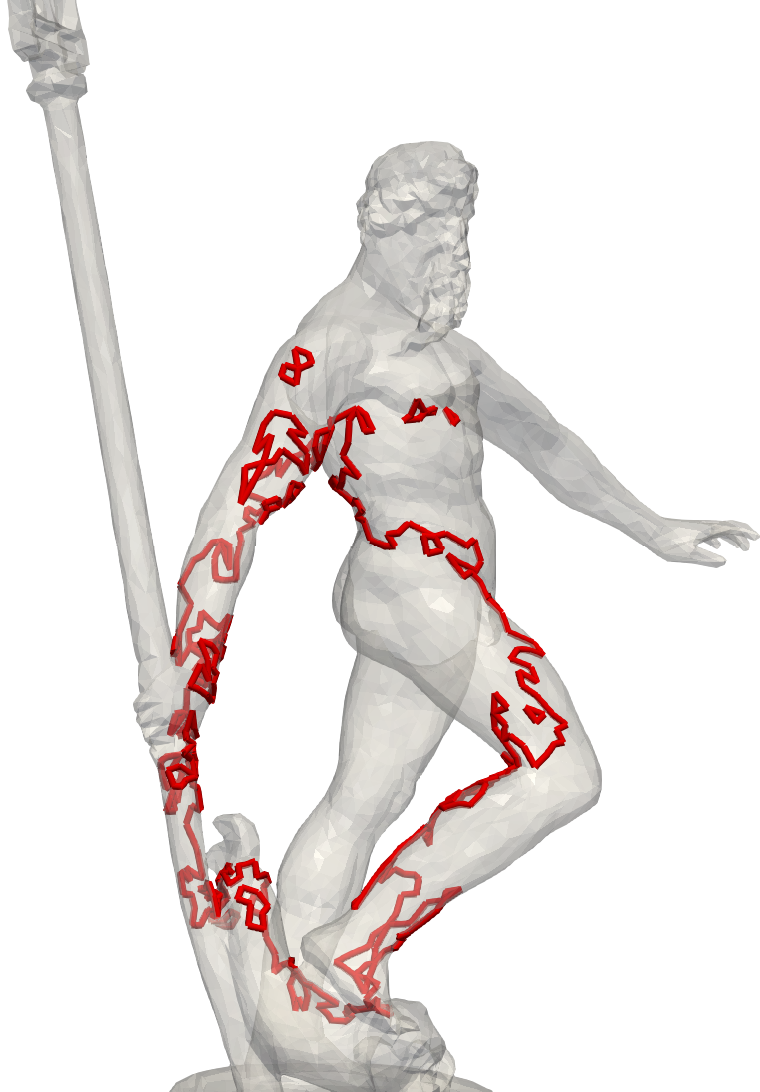} &
\includegraphics[height=1.7in]{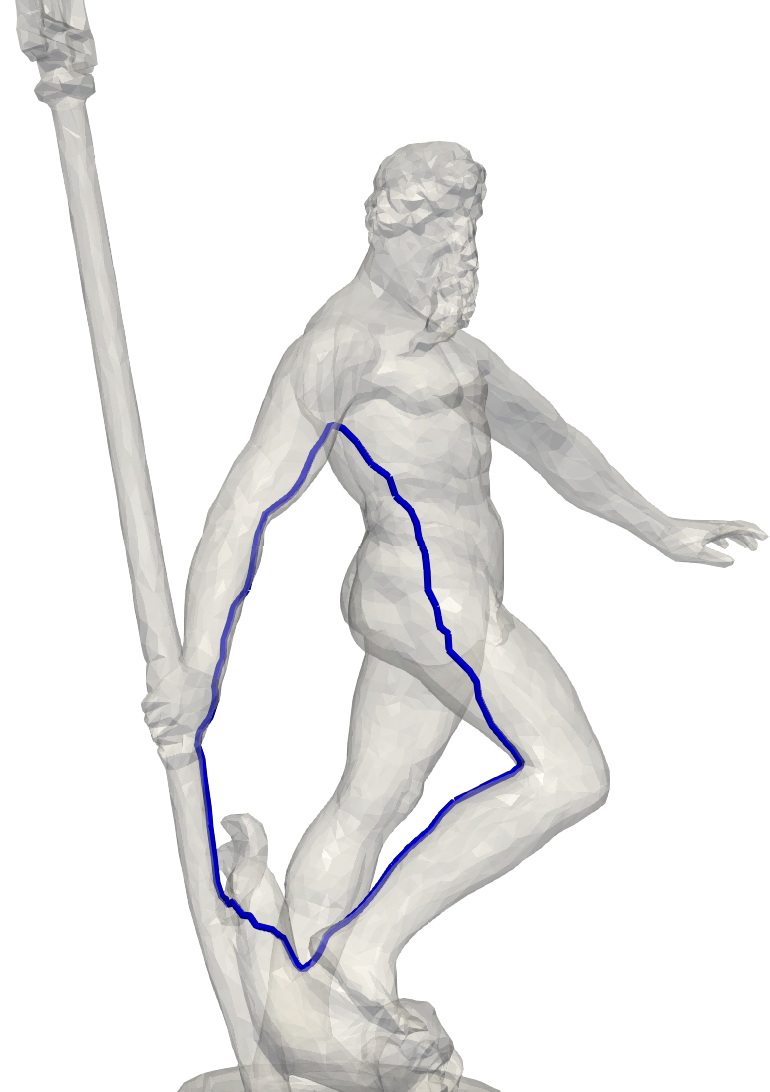}
\end{tabular}
\caption{The localization algorithm computes optimal (blue) 1-cycles that are homologous to input 1-cycles (red).}
    \label{fig:hom_localization}
\end{figure*}

\newpage 

\subsection{Optimal homology basis}
\begin{figure*}[!hbt]
\centering
\fbox{\includegraphics[height=1.8in]{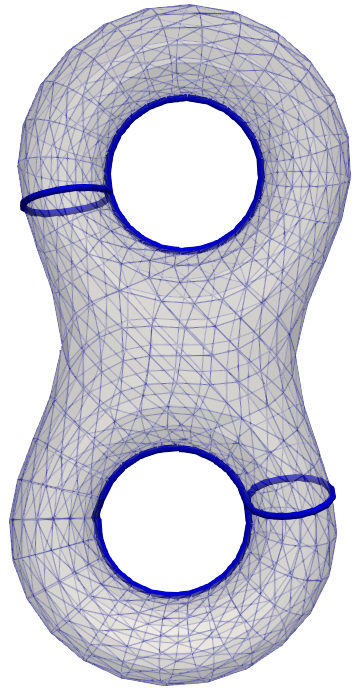}
\includegraphics[height=1.8in]{Images/HomBasis/670_hombasis.png}
\includegraphics[height=1.8in]{Images/HomBasis/1457_hombasis.png}
}  
\caption{Optimal 1-homology basis computed using the localization algorithm.}
\label{fig:optimal_homology_basis}
\end{figure*}

\subsection{Optimal Persistent Homology Representatives}
\begin{figure*}[!htb]
\centering
\begin{tabular}{ccc}
\includegraphics[height=1.8in]{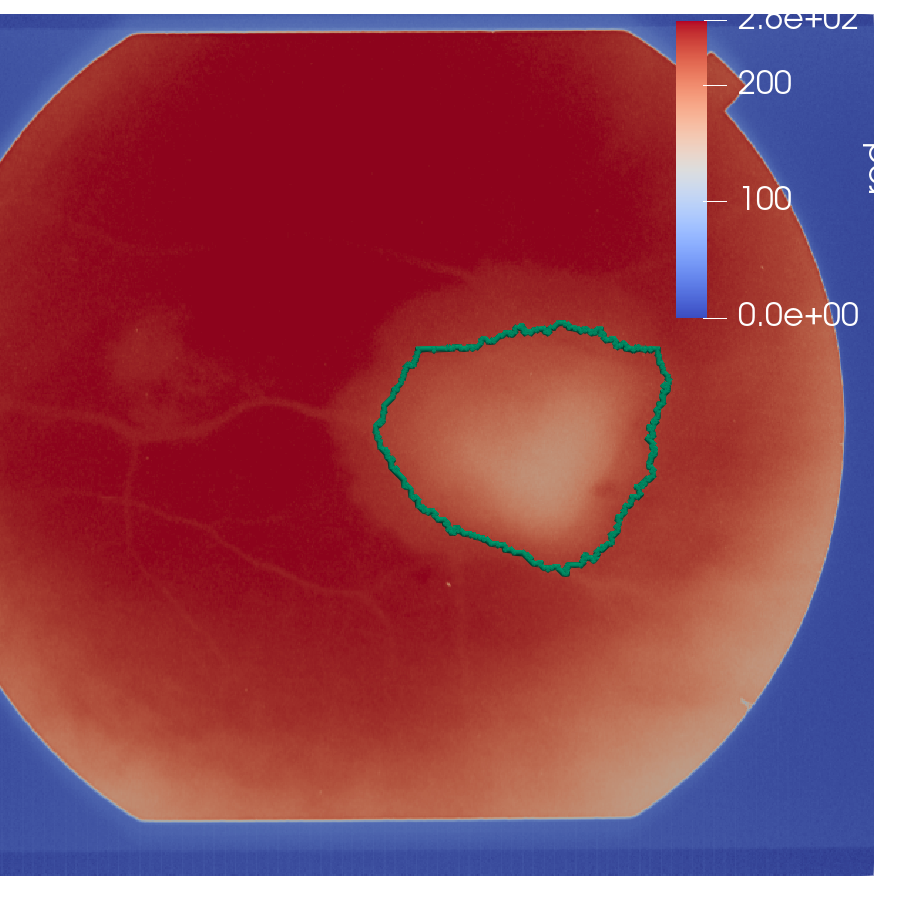} &
\includegraphics[height=1.8in]{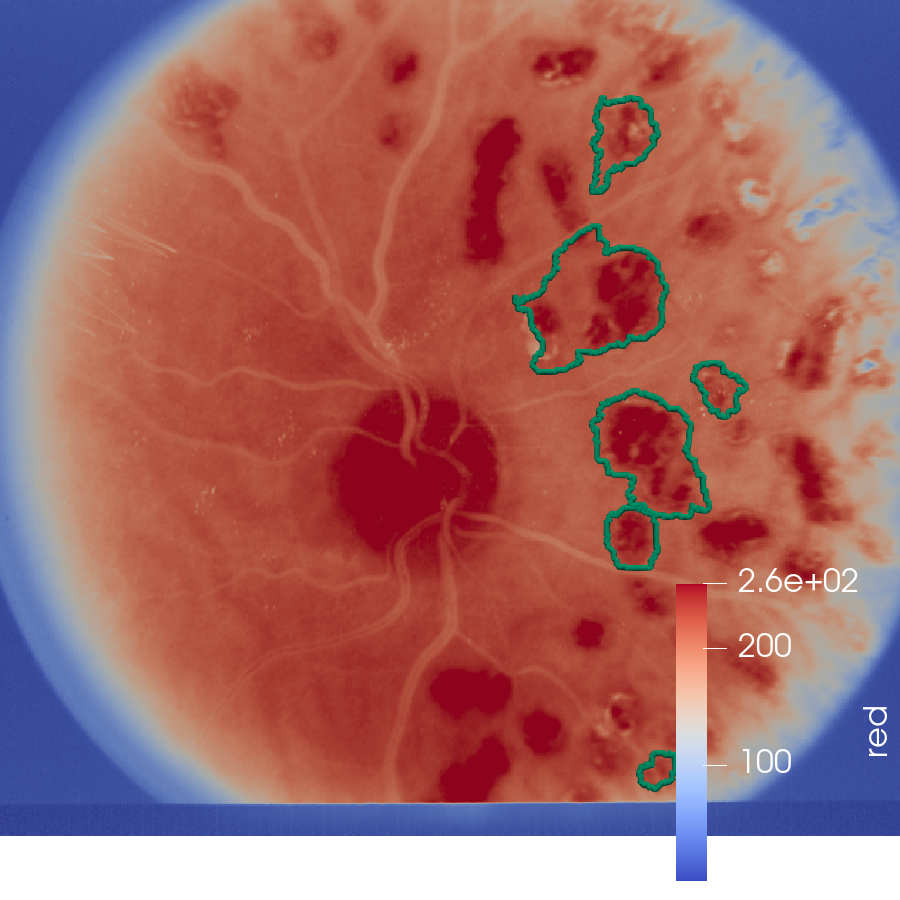} &
\includegraphics[height=1.8in]{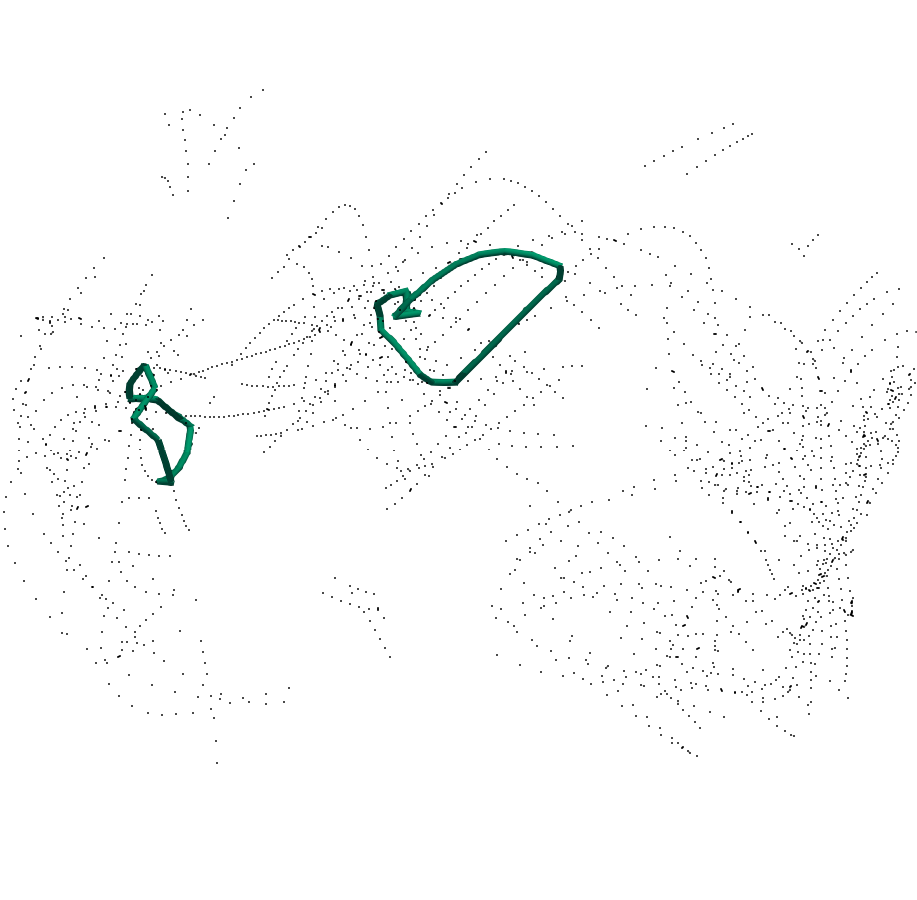}\\ 
\textbf{(a)} & \textbf{(b)} & \textbf{(c)}
\end{tabular}
\caption{\textbf{(a)}Persistent homology representative of longest bar bounding a region of disorder in the retinal image. \textbf{(b)} Representatives of few of the top-most bars indicating regions of disorder in the retinal image. \textbf{(c)} Top 2 representatives of the Lorenz'96 dataset(20\%)}
    \label{fig:suppl_persmhb}
\end{figure*}

\end{document}